\documentclass[twocolumn]{aastex62}%
\usepackage{amssymb}
\usepackage{textcomp}
\usepackage{tabularx}
\usepackage{longtable}
\usepackage{multirow}
\usepackage{array}
\usepackage{color}
\usepackage{longtable}
\usepackage{subfigure}
\usepackage{times}
\usepackage{epsfig}
\usepackage[T1]{fontenc}
\usepackage{ae,aecompl}

\received{January 1, 2020}
\revised{January 1, 2020}
\accepted{\today}

\submitjournal{ApJ}

\usepackage{multirow,hhline,graphicx,array}
\newcolumntype{M}[1]{>{\centering\arraybackslash}m{#1}}
\usepackage{rotating} 

\graphicspath{{./}{figures/}}

\newcommand{\Teff}{\mbox{\,\em T$_{\rm eff}$}}         % effective temperature
\newcommand{\logg}{\mbox{\,log $g$}}                   % surface gravity
\newcommand{\FeH}{\mbox{\,${\rm [Fe/H]}$}}             % relative iron abundance
\newcommand{\Msolar}{\mbox{\,$\rm M_{\odot}$}}         % solar mass
\newcommand{\kelvin}{\,\mbox{K}}                       % K Kelvin

\begin{document}

\title[Metal-poor Stars in the Solar Neighborhood]{On the Origin of Metal-poor Stars in the Solar Neighborhood}
\author{Timur \c Sahin}
\affil{Akdeniz University, Faculty of Science, Department of Space Sciences and Technologies \\
07058, Antalya, TURKEY}
 \author{Sel\c cuk Bilir}
\affil{Istanbul University, Faculty of Science, Department of Astronomy  and Space Sciences \\
34119, Beyaz\i t, Istanbul, TURKEY}
 
\begin{abstract}

We determined the ages, the kinematic parameters and Galactic orbital parameters of six metal-poor (-$2.4<$ [Fe/H] $<-1.0$
dex) F-type high proper motion (HPM) stars to investigate their HPM nature and origin. For the kinematical procedure, the
astrometric data from the {\it Gaia} DR2 was used. High resolution ELODIE spectra of the six dwarfs were also used to obtain
accurate [Fe/H] abundances and up-to-date [$\alpha$/Fe] abundances. The calculations for stellar ages were based on Bayesian
statistics, with the computed ages falling in the range 9.5-10.1 Gyrs. On the basis of the metallicities and ages, six HMP
stars are members of halo (HD6755, HD84937, BD +42 3607) or members of the low-metallicity tail of the thick disk (HD 3567, HD
194598, HD 201891). However, Galactic orbital parameters suggest thin disk (HD 84937, HD 194598), thick disk (HD 3567, HD
201891), and halo (HD 6755, BD +42 3607) population. The dynamical analysis was also performed for the escape scenario from
the candidate GCs. The tidal disruption of a dwarf galaxy was also considered to be as an alternative origin. HD 6755,
HD 194598, and HD 3567 with their retrograde orbital motions are likely candidate stars for a tidally disrupted dwarf galaxy 
origin. However, HD 194598 relationship with NGC\,6284 presents an interesting case. Its encounter velocity is low 
(16 $\pm$ 28 km s$^{\rm -1}$) and their ages and metallicities are very nearly consistent with each other at the 
1$\sigma$ level. The rest of the HPM sample stars have a 4$\%$ to 18$\%$ probability of encountering with selected GCs for 
1.5 tidal radii. This indicates that a globular cluster origin for the program stars is unlikely.

\end{abstract}


\section{Introduction}

The unevolved late-type stars of different Galactic populations with large proper motions play an important
role for our knowledge of the Galaxy's stellar content, and for reconstructing the Galactic chemical evolution. Late-type, metal-poor stars (the F to K type dwarfs) with narrow and less blended spectral lines in their spectra are excellent probes at medium resolution for the chemical history of the earliest stellar populations. It is equally important to note that the ongoing surveys of such stars, such as GALactic Archaeology with HERMES (GALAH) (Heijmans et al. 2012, De Silva et al. 2015), {\it Gaia}-ESO Public Spectroscopic Survey (GES) (Gilmore et al. 2012), the Large Sky Area Multi-Object Fibre Spectroscopic Telescope (LAMOST) (Zhao 2012), the APO Galactic Evolution Experiment (APOGEE) (Allende Prieto et al. 2008), Sloan Extension for Galactic Understanding and Exploration (SEGUE) (Yanny 2009), the RAdial Velocity Experiment (RAVE) (Steinmetz et al. 2006) greatly depend on accurately calibrated  stellar parameters from such stars. In addition, discovery of several halo substructures (Grillmair 2006; Sesar et al. 2007; Belokurov et al. 2007; Kepley et al. 2007; Klement 2010; Helmi et al. 2017; Li et al. 2019) in the literature also requires up-to-date stellar abundances for halo stars. However, an almost distinct abundance pattern in $\alpha$-elements is present for the thin and the thick disk member of the Galaxy (Bensby et al. 2003, 2005, 2014; Reddy et al. 2003, 2006; Nissen \& Schuster 2011; Adibekyan et al. 2012; Karaali et al. 2016, 2019; Yaz G{\"o}k{\c{c}}e et al. 2017; Plevne et al. 2020). The latter group tend to be slightly metal-poor but have enhanced $\alpha$-element abundances. On the other hand, while the majority of the stars in the solar neighborhood appear to be members of the thin disk, several of them are found to be associated with the Galactic thick disk. Furthermore, a small percentage of these nearby stars with very old ages and low metallicities are believed to belong to the halo component of the Galaxy (Bilir et al. 2006, 2012). Altogether, a concise definition of stellar populations is necessary not only in testing and constraining assumptions but also in modeling early nucleosynthesis. For this, the metal-poor dwarf stars are certainly the best candidates.

In this work, we provide fresh metallicities and $\alpha$-element abundances for those F-type high proper motion (here after HPM) stars from the ELODIE library (Prugniel \& Soubiran 2001). We determined their ages, kinematic parameters and Galactic orbital parameters. Since a concise description of their origin is unknown, correlating metallicities and $\alpha$-element abundances as well as abundances for a large number of elements with their kinematics and computed Galactic orbital parameters may shed light on the nature of those metal-poor dwarf stars of HPM. Although most of the stars that are subject of this work have been extensively studied in the literature, the reported stellar parameters of some were seen to present large variations. At the very least, a new measurements of metallicities was seen to be necessary since the metallicity is also of importance for accurate determination of ages. Even a crude discussion of space motions with their chemical compositions in the context of their origin could have been proven to be a useful tool. If far from descriptive, it might, at least, spur the discussion on the HPM nature of the stars. In fact, a recent study on s-process enriched metal-poor star HD\,55496 by Pereira et al. (2019) commented on the possible origins of this field halo object as a second generation globular cluster star. However their estimate was based on the past encounter probabilities (noted to be very small, i.e. $\le 6\%$) with the known globular clusters only. Their speculative conclusion for its origin was that the star possibly originated from the tidal disruption of a dwarf galaxy, interestingly with no justification on abundance. 

This paper is organized as follows. Section 2 provides information concerning the data selection process and the observations. Section 3 describes the methods employed in chemical abundance analysis and model parameter determination. In Section 4, we present findings on their kinematics and computed Galactic orbits. We detail our age derivation in Section 5. We discuss our results and their implications in Section 6 which also includes discussion on possible origin of HPM stars and on our dynamical analysis based on calculation of encounter probabilities with candidate globular clusters.

\section{Selection of program stars}
The {\sc ELODIE} (Soubiran 2003) library contains 1953 spectra of 1388 stars. Our first selection criterion was spectral type. Among the 625 F spectral type stars, we chose the stars with an HPM designation in the library. As a second criterion, binarity was inspected using the {\sc SIMBAD} database\footnote{No comments on binarity was reported in the {\sc ELODIE} archive.}. The spectroscopic binaries from the sample were removed. In consequence the whole HPM\footnote{They cover a range from $-$122 to +633 mas.yr$^{\rm -1}$ in RA and from $-$899 to +340 mas.yr$^{\rm -1}$ in DEC.) Atmospheric parameters are from the {\sc ELODIE} archive at http://atlas.obs-hp.fr/{\sc ELODIE}/; Moultaka et al. (2004)} sample consists of 54 F-type stars, together with the {\sc ELODIE} library (Prugniel \& Soubiran 2001) with effective temperatures $4900 <$ $T_{\rm eff}$ $< 6900$ \kelvin, surface gravities $1.7 <$ log $g$ (cgs) $< 5$, metallicities -$2.4 <$ [Fe/H] $< 1$ dex, and distances $9 <$ $d$ $< 283$ pc. The final selection was done for halo candidate stars with low metallicities. We specifically focussed on metal-poor candidate stars with a metallicity in the range -$2.4 <$ [Fe/H] $< -1$ dex as they have not been considered in previous studies. The total number of stars that satisfies these above criteria was six.

 \begin{figure*}
 \centering 
 \includegraphics*[width=17cm,height=12cm,angle=0]{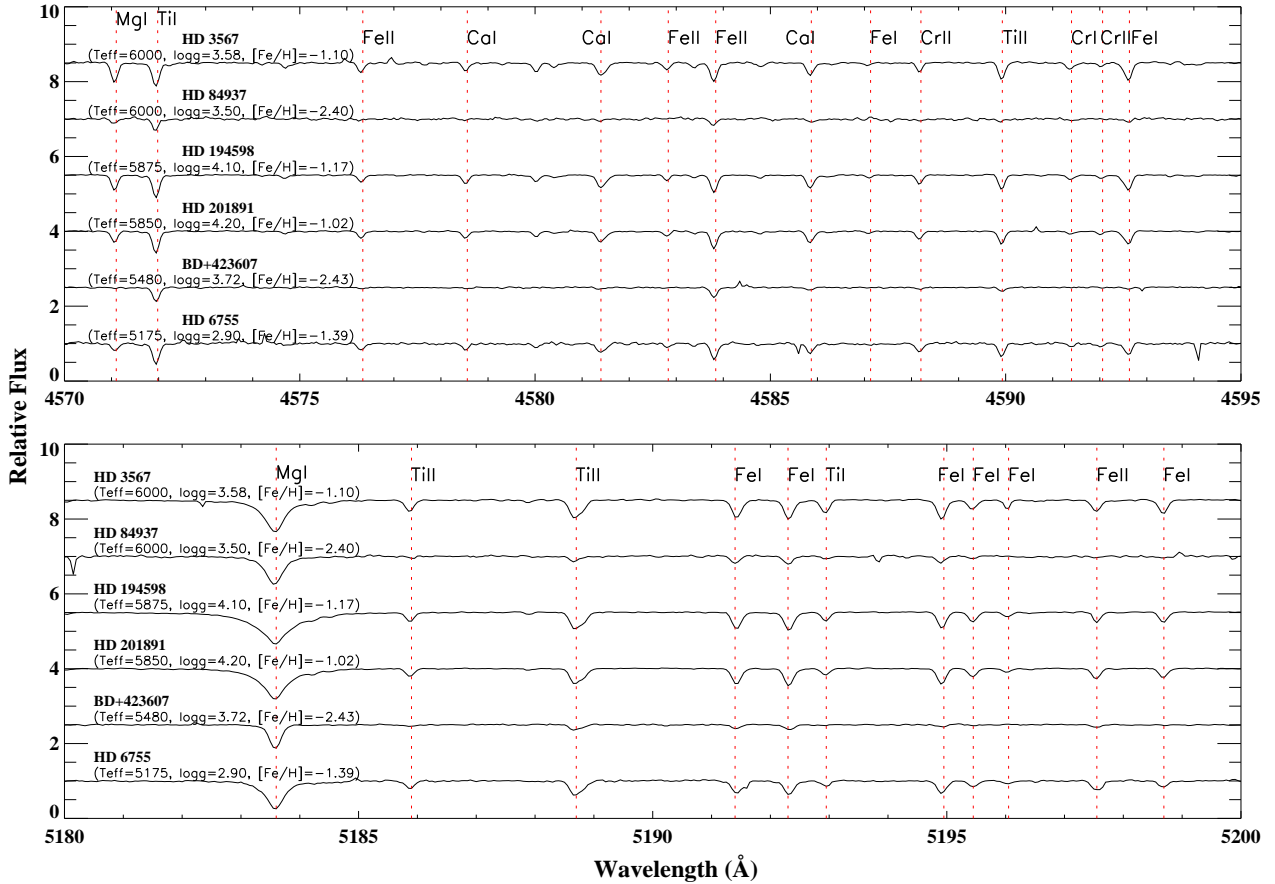}
 \caption{A small region of the spectrum for six HPM sample stars. Identified lines are also indicated.}
 \label{toomre}
 \end{figure*}

\subsection{Observations}

High resolution ($R=42\,000$) and high signal-to-noise ratio (S/N; upto 140) spectra of the stars were obtained on the 1.93 m telescope of the Haute Provence Observatory equipped with the {\sc ELODIE} fiber-fed cross-dispersed echelle spectrograph that provided a spectral coverage from 3900 to 6800 \AA. The log of observations is given in Table 1. The spectra were continuum normalized, wavelength calibrated and the radial velocity corrected by the data-reduction pipeline run at the telescope. Since some problems were seen in continuum normalization of the spectra from the library, the {\sc ELODIE} spectra were re-normalized, using a in-house developed interactive normalization code LIME (\c{S}ahin 2017) in {\sc IDL} prior to the
abundances analysis. The character of the spectra of the program stars is displayed in Figure 1. Many lines in the spectra were suitable for abundance analysis are apparently unblended.

LIME was also employed for the line identification process. It provides the most probable identifications for the line of interest and lists the recent atomic data (e.g. Rowland Multiplet Number-RMT, log$gf$, and Lower Level Excitation Potential-LEP) that are compiled from the literature (e.g. from {\sc NIST} database). Equivalent widths (EWs) are obtained using both SPECTRE (Sneden 1973) and LIME codes. The results for a representative sample of weak and strong lines agreed well within $\pm$ 4 m\AA.

\begin{table*}
\small
\caption[]{Log of observations for the HPM sample stars. The S/N values in the raw spectra are reported near 5000 \AA. Spectral types taken from INCA.}
\centering
\begin{tabular}{l|c|c|c|c|c|c|c|c|r}
\hline\hline
Stars & $\alpha$ & $\delta$  & $l$ & $b$ &Sp.Type& Exposure & S/N & MJD &V$_{\rm HEL}$\\
\cline{2-5}
\cline{7-7}
\cline{9-10}
               &(h:m:s)& ($^\circ$: ' : " ) & ($^\circ$)& ($^\circ$) & & (s)  && (2400000+)& (km s$^{-1}$) \\
\hline 
\hline
HD\,6755    & 01:09:43 & +61:32:50   & 125.11 &$-$1.25  &F8V & 902   &89   & 50360.1 &-312.17 \\ 
BD\,+423607 & 20:09:01 & +42:51:54   & 78.95  & +5.37   &F3  & 3602  &60   & 50682.9 &-195.03 \\ 
HD\,201891  & 21:11:59 & +17:43:39   & 66.71  &$-$20.43 &F8V & 2401  &132  & 50358.8 &-44.47  \\ 
HD\,194598  & 20:26:11 & +09:27:00   & 52.82  &$-$16.13 &F7V & 3000  &139  & 50359.8 &-247.15 \\ 
HD\,3567    & 00:38:31 & $-$08:18:33 & 113.12 &$-$70.93 &F5V & 3600  &74   & 50684.1 &-47.62  \\ 
HD\,84937   & 09:48:56 & +13:44:39   & 220.99 &+45.47   &sdF5& 2700  &95   & 50188.8 &-15.17  \\ 
\hline  
\hline
\end{tabular}
\label{}
\end{table*}

\section{The Abundance Analysis}
For abundance analysis of halo sample stars in this study, we have acquired {\sc ATLAS9} model atmospheres (Castelli \& Kurucz 2004) computed in local thermodynamic equilibrium ({\sc LTE}) (ODFNEW). The elemental abundances for the program stars were computed by using an {\sc LTE} line analysis code MOOG (Sneden 1973)\footnote{The source code of the MOOG can be downloaded from http://www.as.utexas.edu/~chris/moog.html}. The details of the abundance analysis and the source of atomic data are the same as in {\c S}ahin \& Lambert (2009), {\c S}ahin et al. (2011, 2016). In the following subsections, we discuss the adopted line list, atomic data, and the derivation of model parameters.

\subsection{The line list}
A systematic search was performed for unblended lines. They were identified using MOOG by calculating the synthetic spectrum for the observed wavelength region. Only for a very small percentage of the accepted lines, the spectrum synthesis was preferred to a direct estimate of EW. For the iron lines, one has to be cautious for their selection in metal-poor stars. Because, 3D (time dependent)
hydrodynamical model atmospheres of cool stars highlight the importance of lower-level energy dependence of 3D abundance corrections for neutral iron lines (e.g. Dobrovolskas et al. 2013). Also, low-excitation (i.e. E$_{exc}<$1.2 eV) Fe\,{\sc i} lines in metal-poor stars may provide higher abundances compared to lines of higher excitation energies (e.g. Lai et al. 2008). Our inspection of the mean abundances from
low-excitation lines of Fe\,{\sc i} indicates $\approx$0.05 dex differences in mean neutral iron abundances of the program stars. Since, in our case, the 3D effect on \Teff\, determination appears to be negligible. The $gf$-values for chosen lines of Fe\,{\sc i} and Fe\,{\sc ii} were taken from compilation of Fuhr \& Wiese (2006). Our selection of iron lines included 254 Fe\,{\sc i} lines and 29 Fe\,{\sc ii} lines.
Chosen lines of Fe\,{\sc i} and Fe\,{\sc ii} are exhibited in Tables A1-A3. The list of identified lines for elements other than Fe with most up-to-date atomic data is provided in Tables A4 - A6.

%\newpage
\begin{table}
\small
\caption{Solar abundances obtained by employing the solar model atmosphere from Castelli \& Kurucz (2004) compared to the photospheric abundances
from Asplund et al. (2009). The abundances presented in bold type-face are measured by synthesis while remaining elemental abundances were
calculated using the line EWs. $\Delta {\rm log} \epsilon_{\rm \odot}{\rm (X) = log} \epsilon_{\rm \odot} {\rm (X)}_{\rm This\,study} - {\rm log} \epsilon_{\rm \odot} {\rm (X)}_{\rm Asplund}$  }
\centering
\begin{tabular}{l|c|c|c|c}
\hline
     	&   This study &  &   Asplund \\
\cline{2-2}
\cline{4-4}
Spac.   &  $\log\epsilon_{\rm \odot}$(X) & n &   $\log\epsilon_{\rm \odot}$(X) & $\Delta\log\epsilon_{\rm \odot}$(X) \\
\cline{2-2}
\cline{4-4}
		&  (dex) &&  (dex)&  (dex)\\
 \hline
%Li\,{\sc i}	&    ---	    &	1     &   1.05 $\pm$0.10  & 	    \\
C\,{\sc i}	&    8.36$\pm$0.00  &	1     &   8.43$\pm$0.05  &   0.07  \\
Na\,{\sc i}	&    6.23$\pm$0.01  &	2     &   6.24$\pm$0.04  &  -0.01  \\
Mg\,{\sc i}	&    7.65$\pm$0.04  &	3     &   7.60$\pm$0.04  &   0.05  \\
Si\,{\sc i}	&    7.60$\pm$0.06  &	5     &   7.51$\pm$0.03  &   0.11  \\
Si\,{\sc ii}&    7.60$\pm$0.09  &	2     &   7.51$\pm$0.03  &   0.11 
22
  \\
Ca\,{\sc i}	&    6.21$\pm$0.16  &	14    &   6.34$\pm$0.04  &  -0.13  \\
Sc\,{\sc i}	&    3.11$\pm$0.00  &	1     &   3.15$\pm$0.04  &  -0.04  \\
Sc\,{\sc ii}&    3.17$\pm$0.10  &	5     &   3.15$\pm$0.04  &   0.02  \\
Ti\,{\sc i}	&    4.93$\pm$0.12  &	35    &   4.95$\pm$0.05  &  -0.02  \\
Ti\,{\sc ii}&    5.10$\pm$0.14  &	18    &   4.95$\pm$0.05  &   0.15  \\
\textbf{V\,{\sc i}}&  \textbf{3.83$\pm$0.11}  &\textbf{2}     &   \textbf{3.93$\pm$0.08}  &  \textbf{-0.10}  \\
Cr\,{\sc i}	&    5.63$\pm$0.10  &	17    &   5.64$\pm$0.04  &  -0.01  \\
Cr\,{\sc ii}&    5.66$\pm$0.07  &	6     &   5.64$\pm$0.04  &   0.02  \\
Mn\,{\sc i}	&    5.64$\pm$0.17  &	9     &   5.43$\pm$0.05  &   0.21  \\
Fe\,{\sc i}	&    7.48$\pm$0.12  &	160   &   7.50$\pm$0.04  &  -0.02  \\
Fe\,{\sc ii}&    7.45$\pm$0.15  &	24    &   7.50$\pm$0.04  &  -0.05  \\
\textbf{Co\,{\sc i}}	&    \textbf{4.93$\pm$0.03}  &	\textbf{3}     &   \textbf{4.99$\pm$0.07}  &  \textbf{-0.06}  \\
Ni\,{\sc i}	&    6.22$\pm$0.13  &	42    &   6.22$\pm$0.04  &   0.00  \\
\textbf{Cu\,{\sc i}}     &    \textbf{4.19$\pm$0.03}  &	\textbf{2}     &   \textbf{4.19$\pm$0.04}  &   \textbf{0.00}  \\
\textbf{Sr\,{\sc i}}	&    \textbf{2.90$\pm$0.00}  &	\textbf{1}     &   \textbf{2.87$\pm$0.07}  &   \textbf{0.03}  \\
\textbf{Sr\,{\sc ii}}	&    \textbf{2.90$\pm$0.05}  &	\textbf{2}     &   \textbf{2.87$\pm$0.07}  &   \textbf{0.03}  \\
\textbf{Y\,{\sc ii}}	&    \textbf{2.24$\pm$0.06}  &	\textbf{4}     &   \textbf{2.21$\pm$0.05}  &   \textbf{0.03}  \\
\textbf{Zr\,{\sc i}}	&    \textbf{2.80$\pm$0.00}  &	\textbf{1}     &   \textbf{2.58$\pm$0.04}  &   \textbf{0.22}  \\
\textbf{Zr\,{\sc ii}}	&    \textbf{2.76$\pm$0.01}  &	\textbf{2}     &   \textbf{2.58$\pm$0.04}  &   \textbf{0.18}  \\
\textbf{Ba\,{\sc ii}}	&    \textbf{2.06$\pm$0.06}  &	\textbf{3}     &   \textbf{2.18$\pm$0.09}  &  \textbf{-0.12}  \\
\textbf{Ce\,{\sc ii}}	&    \textbf{1.60$\pm$0.00}  &	\textbf{1}     &   \textbf{1.58$\pm$0.04}  &   \textbf{0.02}  \\
\textbf{Nd\,{\sc ii}}	&    \textbf{1.66$\pm$0.14}  &	\textbf{3}     &   \textbf{1.42$\pm$0.04}  &   \textbf{0.24}  \\
\textbf{Sm\,{\sc ii}}    &    \textbf{1.18$\pm$0.00}  &	\textbf{1}     &   \textbf{0.96$\pm$0.04}  &   \textbf{0.22}  \\
\hline
\end{tabular}
\end{table}

To check the uncertainties in $\log gf$ -values and to minimize systematic errors, we have derived solar abundances using stellar lines. The lines were measured off the solar flux atlas of Kurucz et al. (1984) analyzed with the solar model atmosphere from the Castelli \& Kurucz (2004) grid for \Teff\,= 5788 \kelvin\, and \logg\,= 4.40 cgs. Our analysis gave a microturbulent of 0.8 km s$^{-1}$ and the solar abundances in Table 2. As can be seen, the solar abundances were recreated fairly well. The solar abundances obtained are compared with those from Asplund et al. (2009) in their critical review. In referencing stellar abundances to solar values, we use our solar abundances. Hence, the analysis in this study is performed differentially with respect to the Sun. Such a differential approach in the analysis will reduce the errors due to uncertainties in the oscillator strengths, the influence of the spectrograph characteristics and also deviations from LTE. Errors in effective temperatures from spectroscopic excitation technique can also stem from systematic errors in oscillator strengths as a function of excitation potential. The same is also true for errors in equivalent widths.

A further health check on $\log gf$ - values was performed by comparing the $\log gf$-values employed in this study to those included in {\it Gaia}-ESO line list v.5 which was provided by the {\it Gaia}-ESO collaboration. The {\it Gaia}-ESO line list includes a list of recommended lines and atomic data (hyperfine structure-hfs- corrected $gf$-values) for the analysis of FGK stars. It is worth to note that several lines in the spectra of FGK stars are still unidentified (Heiter et al. 2015). Over 228 common Fe\,{\sc i} lines, the difference in $\log gf$ -values (log$gf_{\rm This study}$-log$gf_{\rm Gaia-ESO}$) is found to be 0.001 dex. There is no difference in $\log gf$ - values for 25 common lines of Fe\,{\sc ii}. Only two Fe\,{\sc ii} lines showed differences at 
$\approx$0.2 dex level (e.g. Fe\,{\sc ii} 4416 \AA\, and Fe\,{\sc ii} 5425 \AA). The differences in $\log gf$-values for elements other than Fe reported in A4 - A6 with the number of lines given in subsequent parenthesis are as follows: 0.004 dex for C\,{\sc i} (1) and Mg\,{\sc i} (7); -0.004 dex for Na\,{\sc i} (4); -0.11 dex for Si\,{\sc i} (6); -0.03 dex for Si\,{\sc ii} (2); -0.09 dex for Ca\,{\sc i} (28), Ni\,{\sc i} (46), and Y\,{\sc ii} (4); 0.02 dex for Sc\,{\sc ii} (8) and Ti\,{\sc i} (37); no difference for Ti\,{\sc ii} (27), Cr\,{\sc i} (25), Sr\,{\sc i} (1), Zr\,{\sc i/ii} (1,3), Ce\,{\sc ii} (1), and Sm\,{\sc ii} (1); -0.07 dex for
Cr\,{\sc ii} (7); 0.06 dex for Sr\,{\sc ii} (2); 0.03 dex for Ba\,{\sc ii} (3); 0.08 dex for Nd\,{\sc ii} (3).  

Since Fe\,{\sc i} and Fe\,{\sc ii} abundances were employed to constrain the model atmosphere parameters in this study, we must consider non-LTE effects on Fe. These effects were considered to be insignificant for Fe\,{\sc ii} lines (Bergemann et al. 2012; Lind, Bergemann \& Asplund 2012; Bensby et al. 2014). Lind et al. (2012) demonstrated that the departures from LTE for Fe\,{\sc ii} lines of low excitation potential ($<$ 8 eV) at metallicities [Fe/H] $>$ -3 dex was negligible. To account for non-LTE effects on Fe\,{\sc i} lines, we used 1D non-LTE investigation of Lind et al. (2012). For this, the non-LTE web tool INSPECT program v1.0 (see Lind et al. 2012) was employed. For common Fe\,{\sc i} (also see Tables A1-A3), the non-LTE corrections computed by us via INSPECT program was of about ~0.1 dex with the exception of 4994 \AA\, Fe\,{\sc i} line. The line required a non-LTE correction ($\Delta$($\log\epsilon$(Fe\,{\sc
i}))\footnote{$\Delta$($\log\epsilon$(Fe\,{\sc i})) = $\log\epsilon$(Fe\,{\sc i})$_{\rm NLTE}$ - $\log \epsilon$(Fe\,{\sc i})$_{\rm LTE}$)} of -0.6 dex for HD\,3567. We obtained comparable departures from the LTE for this line for other HPM stars. For the 5198 \AA\, Fe\,{\sc i} line in the spectrum of HD\,201891, we predict $\Delta$($\log\epsilon$(Fe\,{\sc i}))=0.4 dex, and it is 0.5 dex for 5216 \AA\, Fe\,{\sc i} line and HD\,3567. Its effect on the model atmosphere parameters or equilibria of Fe is not discernible, i.e. the magnitudes of the slopes of the relationship between the abundance of iron from Fe\,{\sc i} lines and the excitation potential of each line (or the reduced EW of each line) showed no significant variation when this line was excluded from the analysis. We also followed the recipe given by Lind, Bergemann \& Asplund (2012; see also their Figure 4, 5, and 7) to scrutiny the role of non-LTE effects in determining the atmospheric parameters from Fe\,{\sc i} and Fe\,{\sc ii} lines in the current analysis. For HD\,201891, HD\,194598, and HD\,3567, the non-LTE effects on the excitation balance in 1D was less than 100 \kelvin. The excitation balance for HD\,6755 required a correction of about 50 \kelvin. For the most metal poor HPM stars in our sample, BD+42\,3607 and HD\,84937, it was found to be $<$150 \kelvin\, and $<$200 \kelvin, respectively. The corrections estimated by Lind, Bergemann \& Asplund (2012) was typically $\le$50 \kelvin.  

Bergemann et al. (2017) reported non-LTE abundance corrections for Mg\,{\sc i} lines (e.g. 4571, 5528, and 5711 \AA). They were computed using 1D hydrostatic model atmospheres. The corrections for a representative model of ($T_{\rm eff}$, $\logg$, $\FeH$) = (6000, 4.0, -2.0), were ranging from 0.04 dex for 5528 \AA\, line to 0.07 dex for 4571 \AA\, line. Bergemann et al. (2017) used wavelengths and oscillator strengths from Pehlivan Rhodin et al. (2017) who provides theoretical transition probabilities for Mg\,{\sc i} lines. Comparison of $\log gf$ -values that are common to both Bergemann et al. (2017) and Pehlivan Rhodin et al. (2017) provided a difference of only -0.10$\pm$0.11 dex (over 7 lines).

Odd-Z iron peak elements such as V (e.g. 6016 \AA), Mn, and Co especially suffer from hfs since the hfs causes additional broadening in a spectral line and its treatment desaturates the line (Prochaska et al. 2000) for strong lines. Therefore, it must be taken into account in the calculation of abundances via spectrum synthesis for certain species. The wavelengths and $\log gf$-values for the hfs components are obtained from Lawler et al. (2014) for V and from Den Hartog et al. (2011) for Mn and from Lawler et al. (2015) for Co. We assumed a solar system isotopic ratios for these elements. More onto this, the study by Bergemann et al. (2019) on 3D non-LTE formation of Mn lines in late type stars provided new experimental transition probabilities for manganese lines. They noted that for some of the manganese lines, the new $\log gf$ - values were typically 0.05-0.1 dex lower than the old values. The difference between our and Bergemann's (hfs included) $gf$ -values is small, i.e. $-0.06\pm0.08$ dex. 

Over a sample of metal-poor stars, the study by Bergemann \& Gehren (2008) for the non-LTE effects on manganese lines shows that non-LTE effects begin to dominate with increasing \Teff\, and decreasing metallicity. Surface gravity was noted to become important at [Fe/H] $\le$-2 dex and \Teff $>$ 6000 \kelvin. However, the program stars in this study have effective temperatures below this temperature limit. Furthermore, a recent study by Bergemann et al. (2019) showed that 3D non-LTE effects would not affect the excitation balance in the atmospheres of very metal-poor dwarfs because they were seen to be similar for all Mn\,{\sc i} lines regardless of their multiplet.
It was also noted that, the Mn abundances could be underestimated via 1D LTE analysis by $\approx$-0.2 dex for the models of dwarf stars and the change in metallicity would not have changed this value. The lines of multiplets 23 (e.g. 4761, 4762, 4765, and 4766 \AA) and 32 (e.g. 6016 and 6021 \AA) were listed to be reliable lines of Mn\,{\sc i} for abundance analysis. We employed these multiplet members for abundance analysis of the HPM stars in this study.

\begin{figure*}
\centering 
\includegraphics*[width=8.66cm,height=7.0cm,angle=0]{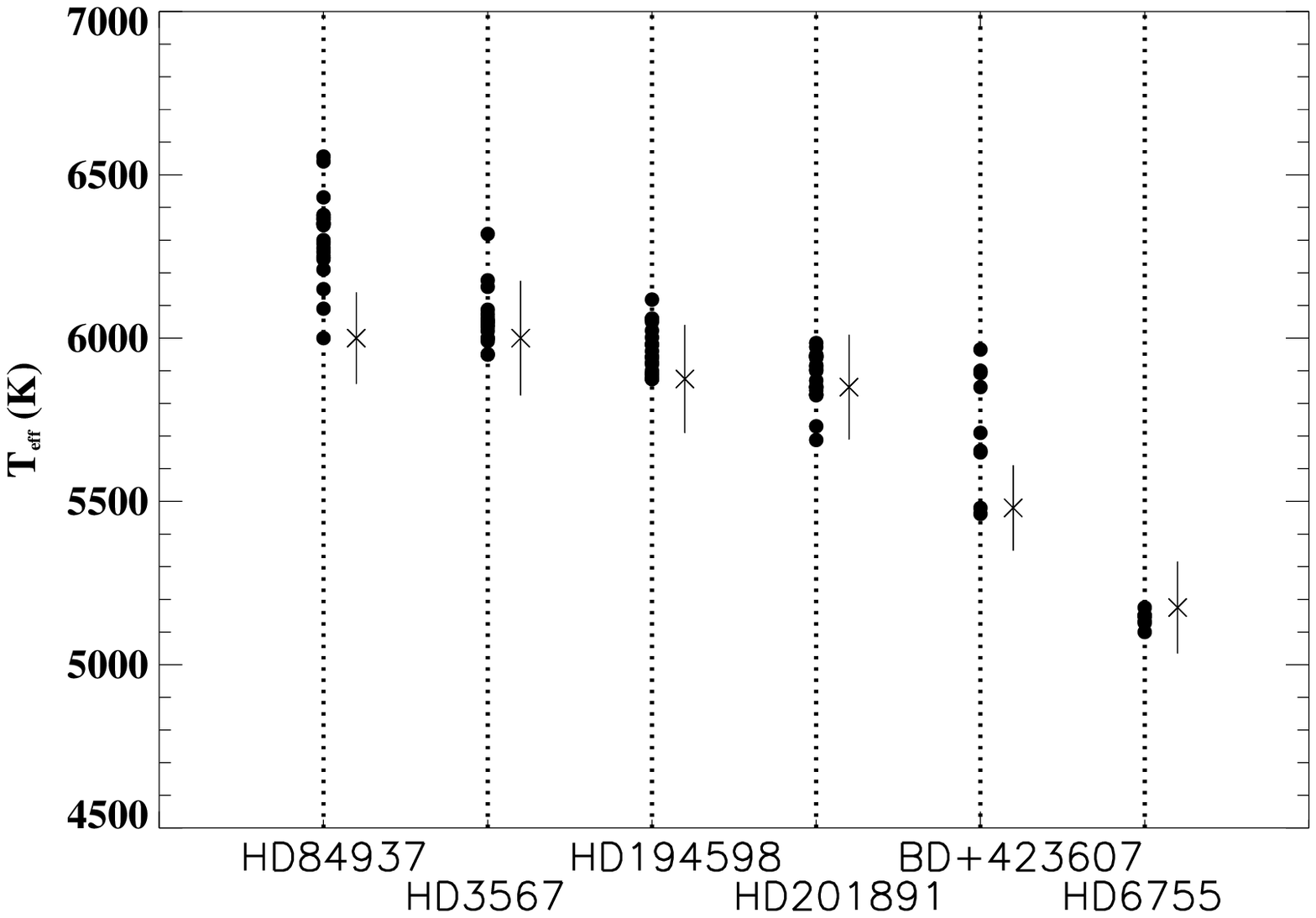}
\includegraphics*[width=8.66cm,height=7.0cm,angle=0]{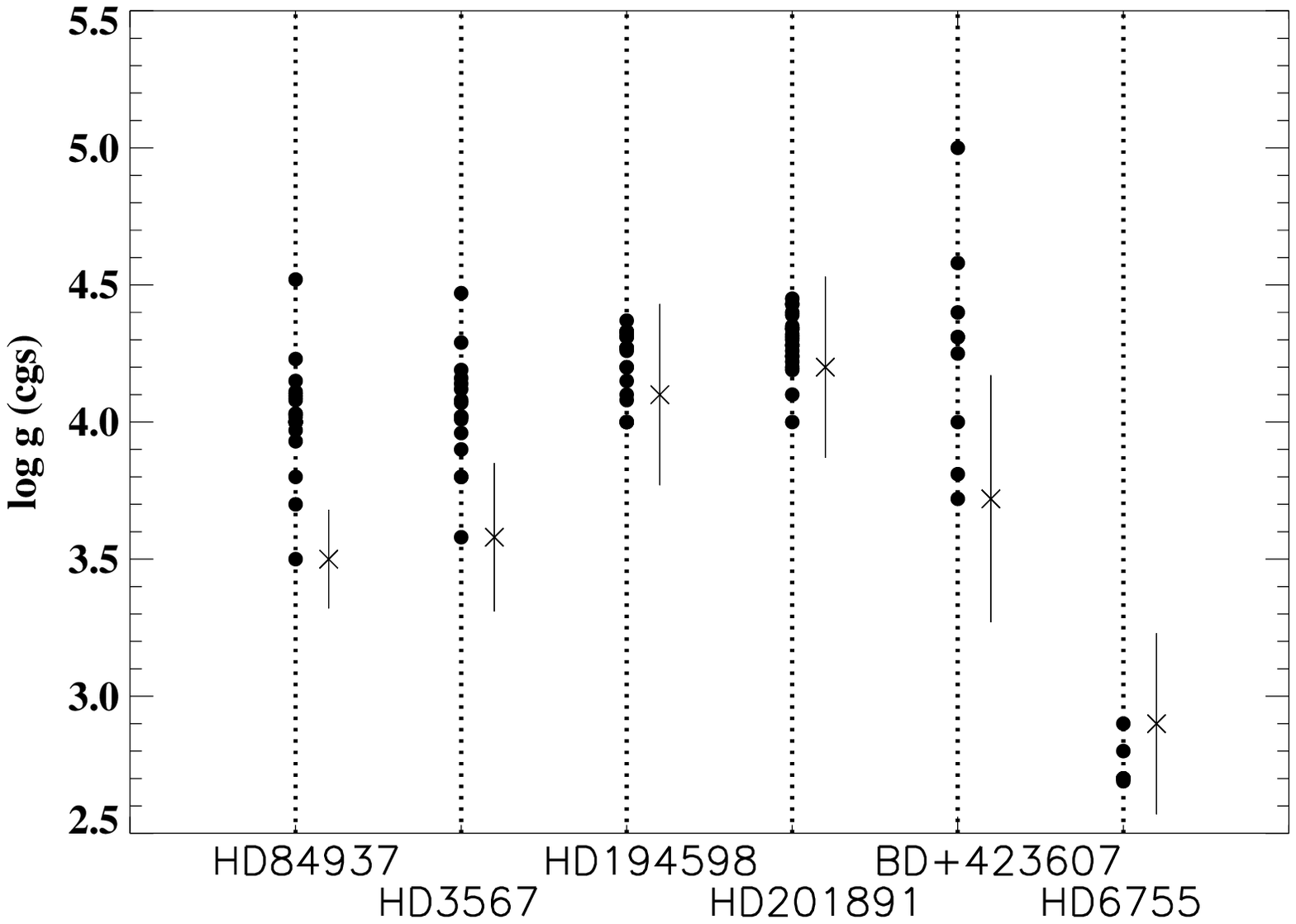}
\includegraphics*[width=8.66cm,height=7.0cm,angle=0]{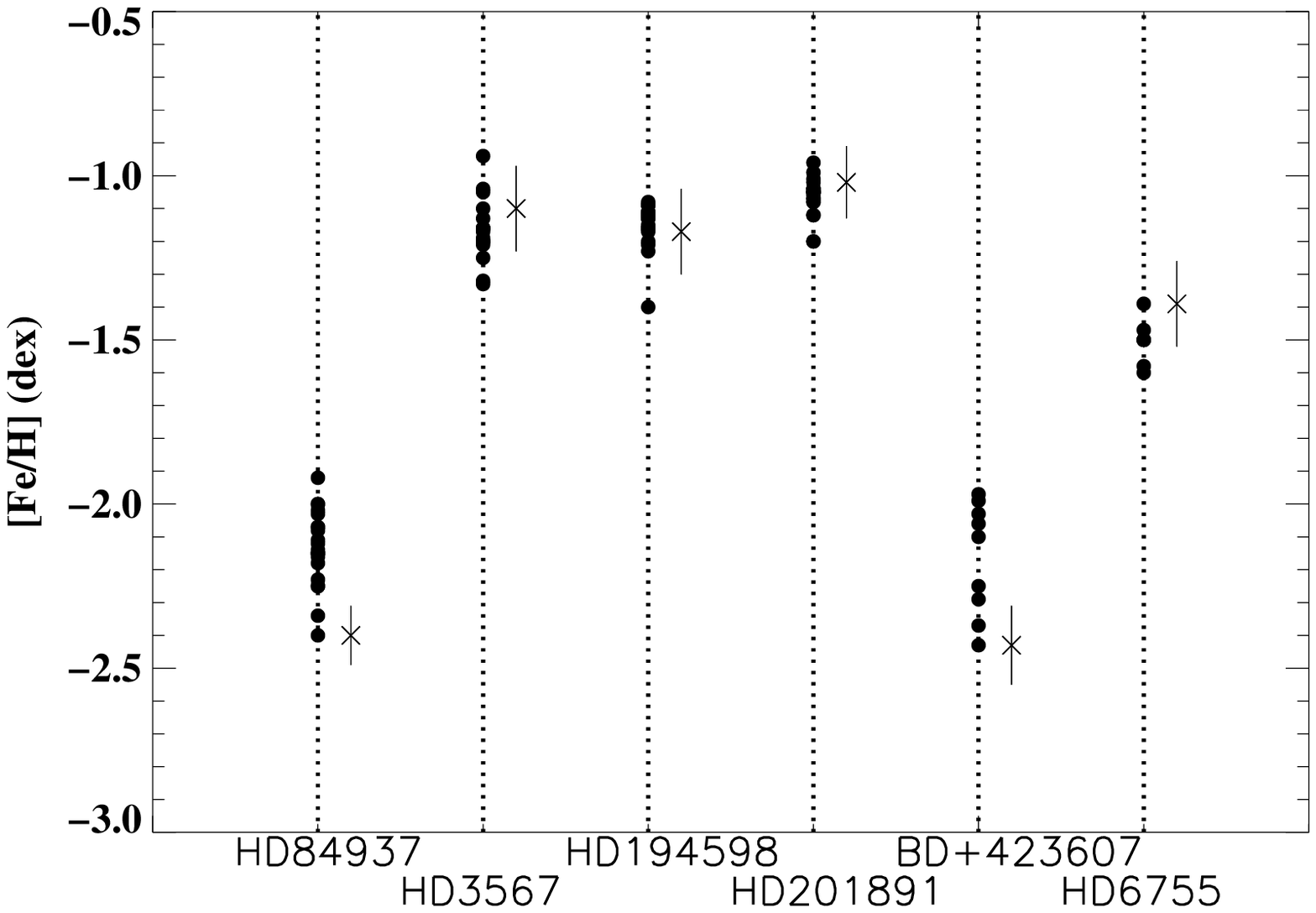}
\caption{Model parameters reported in the literature for the program stars (filled circle). The new measurements of stellar
parameters ($T_{\rm eff}$, $\logg$, [Fe/H]) as well as estimated errors in those was also indicated with a cross symbol (see also
Table A9).}
\label{feh_w}
\end{figure*}

Zhang et al. (2008) investigated non-LTE effects on the scandium for the Sun and reported that the non-LTE effects were negligible for Sc\,{\sc ii}. However, the strong non-LTE effects were observed for Sc\,{\sc i}. For the Sun, the non-LTE effect on solar Sc\,{\sc i} abundance was +0.18 dex and the Sc\,{\sc i} abundance did not present any change.

Bergemann et al. (2010) reported non-LTE corrections for cobalt and noted the metallicity as the main stellar parameter that determines the magnitude of non-LTE correction to be applied on cobalt abundance. In their Table 3, they listed non-LTE abundance corrections for the common Co\,{\sc i} line at 4121 \AA\, for selected model atmospheres (see also their Figure 4). Inspection of their table provided the
non-LTE abundance correction for the  4121 \AA\, line. It is ranging from $\approx$0.2 dex for HD\,6755 to $\approx$0.6 dex for two lowest metallicity HPM stars, HD\,84937 and BD+42\,3607. It is at $\approx$0.5 dex level for HD\,3567, HD\,194598\ and HD\,201891.

\begin{table}
\scriptsize
\caption{Model atmosphere parameters of HPM stars from this study.}
\centering
\begin{tabular}{l|c|c|c|c}
\hline
 Stars 	&\Teff	& \logg  &[Fe/H]  & $\xi$	\\
\cline{2-5}
        &(\kelvin)& (cgs)  & (dex) & (km s$^{\rm -1}$) \\    
\hline 
\hline
HD\,6755    &   5175 $\pm$140   & 2.90$\pm$0.33  &  -1.39$\pm$0.18  & 1.30  \\
BD+42\,3607 &   5480 $\pm$130   & 3.72$\pm$0.45  &  -2.43$\pm$0.17  & 0.90  \\
HD\,201891  &   5850 $\pm$160   & 4.20$\pm$0.33  &  -1.02$\pm$0.16  & 0.85  \\
HD\,194598  &   5875 $\pm$165   & 4.10$\pm$0.33  &  -1.17$\pm$0.18  & 0.83  \\
HD\,3567    &   6000 $\pm$175   & 3.58$\pm$0.27  &  -1.10$\pm$0.18  & 0.93  \\
HD\,84937   &   6000 $\pm$140   & 3.50$\pm$0.18  &  -2.40$\pm$0.15  & 1.60  \\
Sun	    &   5788            & 4.40           &   0.00           & 0.80  \\
\hline 
\hline
\end{tabular}
\end{table}

\begin{figure*}[ht]
\centering 
\includegraphics*[width=17cm,height=12cm,angle=0]{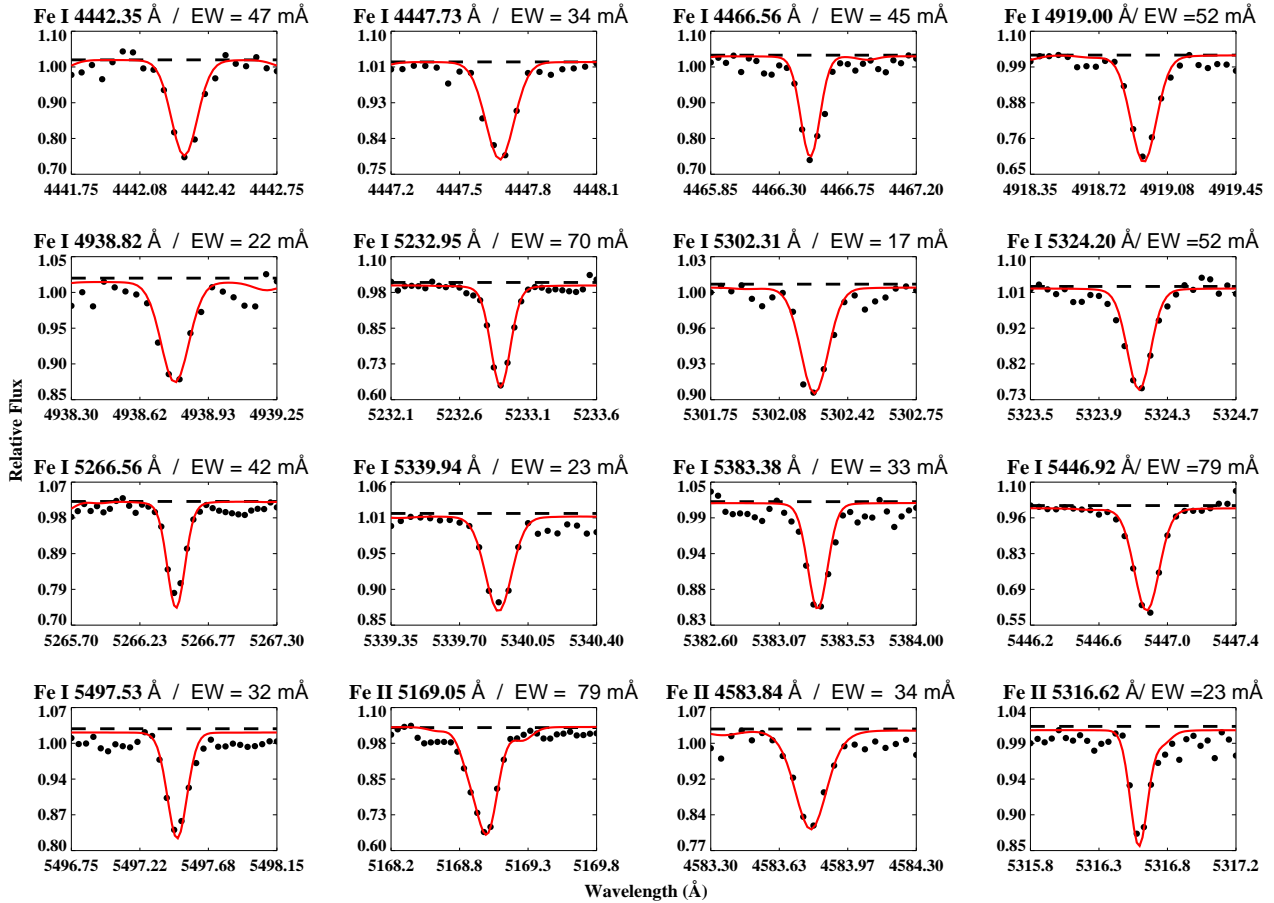}
\caption{The observed (filled circles) and computed (full red line) line
profiles for Fe\,{\sc i} and Fe\,{\sc ii} lines. The measured EWs and their wavelengths
are indicated at the top of each panel. The computed profiles show synthetic
spectra for the abundances reported for BD+42\,3607 in Table A1-A3.}
\label{feh_w}
\end{figure*}

\subsection{Model atmosphere parameters}
Since the reported stellar parameters in the literature for some of the program stars were seen to present large variations (see
Figure 2), we decided to obtain new measurements of model atmosphere parameters spectroscopically in this study.

We determined model atmosphere parameters -- effective temperature, surface gravity, microturbulence, and metallicity -- by using neutral and ionized Fe lines (Table A1-A3), and the full line list can be found online. A selection of observed and computed line profiles of the chosen Fe\,{\sc i} and Fe\,{\sc ii} lines for BD+42\,3607 are shown in Figure 3.
 
The effective temperature was estimated first through imposing the condition that the derived abundance be independent of the lower level excitation potential (LEP). If all lines have the same LEP and a similar wavelength, the microturbulence ($\xi$) is found by requiring that the derived abundance be independent of the reduced equivalent width (EW). The precision in determination of the microturbulent velocity is $\pm$0.5 km s$^{-1}$. We determined the surface gravity (\logg) by requiring ionization equilibrium, e.g. that Fe\,{\sc i} and Fe\,{\sc ii} lines produce the same iron abundance (Figure A1). Iron lines are quite numerous even in very metal-poor sample stars. Since these model atmosphere parameters are interdependent, an iterative procedure is necessary. Small changes are made to the model parameters between each above mentioned step. We also verified that no significant trend of iron abundances with wavelength was present. The resulting stellar model parameters along with our
determination of model parameters of the Sun are listed in Table 3. 

The program stars are also listed in the {\it Gaia} DR2 (Gaia collaboration, 2018). Our spectroscopic temperatures are in excellent agreement with \Teff\, values reported by the {\it Gaia} consortium for HD\,6755, HD\,201891, and HD\,194598. 
 
\section{Kinematic properties of the HPM stars}
The space velocity components of the six HPM stars were calculated using the algorithm of Johnson \& Soderblom (1987) for the J2000 epoch. The $U$, $V$ and $W$ are the
components of a space velocity vector for a given star with respect to the Sun. The transformation matrices use the notation of right-handed system. Hence, $U$ is defined to be positive towards the Galactic centre, $V$ is positive in the direction of the Galactic rotation and $W$ is positive towards the North Galactic Pole.

Equatorial coordinates ($\alpha$, $\delta$), radial velocities (RV), proper motions ($\mu_{\alpha} \cos \delta$, $\mu_{\delta}$) and trigonometric parallax ($\pi$) of the program stars were taken into account in the calculations (Table 4). The proper motions and trigonometric parallaxes of the stars are taken from the {\it Gaia} DR2 catalogue (Gaia collaboration, 2018) while the radial velocities were obtained from the {\sc ELODIE} spectra. Since no astrometric data for HD\,84937 was available from the {\it Gaia} DR2 catalogue, the proper motions and trigonometric parallax of the star were taken from the re-reduced {\it Hipparcos} catalogue (van Leewuen,
2007). Errors in the proper motion components from two different catalogues were compatible with each other, however, the errors in the trigonometric parallaxes were not consistent. The relative parallax errors calculated for five stars from the {\it Gaia} DR2 catalogue are smaller than 0.006, whereas the relative parallax
error calculated from the {\it Hipparcos} catalog for HD\,84937 is in the order of 0.06. In space velocity calculations, the biggest uncertainty comes from the errors in distances. Even if the relative parallax error calculated for HD\,84937 is approximately 0.06, it is expected that the errors in space velocity components will not be large because HD\,84937 is close to the Sun (i.e. $d=73$ pc).

To obtain the precise space velocity components, the first order Galactic differential rotation correction by Mihalas \& Binney (1981) was applied. The corrections for $U$ and $V$ space velocity components were found to be $-0.91\le du \le3.73$ and $-0.23\le dv \le0.08$ km s$^{\rm -1}$, respectively, while the $W$ space velocity component is not affected by a first order approximation. Following the differential rotation correction, the space velocity components were also corrected for the peculiar velocity of the Sun (LSR), which is ($U_{\rm \odot}$, $V_{\rm \odot}$,
$W_{\rm \odot})_{\rm LSR}$ = (8.50$\pm$0.29, 13.38$\pm$0.43, 6.49$\pm$0.26) km s$^{\rm -1}$ (Co\c skuno\u glu et al. 2011). The total space velocities of the stars ($S$) were calculated via the square root of the sum of the squares of the space velocity components. The uncertainties in the space velocity components ($U_{\rm err}$, $V_{\rm err}$, $W_{\rm err}$) were computed by propagating the uncertainties in proper motions, distances and radial velocities using the algorithm by Johnson \& Soderblom (1987). The corrected space velocity components, total space velocities and their errors for six HPM stars are listed in Table 5.

\begin{table*}
\tiny
\caption{The equatorial coordinates, radial velocities, proper motion components, trigonometric parallaxes, and
estimated distances from the Sun for six HPM stars.}
\centering
\begin{tabular}{l|c|c|c|c|c|c|c|c|c}
\hline\hline
Star& $\alpha$& $\delta$ &  RV~$\pm$~err &$\mu_{\rm \alpha} cos\delta~\pm$~err &$\mu_{\rm \delta}~\pm$~err & $\pi~\pm$~err &$d~\pm$~err &$d~\pm$~err$^{\rm *}$ & Ref.   \\
\cline{1-10}
    &  $(^h:^m:^s)$ &  ($^\circ$: ' : " ) & (km s$^{\rm -1}$) & (mas yr$^{-1}$)  &     (mas yr$^{-1}$)   &  (mas)   &(pc)  &(pc) &(RV/$\mu$/$\pi$) \\	       
\cline{1-10}
HD\,6755   &01 09 43.06  &61 32 50.29  &-317.86 $\pm$ 1.44  & 633.010  $\pm$ 0.064 & 65.303  $\pm$ 0.067 &  5.969 $\pm$  0.049 & 167 $\pm$ 1.36 & 166 $\pm$ 0.76  & {\it Gaia}\\
BD+42\,3607&20 09 01.41  &42 51 54.93  &-195.44 $\pm$ 0.68  & 119.536  $\pm$ 0.043 & 340.224 $\pm$ 0.043 & 11.766 $\pm$  0.025 & ~85 $\pm$ 0.18 & ~85 $\pm$ 0.05 & {\it Gaia}\\
HD\,201891 &21 11 59.03  &17 43 39.89  &-44.15  $\pm$ 0.19  &-122.216  $\pm$ 0.060 &-899.263 $\pm$ 0.060 & 29.788 $\pm$  0.033 & ~34 $\pm$ 0.04 &  ~34 $\pm$ 0.01 & {\it Gaia}\\
HD\,194598 &20 26 11.92  &09 27 00.43  &-246.76 $\pm$ 0.18  & 116.508  $\pm$ 0.070 &-548.329 $\pm$ 0.058 & 18.479 $\pm$  0.048 & ~54 $\pm$ 0.14 &  ~54 $\pm$ 0.01 & {\it Gaia}\\
HD\,3567   &00 38 31.95  &-08 18 33.40 &-46.79  $\pm$ 0.27  & ~20.340  $\pm$ 0.093 &-546.373 $\pm$ 0.065 &  8.371 $\pm$  0.043 & 119 $\pm$ 0.61 & 119 $\pm$ 0.20 & {\it Gaia}\\
HD\,84937  &09 48 56.10  &13 44 39.32  &-15.17  $\pm$ 1.00  & 373.050  $\pm$ 0.910 &-774.380 $\pm$ 0.330 & 13.740 $\pm$  0.78  & ~73 $\pm$ 4.13 & -- & ELODIE/{\it Hipparcos}\\
\hline 
\hline
\end{tabular}
\begin{list}{}{}
\item *: Sch\"{o}nrich et al. (2019).
\end{list}

\end{table*}

\begin{table*}
\setlength{\tabcolsep}{3.5pt}
\tiny
\caption{The calculated $U, V, W$ space velocity components and $S$ total space velocities of six HPM stars along with their membership probabilities with different Galactic populations. The Galactic orbital parameters and their errors are also presented. The space velocity
components and total space velocities are given according to LSR.}
\centering
\begin{tabular}{l|c|c|c|c|c|c|c|c|c|c}
\hline\hline
Star	  & $U~\pm$~err          &  $V~\pm$~err          & $W~\pm$~err    & $S~\pm$~err   & \multicolumn{2}{c|}{\mbox{Prob.}} & $Z_{\rm max}$ & $R_{\rm a}$ & $R_{\rm p}$ & $e_{\rm p}$ -- $e_{\rm v}$\\
\cline{2-10}
          &  (km s$^{\rm -1}$) & (km s$^{\rm -1}$)   &  (km s$^{\rm -1}$)& (km s$^{\rm -1}$)                 &$TD/D$ & $TD/H$ &       (kpc)   & (kpc) & (kpc) &                        \\
\cline{1-11}
HD\,6755   & -220.78$\pm$ 3.45  & -531.09$\pm$2.62   & 100.78$\pm$ 0.72 & 583.92$\pm$4.39  & 1.62E+120&   0.00  & 13.22 $\pm$ 0.74  & 58.07 $\pm$ 3.15 & 6.33$\pm$0.01 & 0.80 $\pm$ 0.06 / 0.82 $\pm$ 0.09 \\
BD+42\,3607& -237.95$\pm$0.46  & -144.52$\pm$0.67   & 64.28 $\pm$ 0.17 & 285.73$\pm$0.83  & 4.19E+15  &   0.04  &  3.07 $\pm$ 0.03  & 16.15 $\pm$ 0.05 & 1.33$\pm$0.01 & 0.85 $\pm$ 0.01 / 0.70 $\pm$ 0.01 \\
HD\,201891 & 94.34  $\pm$0.13  & -97.51 $\pm$0.18   & -48.42$\pm$ 0.10 & 144.06 $\pm$0.25 & 8.34E+03  & 114.88  &  1.13 $\pm$ 0.01  & ~9.16 $\pm$ 0.01 & 2.84$\pm$0.01 & 0.53 $\pm$ 0.01 / 0.38 $\pm$ 0.01 \\
HD\,194598 & -70.22 $\pm$0.20  & -260.18$\pm$ 0.26   & -20.98$\pm$ 0.25 & 270.31$\pm$0.42  & 3.72E+24 &   0.00  &  0.35 $\pm$ 0.01  & ~8.39 $\pm$ 0.01 & 0.76$\pm$0.01 & 0.83 $\pm$ 0.01 / 0.15 $\pm$ 0.01 \\
HD\,3567   & 160.80 $\pm$0.76  & -253.80$\pm$1.30   & -49.74$\pm$ 0.58 & 304.51$\pm$1.61  & 2.29E+27  &   0.00  &  1.49 $\pm$ 0.03  & 10.78 $\pm$ 0.04 & 0.60$\pm$0.03 & 0.89 $\pm$ 0.01 / 0.52 $\pm$ 0.01 \\
HD\,84937  & 205.96 $\pm$10.66 & -209.09$\pm$13.04  & 2.58  $\pm$ 0.74 & 293.50$\pm$16.87 & 6.70E+19  &   0.00  &  0.44 $\pm$ 0.25  & 12.74 $\pm$ 0.80 & 0.23$\pm$0.15 & 0.96 $\pm$ 0.10 / 0.14 $\pm$ 0.08 \\
\hline 
\hline
\end{tabular}
\end{table*}

To determine their population types, we employed the kinematic method of Bensby et al. (2003). The kinematic method
assumes a Gaussian distribution for the Galactic space velocity components in the following form. 

\begin{equation}
f = \frac{1}{(2\pi)^{3/2}\sigma_{\rm U}\sigma_{\rm V}\sigma_{\rm W}} \exp\left[-\frac{U^{2}}{2\sigma_{\rm
U}^{2}}-\frac{(V-V_{\rm asym})^{2}}{2\sigma_{\rm V}^{2}}-\frac{W^{\rm 2}}{2\sigma_{\rm W}^{2}} \right] \label{eq:f}
\end{equation} 
Here, $\sigma_{U}$, $\sigma_{V}$, and $\sigma_{W}$ are the characteristic velocity dispersions for different Galactic
populations: 35, 20 and 16 km s$^{-1}$ for thin disk (D); 67, 38 and 35 km s$^{-1}$ for thick disk (TD); 160, 90 and
90 km s$^{-1}$ for halo (H). $V_{\rm asym}$ is the asymmetric drift velocity: -15, -46 and -220 km s$^{-1}$ for thin
disk, thick disk and halo, respectively (Bensby et al. 2003; 2005).

The probability for a star of a given population with respect to another population is defined as the ratio of the Gaussian distribution functions (see Eq. 1) multiplied by the ratio of the local space densities for these two populations. For each star, the relative probabilities for a certain Galactic population was calculated via the
following equations:

\begin{equation}
TD/D = \frac{X_{\rm TD}}{X_{\rm D}} \frac{f_{\rm TD}}{f_{\rm D}},~~~TD/H = \frac{X_{\rm TD}}{X_{\rm H}} \frac{f_{\rm
TD}}{f_{\rm H}}  \label{eq:TD}
\end{equation}
$X_{\rm D}$, $X_{\rm TD}$ and $X_{\rm H}$ are the local space densities for thin-disk, thick-disk and halo, i.e. 0.94, 0.06, and 0.0015, respectively (Robin et al. 1996, Buser et al. 1999, Cabrera-Lavers et al. 2007). Bensby et al. (2003, 2005) proposed four categories to determine the Galactic population memberships of the stars and they are as follows: TD/D $\le$0.1 for high probability thin-disk stars, $0.1< TD/D\le1$ for low probability thin-disk stars, $1<TD/D\le10$ for low probability thick-disk stars and $TD/D>10$ for high probability thick-disk stars. The computed $TD/D$ and $TD/H$ values for each program star are given in Table 5. If $TD/D$ is equal to a small number, the probability of the star in question of being thick-disk star relative to the thin-disk is relatively low. 

{\it galpy}, a python library developed by Bovy (2015) for Galactic dynamics, was used to calculate the Galactic orbital parameters of the HPM stars. We assumed $R_{\rm gc}=$ 8 kpc (Majewski 1993) and $Z_{\rm \odot}=$ 27$\pm$4 pc (Chen et al. 2001). We further assumed that the Milky Way is well represented by the {\it galpy} potential {\it MilkywayPotential2014} which is composed of three potentials that make up the gravitational field of the bulge, disk and halo components of the Milky Way. The bulge component is represented as a spherical power law density profile by Bovy (2015): 

\begin{equation}
\rho (r) = A \left( \frac{r_{\rm 1}}{r} \right) ^{\alpha} \exp \left[-\left(\frac{r}{r_{\rm c}}\right)^2 \right] \label{eq:rho}
\end{equation} 
Here $r_{\rm 1}$ and $r_{\rm c}$ present reference and the cut-off radius, respectively. $A$ is the amplitude that is
applied to the potential in mass density units and $\alpha$ presents the inner power. For the the Galactic disk component, we used
the potential proposed by Miyamoto \& Nagai (1975).

\begin{equation}
\Phi_{\rm disk} (R_{\rm gc}, z) = - \frac{G M_{\rm d}}{\sqrt{R_{\rm gc}^2 + \left(a_{\rm d} + \sqrt{z^2 + b_{\rm d}^2 } \right)^2}} \label{eq:disc}
\end{equation}
Here $z$ is the vertical distance from the Galactic plane, $R_{\rm gc}$ is the distance from the Galactic centre, $M_{\rm d}$, is the mass of the Galactic disk, $G$ is the Universal gravitational constant, $a_{\rm d}$ and $b_{\rm d}$ are the scale length and scale height of the disk, respectively. The potential for the halo component was obtained from Navarro et al. (1996) and it is as follows.

\begin{equation}
\Phi _{\rm halo} (r) = - \frac{G M_{\rm s}}{R_{\rm gc}} \ln \left(1+\frac{R_{\rm gc}}{r_{\rm s}}\right) \label{eq:halo}
\end{equation} 
where $r_{\rm s}$ and $M_{\rm s}$ are the radius and mass of the dark matter halo of the Galaxy.

The orbits of the six stars around the Galactic center were determined with 1 Myr time steps and for a 10 Gyr integration
time. We used the same input data that are used in calculations of space velocity components in estimating orbital
parameters. The apo and peri Galactic distances ($R_{\rm a}$, $R_{\rm p}$), the
mean Galactocentric distance ($R_{\rm m}=(R_{\rm a}+R_{\rm p})/2$), planar and vertical eccentricities ($e_{\rm
p}$, $e_{\rm v}$) and maximum and minimum distances above the Galactic plane ($Z_{\rm max}$, $Z_{\rm min}$) were
obtained for each program star. In the calculation of $e_{\rm p}$ and $e_{\rm v}$ eccentricities, $e_{\rm p}=(R_{\rm a}-R_{\rm
p})/(R_{\rm a}+R_{\rm p}$) and $e_{\rm v}= (|Z_{\rm max}|+|Z_{\rm min}|)/R_{\rm m}$ were used, respectively. $Z_{\rm max}$ values are very close to $Z_{\rm min}$ values because the
axisymmetric approach is applied in the solutions of the Galactic potentials. The calculated orbital parameters of the stars with {\it
galpy} code are listed in Table 5 and
Galactic orbits of six stars as projected onto $X-Y$ and $X-Z$ planes are shown in Figure 4.

\begin{figure*}
\centering
\setlength{\tabcolsep}{0pt}
 \begin{tabular}{cc}
\includegraphics*[trim=3cm 0cm 3cm 0cm, width=95mm]{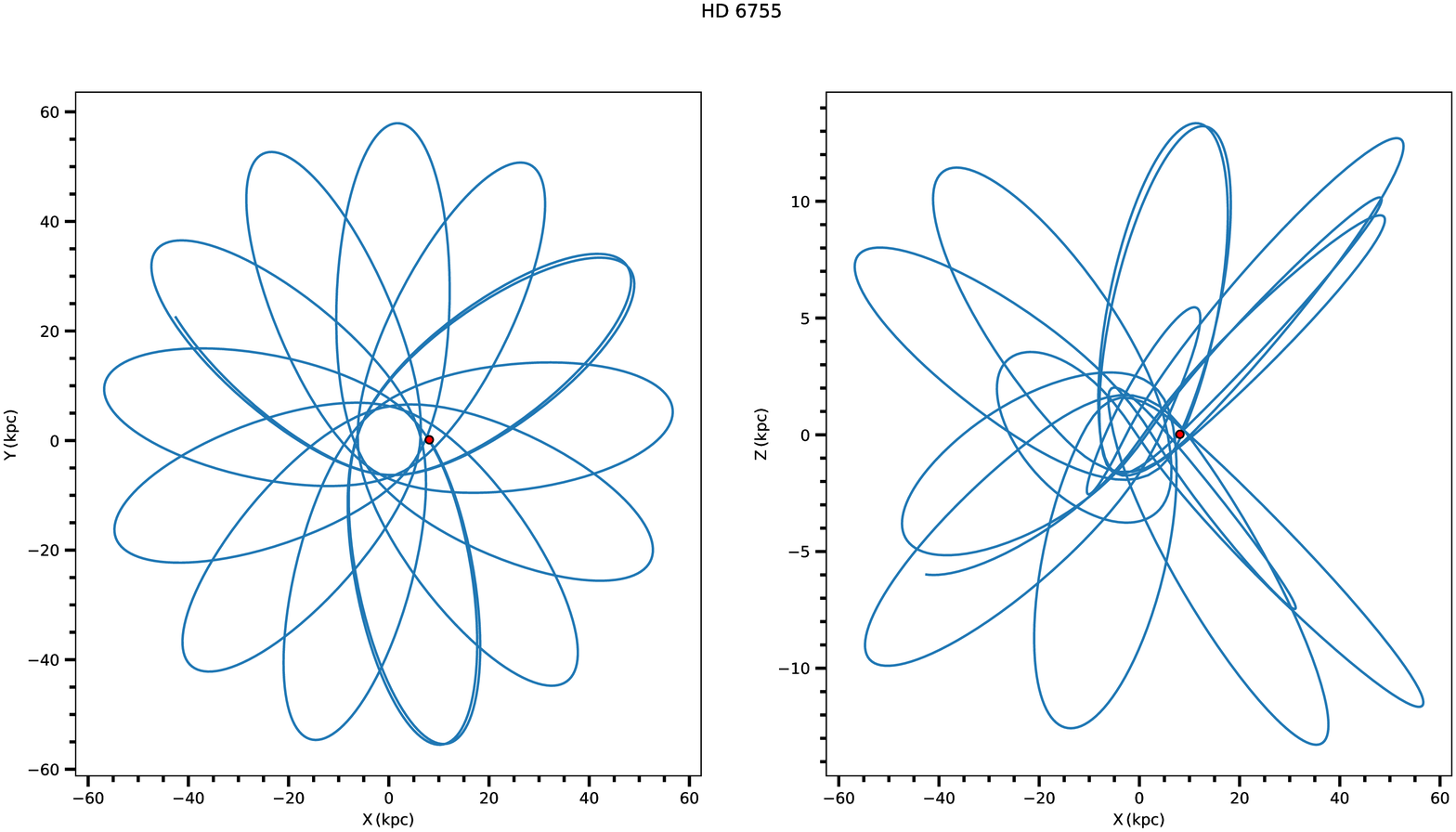} & 
\includegraphics*[trim=3cm 0cm 3cm 0cm, width=95mm]{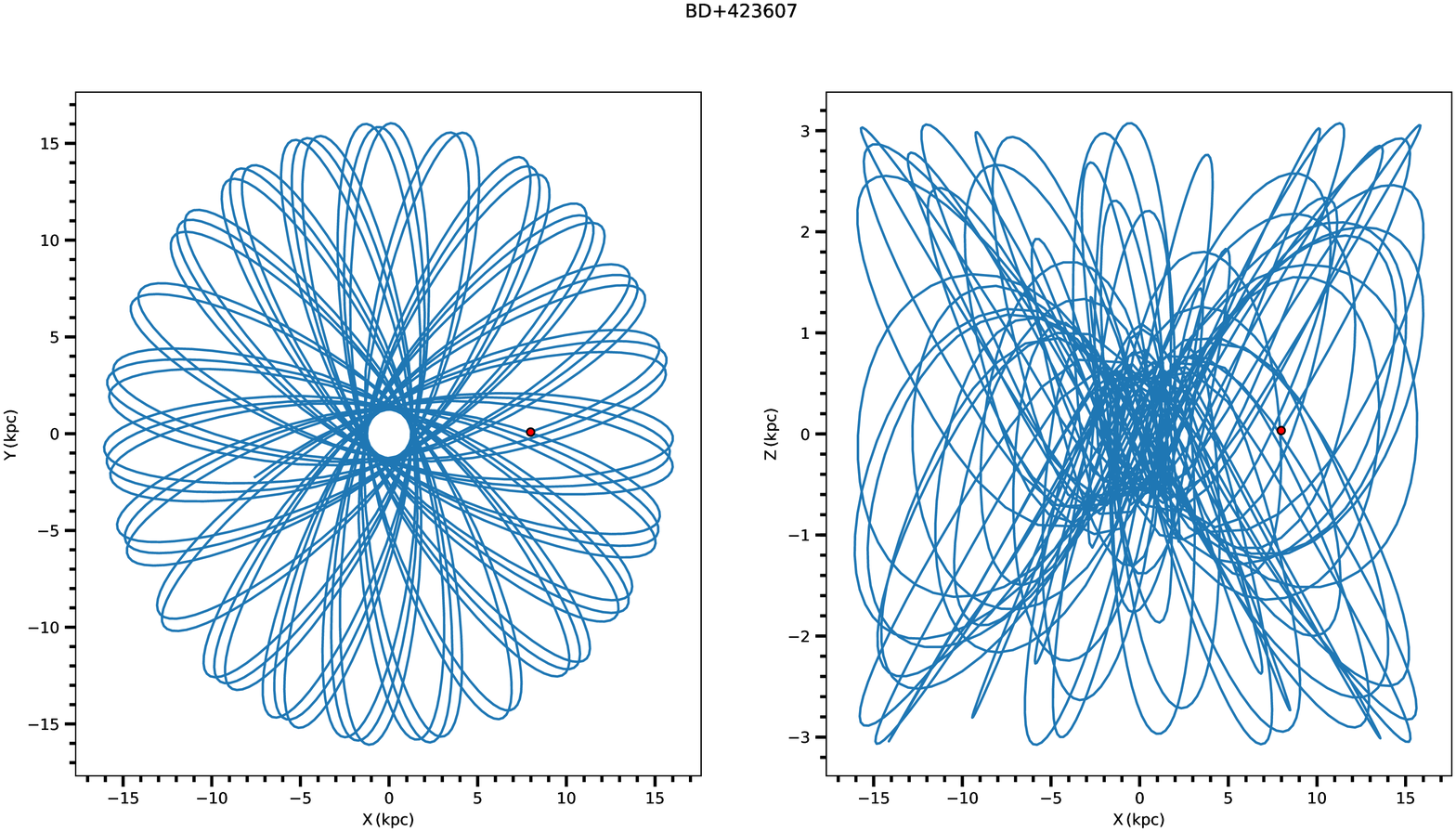} \\
\includegraphics*[trim=3cm 0cm 3cm 0cm, width=95mm]{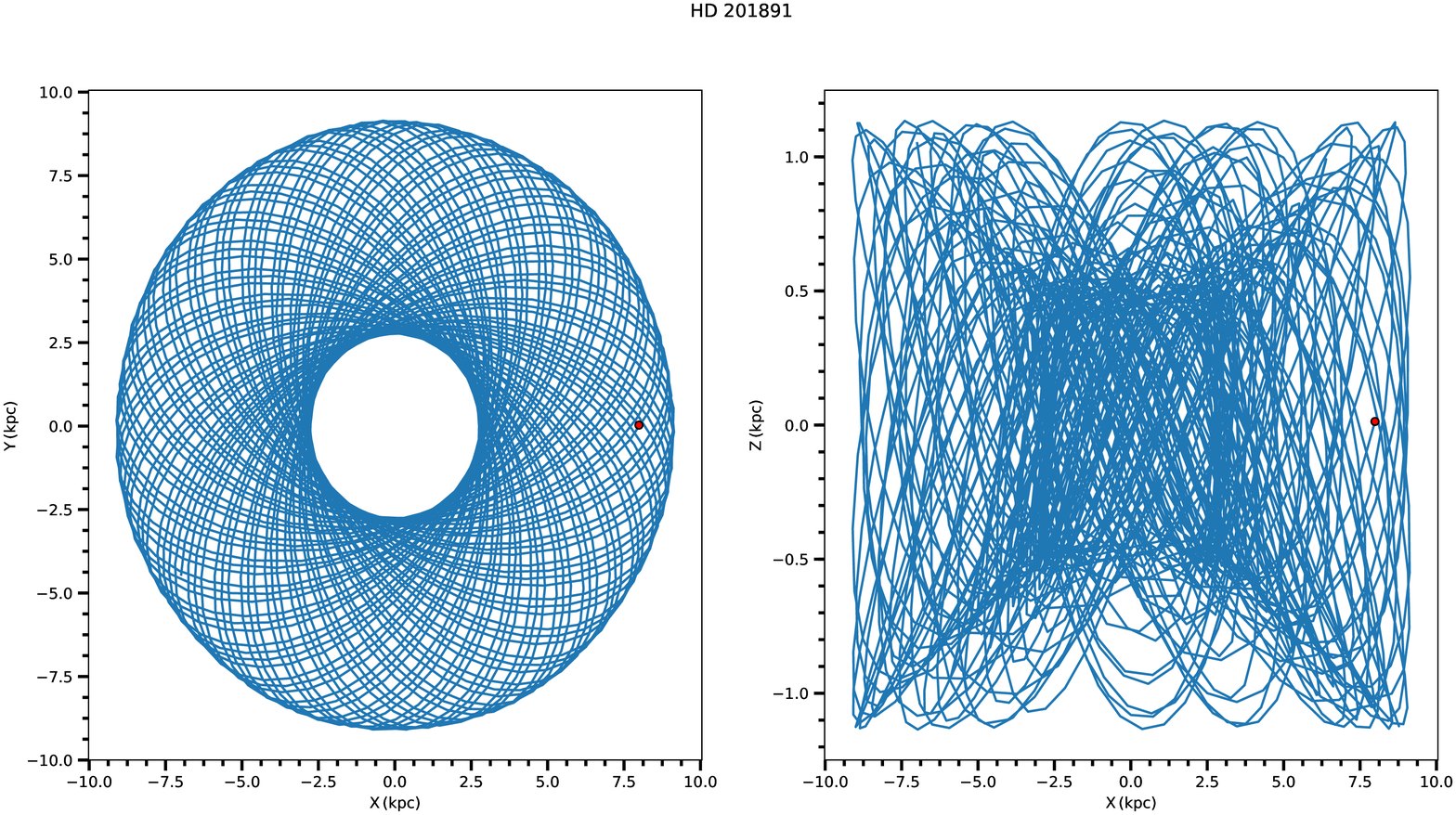} & 
\includegraphics*[trim=3cm 0cm 3cm 0cm, width=95mm]{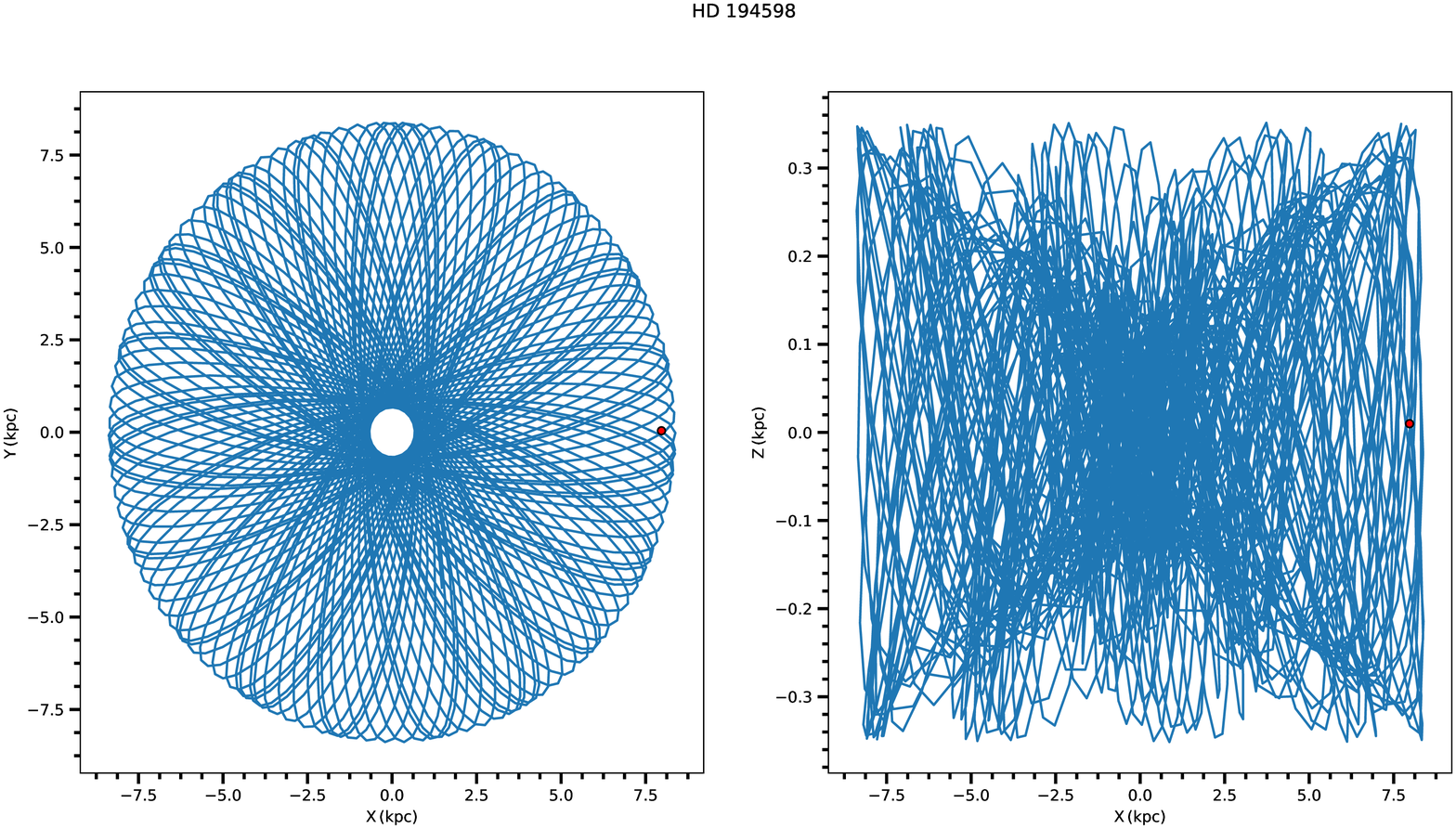} \\
\includegraphics*[trim=3cm 0cm 3cm 0cm, width=95mm]{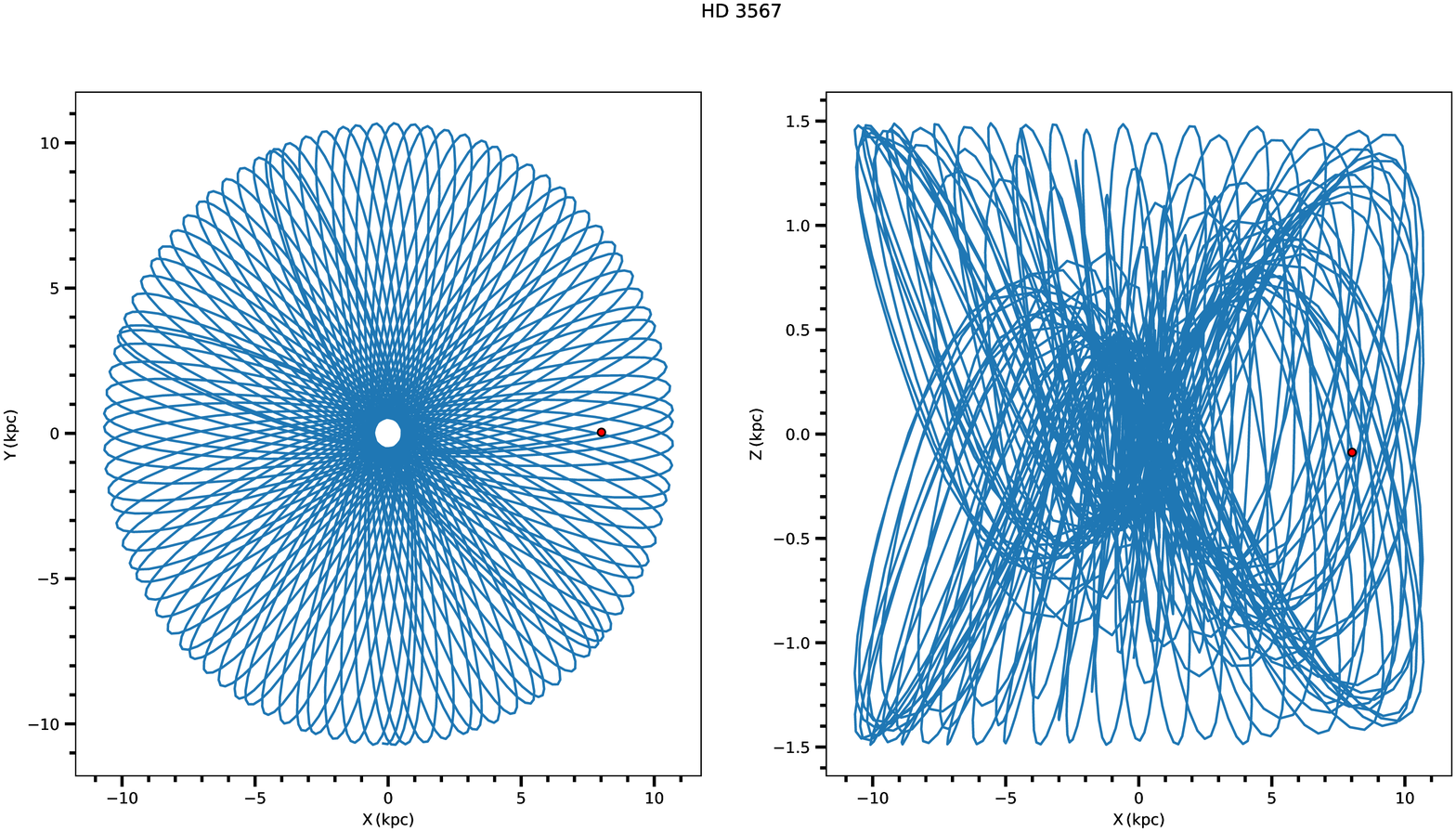} & 
\includegraphics*[trim=3cm 0cm 3cm 0cm, width=95mm]{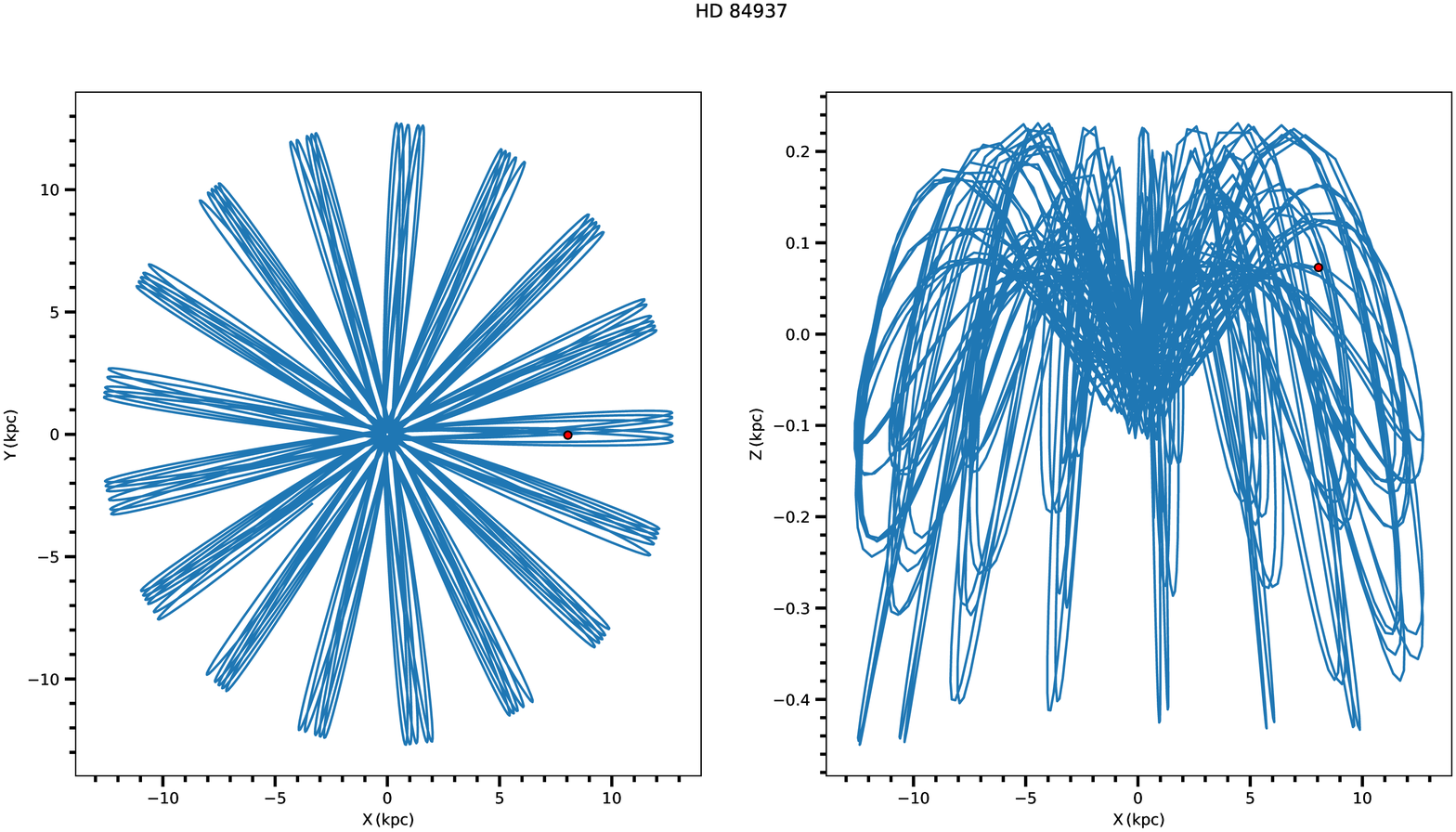} \\

\label{feh_w}
 \end{tabular}
 \caption{The computed meridional Galactic orbits and projected onto the Galactic $X-Y$ and $X-Z$ planes for six HPM stars. The filled red circles shows the present observed position for each program stars.}
\end{figure*}

\begin{figure}
\centering 
\includegraphics*[width=8.66cm,height=4.61cm,angle=0]{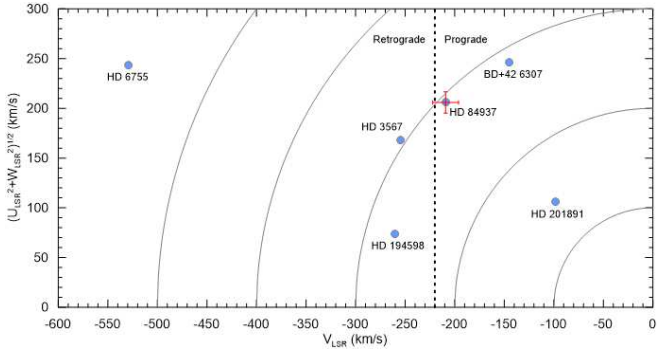}
\caption{Toomre diagram computed for six HPM stars. Solid lines (iso-velocity curves) represent constant total space velocities $S=100$, 200, 300, 400, and 500 km s$^{\rm -1}$. The uncertainties are also presented; however, they are smaller than the symbol size for five HPM stars. The dashed vertical line shows the boundary of the objects that make retrograde ($V_{LSR}<-220$ km s$^{-1}$) and prograde ($V_{LSR}\geq -220$ km s$^{-1}$) motion.}
\label{feh_w}
\end{figure}
 
\begin{figure}
\centering 
\includegraphics*[width=8.66cm,height=4.75cm,angle=0]{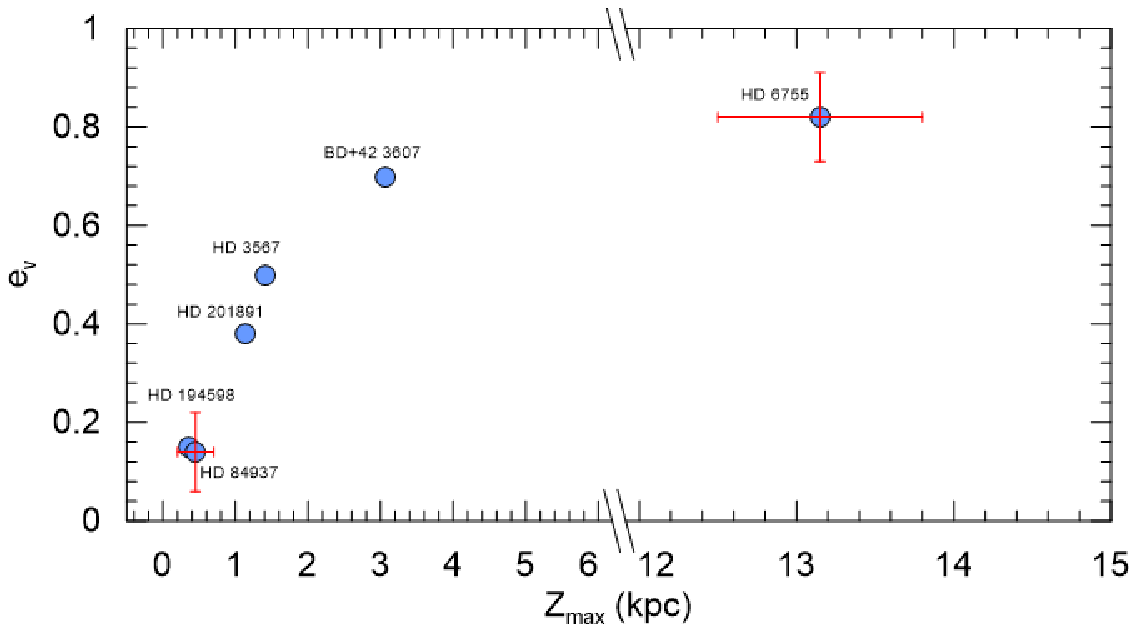}
\caption{$Z_{\rm max} \times e_{\rm v}$ diagram for six HPM stars. The uncertainties on $Z_{\rm max}$ and $e_{\rm v}$ are also indicated. In some cases, they are smaller than the symbol size.}
\label{feh_w}
\end{figure}

In order to study the radial and vertical kinetic energies of the stars, a Toomre diagram was compiled (Figure 5). In the same figure, the six stars are also marked. It is apparent that the total space velocity ($S$) of five stars in the sample is greater than 250 km s$^{\rm -1}$ and is in the range of $100<S<150$ km s$^{\rm -1}$ for HD\,201891. According to Nissen (2004), thin-disk stars have total space velocities of $S<60$ km s$^{\rm -1}$, while the total space velocities of thick-disk stars were reported to show a larger interval in velocity, i.e. $80<S<180$ km s$^{\rm -1}$. The total space velocity of the halo stars in the Solar neighbourhood is greater than 180 km s$^{\rm -1}$. On the basis of Nissen's (2004) kinematic criteria, five stars in the sample are members of the halo population and one star is a member of the thick disk population. When considering the kinematic criteria of Benbsy et al. (2003)'s, $TD/D$ values of all six stars are greater than 10 (see Table 5), indicating that the stars in the sample are members of the halo population. In the same diagram, one can also distinguish between disk and halo stars and between retrograde and prograde halo stars. It is apparent from Figure 5 that HD\,6755, HD\,3567, and HD\,194598 have retrograde motions. This may be indicating that they originated in a satellite (Sakari et al. 2018). In terms of their metal abundances, these three stars are apparently in the metal-poor tail of the thick disk.

\begin{figure*}%[!ht]
\centering \includegraphics*[width=17cm,height=12cm,angle=0]{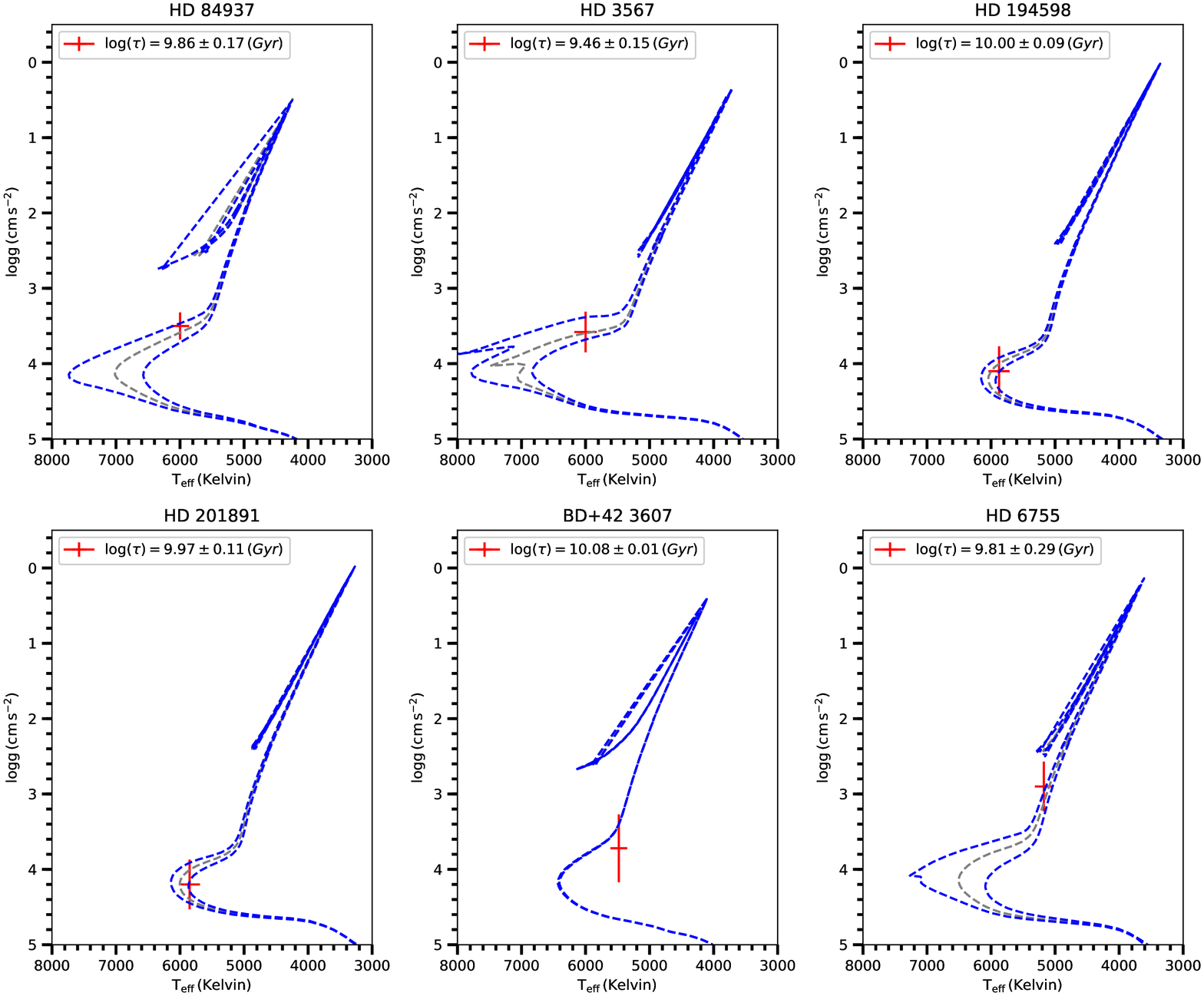}
\caption{Hertzsprung-Russell diagram for the six HPM stars, plotted with PARSEC isochrones. LTE parameters are shown as cross symbols. The errors in computed ages are also indicated.}
\label{feh_w}
\end{figure*}

\begin{figure*}%[!ht]
\centering 
\includegraphics*[trim=3cm 0cm 0cm 0cm,scale=0.42,angle=0]{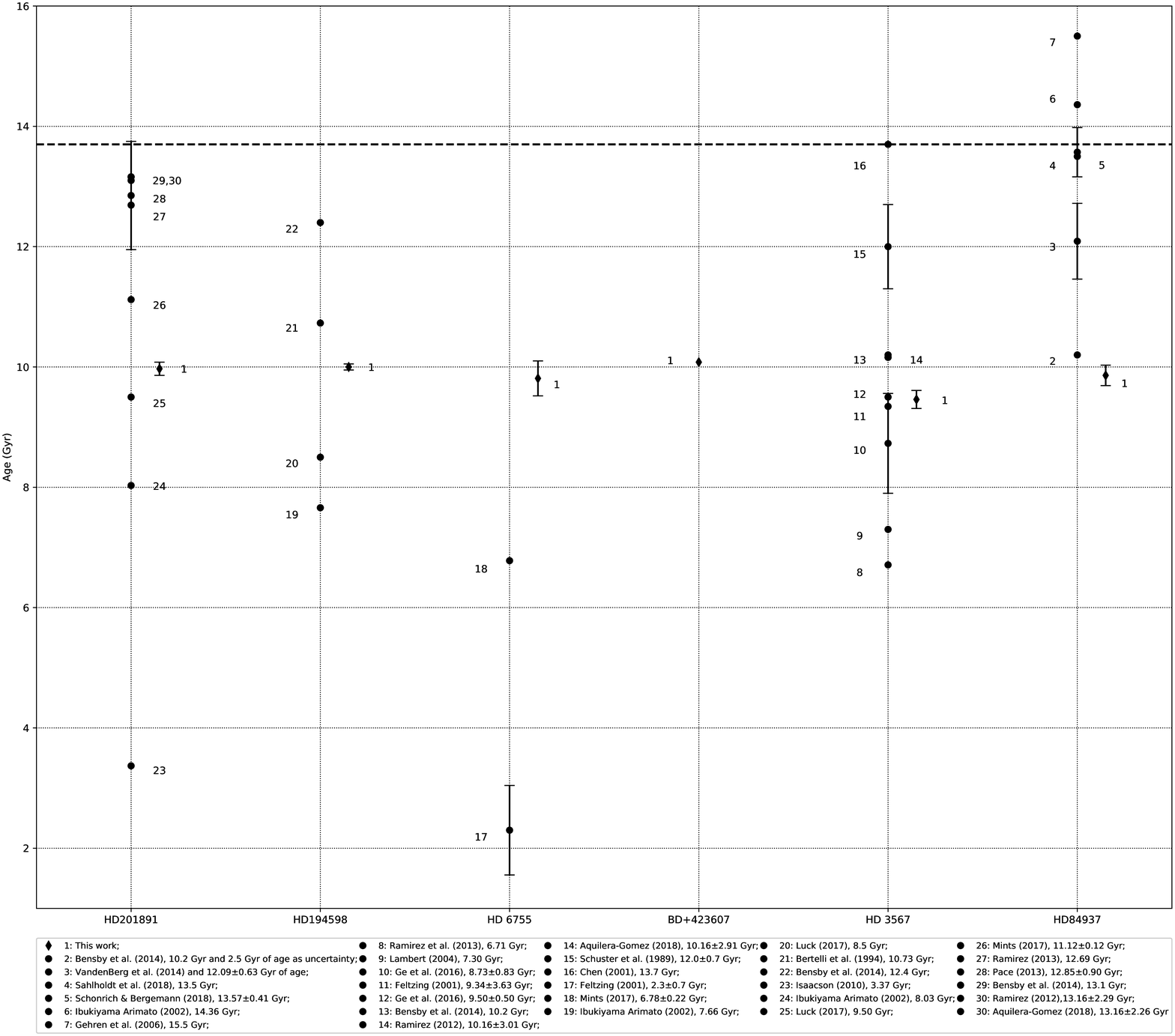}
\caption{Ages estimated in this work based on Bayesian isochrone fitting to the PARSEC models with the ages collected from the literature. The uncertainties on ages are also presented for all available age measurements in the literature;  however, they are smaller than the symbol size in some cases (e.g. 0.01 Gyr for BD+42\,3607). The horizontal dashed line indicates the age of the Universe of 13.7 Gyr as determined by WMAP (Bennett et al. 2013).}
\label{ab_gc_vs_star}
\end{figure*}

In this study, Galactic orbital parameters were also used to check the population classes of the stars. For this, stars in the sample are marked on the $Z_{\rm max} \times e_{\rm v}$ plane where their $Z_{\rm max}$ values increase towards the increasing vertical eccentricities as seen in Figure 6. Also, from Galactic structure studies (e.g. Karaali  et al. 2003; Bilir et al. 2008; G\"u{\c c}tekin et al. 2019), stars located within a distance of 2 kpc from the Galactic plane are members of the thin disk while stars within 2 and 5 kpc are members of the thick disk. The stars with distances larger than 5 kpc are considered to be members of the halo population. When stars in the sample were analysed on the basis of their spatial distribution in the Galaxy, four stars were found to be members of the thin disk, one star belongs to the thick disk and one star is a member of the halo. As seen in Table 5, the four stars with $e_{\rm v}<0.55$ have $e_{\rm p}$ values are quite large, i.e. $e_{\rm p}> 0.50$. Hence the assigned population class for these four stars as the thin-disk, is surprising. 
\\
\section{Ages of the HPM stars}
For computation of ages in this paper, we used the stellar age probability density function calculated using Bayesian statistics to estimate a precise
age with isochrone matching. The probability density functions to be used for the age computations are provided by Jorgensen \& Lindegren (2005) and
given in the following form.

\begin{equation}
f (\tau , \zeta , m) \propto f_0  (\tau, \zeta, m) \times {\cal L}(\tau, \zeta, m)
\label{eq:bayesian}
\end{equation}
where $f_0$ presents the initial probability density function given as

\begin{equation}
f_0=\psi(\tau) \varphi (\zeta | \tau) \xi(m | \zeta , \tau)
\label{eq:priors}
\end{equation}

In Equation 6, $\tau$, $\zeta$ and $m$ are the theoretical model parameters which represents age, metallicity, and mass, respectively and they are independent from each other. In Equation 7, $\psi (\tau)$, $\varphi (\zeta)$ and $\xi (m)$ represent the star formation rate, the metallicity distribution and the initial mass function, respectively. Inserting Equation 6 in Equation 7 when integrated, one obtains $G(\tau)$ function which is the probability density function provided by comparison of theoretical model parameters and observationally obtained model atmosphere parameters. It is calculated for large range of ages. In this calculation, the model parameters are compared with the available isochrones. The maximum of $G$ provides the most likely age for a given star. The isochrones are from the {\sc PARSEC} stellar evolution models for $-2.25<{\rm [Fe/H](dex)}<+0.5$ and $0<\tau {\rm (Gyr)}<13$ with 0.05 dex and 0.1 Gyr steps (in $\zeta$), respectively (Bressan et al. 2012). Figure 7 shows {\sc PARSEC} isochrones for six program stars along with the Bayesian age estimations. The further details of the method for age calculations can be found in \"{O}nal Ta\c{s} et al. (2018). The age determination via Bayesian approach provided the following range in ages for the program stars, $9.5<\tau {\rm (Gyr)}<10.1$ (see Table 6).

The isochrones can also be used to infer stellar masses. On the basis of {\sc PARSEC} isochrones, we report following mass values: 1.06$\pm$0.30\Msolar\,for HD\,6755; 0.801$\pm$0.003\Msolar\,for BD\,+42 3607; 0.982$\pm$0.071\Msolar\,for HD\,201891;
0.935$\pm$0.061\Msolar\,for HD\,194598; 1.280$\pm$0.204\\ \Msolar\,for HD\,3567 and 0.948$\pm$0.109\Msolar\,for HD\,84937 which is a {\it Gaia} benchmark star. 

Sahlholdt et al. (2019) also used Bayesian isochrone fitting method, as we did in this study, to derive age estimates for the benchmark stars including HD\,84937. This closest dwarf star with such a low metallicity has neither interferometric nor asteroseismic data that would be extremely useful to provide a precise value for $\logg$. It is important to remind the reader that although {\it Gaia} benchmark stars have been widely analysed in the literature, different authors reported different model atmosphere parameters. In fact, this was the case for all HPM stars in this study and was also main source of our motivation to derive their updated model parameters. We determined the age of HD\,84937 as 9.86$\pm$0.17 Gyr. This value is in excellent agreement with Bensby et al. (2014) who reported 10.2$^{+1.9}_{\rm -2.5}$ Gyr for the star. Earlier studies (e.g. Casagrande et al.
2011; VandenBerg et al. 2014) reported somewhat higher values of age such that they were comparable to the age of the Universe (13.7 Gyr by WMAP; Bennett et al. 2013). For instance, VandenBerg et al. (2014) reported 12.09$\pm$0.63\footnote{The error in age is derived from the \Teff\, error bar.} Gyr of age for the star. It is interesting to note that the ages reported by Gehren et al. (2006) and for instance, by Ibukiyama \& Arimoto (2002) was even larger than 13.7 Gyr. Sahlholdt et al. (2019) reported an age of 13.5 Gyr for HD\,84937 and provided 11.0 - 13.5 Gyr as a range of age for the star. In this context, it should be mentioned that the isochrones employed in Sahlholdt et al. (2019) and in this study were created with the same resolutions in age and in metallicity. As a health check, we repeated the calculation by Sahlholdt et al. (2019) by using the model atmosphere parameters that they reported for the star. We confirmed their computed age for HD\,84937. However, by using up-to-date model parameters obtained in this study via spectroscopy, the star seemed to be $\approx$3.6 Gyr younger. Their age estimate of 13.5 Gyr remains same for {\sc PARSEC}-isochrones when fitting to the magnitude or the $\log g$ (see also their Figure A1) and the alternative grids of isochrones that they preferred provided somewhat larger values for age. In Figure 8, we provide age estimates for all HPM stars in this work and from the literature. 

Luck (2017) reported masses and ages for HD\,201891 and HD\,194598. The reported mass and age by Luck (2017) for HD\,201891 from BaSTI\footnote{2016, BaSTI Ver.5.0.1: http://basti.oa-teramo.inaf.it/} isochrones are 0.86 \Msolar\, and 9.50 Gyr, respectively. We reported a mass of 0.98$\pm$0.07 \Msolar\, for the star from the {\sc PARSEC} isochrones. Our determination of age for HD\,201891 from the {\sc PARSEC} isochrones is 9.97$\pm$0.11 Gyr. The agreement is satisfactory. For HD\,194598, we obtained a mass of 0.94$\pm$0.06\Msolar\, and 10.0$\pm$0.05 Gyr of age from the {\sc PARSEC} isochrones. Luck's (2017) reported age for the star from the BaSTI  isochrones is 1.5 Gyr younger. An alternative isochrones from Bertelli et al. (1994) by Luck (2017) provided 10.73 Gyr of age for HD\,194598. To validate Luck's (2017) result, we also computed mass and age of the star using Bertelli's isochrones but with the model parameters determined in this study. The results are 0.82$^{\rm +0.04}_{\rm -0.01}$ \Msolar\, and 8.46$^{\rm +0.82}_{\rm -0.89}$ Gyr, respectively. When Luck's (2017) model atmosphere parameters used (agreed with ours within error limits) for the star, we obtained a mass of 0.86 \Msolar\, which is in excellent agreement with Luck's (2007) reported mass from Bertelli's isochrones (1994) for HD\,194598.
\newpage
\section{Results and discussion}
In addition to kinematics and Galactic orbits, since we also aim to spur discussion on the origin of the HPM stars under scrutiny in the context of abundances, we derived up-to-date photospheric abundances of 29 species including not only $\alpha$-elements, but also slow (s)- and rapid (r)-process elements from Y to Sm. The final elemental abundances log $\epsilon$(X) averaged over the sets of measured lines for the stars are listed in Tables A7 and A8, where the first column shows species, the second - logarithmic elemental abundances and the third - element over iron ratios\footnote{[X/Fe]=[log$\epsilon$(X)-log$\epsilon$(Fe)]$_{\rm star}$ - [log$\epsilon$(X)-log$\epsilon$(Fe)]$_{\rm \odot}$}. The number of lines used in this analysis are also given. The last columns in those tables present computed solar abundances by us in this study. The errors reported in logarithmic abundances present 1$\sigma$ line-to-line scatter in abundances. The error in [X/Fe] is the square root of the sum of the quadrature of the errors in [X/H] and [Fe/H]. The formal errors for the abundances arising from uncertainties of the atmospheric parameters $T_{\rm eff}$, $\logg$ and $\xi$ are summarized in Table A10 for changes with respect to the model. For a further discussion on chemical abundances including up-to-date photospheric abundances of 29 species in the spectra of HPM stars, the reader is referred to the appendix section.

\begin{figure*}%[!ht]
\centering \includegraphics*[width=19cm,height=13cm,angle=0]{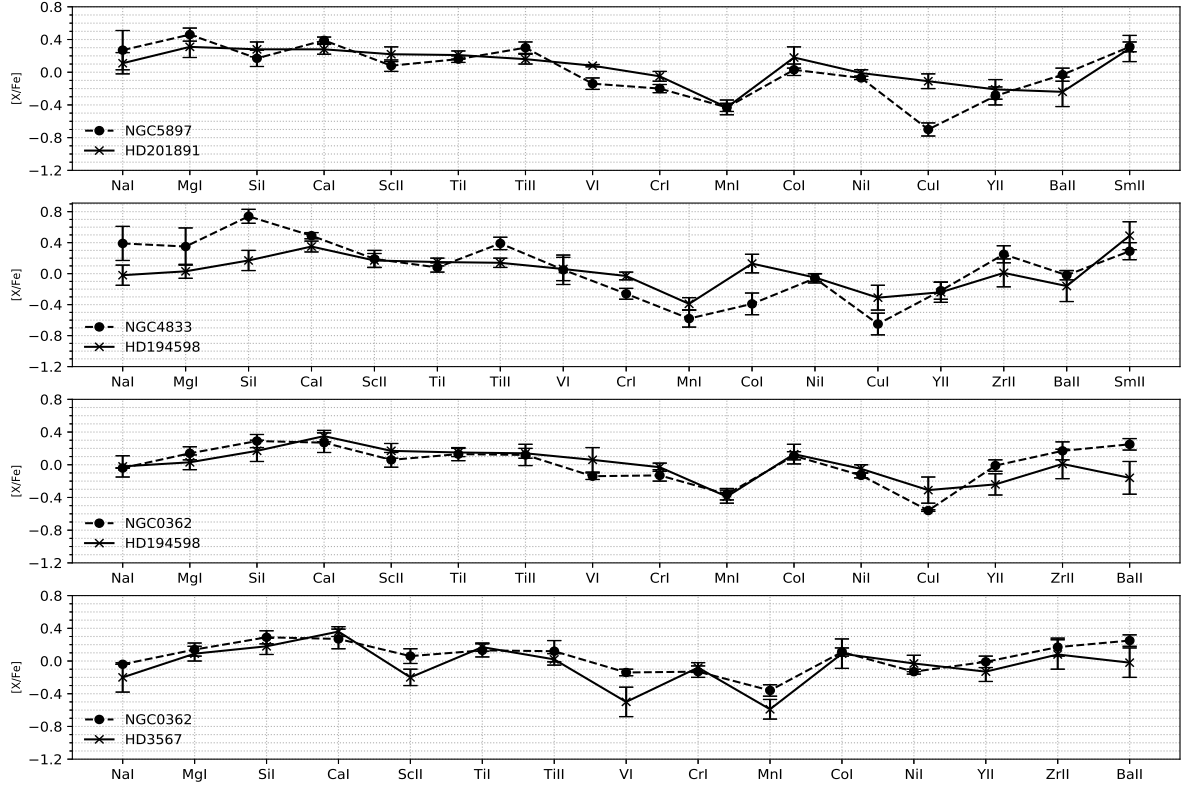}
\caption{Abundances of NGC\,5897, NGC\,4833, and NGC\, 362 along with the abundances of HD\,201891, HD\,194598, and HD\,3567.}
\label{ab_gc_vs_star}
\end{figure*}

\begin{table*}
\setlength{\tabcolsep}{3pt}
\scriptsize
\caption{Parent globular cluster candidates for the program stars.
The Galactic orbital parameters, metallicities, ages as well as their errors, the encountering probabilities ($P$($\%$)) and the average encounter velocities for the program stars for 5 and 1.5 tidal radii are also presented. The uncertainties available from the literature for the parent GCs candidates were also included (e.g. Helmi et al. 2018). For some, no uncertainty was reported.}
\centering
\begin{tabular}{l|c|c|c|c|c|c|c|c|c|c|c||c|c|c|}
\hline\hline
	  & \multicolumn{2}{c|}{$R_{\rm apo}$} & \multicolumn{2}{c|}{$R_{\rm peri}$}  & \multicolumn{2}{c|}{$e_{\rm p}$} 
	  &\multicolumn{2}{c|}{[Fe/H]}&\multicolumn{2}{c|}{Ages} & & GC&$P$($\%$)$_{\rm 5.0}$ & $V_{\rm enc}$\\
\cline{2-12}
 Star     &Star&GC& Star&GC& Star   & GC &    Star   &GC$^{\rm *}$  & Star& GC & Ref. &  & $P$($\%$)$_{\rm 1.5}$ & \\
\cline{2-15}
%&&& &&    & &    & & & &  & &  \\
          &\multicolumn{2}{c|}{(kpc)}& \multicolumn{2}{c|}{(kpc)}&   \multicolumn{2}{c|}{} &\multicolumn{2}{c|}{(dex)} & \multicolumn{2}{c|}{(Gyr)} & & & & (km s$^{-1}$)\\
\hline \hline
HD 6755   &58.07& {\bf 48.10$^{+3.80}_{-2.80}$}&6.33& {\bf 6.17$^{+0.58}_{-0.46}$}&0.80 &{\bf 0.77$^{+0.01}_{-0.01}$}&-1.39 &{\bf -2.31}& 9.81&{\bf 12.50}&1&{\bf NGC 5466}&{\bf 15.92} & {\bf 11$\pm$17}\\
& (3.15) &  & (0.01)  & &(0.06)  & &(0.18)   & {\bf (0.09)} & (0.29) &{\bf (0.25)}&  &  &{\bf 5.45} &\\
\cline{1-15}
& & {\bf 17.82$^{+0.35}_{-0.30}$}& & {\bf 1.68$^{+0.19}_{-0.16}$}& & {\bf 0.83$^{+0.01}_{-0.01}$} & &{\bf -1.96} & &{\bf 10.32} & 2 &{\bf NGC 2298}& {\bf 5.60} & {\bf 76$\pm$62}\\
& & &         &                             &        &                              &      &{\bf (0.04)}&                &        {\bf (1.54)} & & & {\bf 4.94}  &\\
\cline{14-14}
BD+423607 &16.15   & 14.29$^{+0.23}_{-0.24}$      & 1.33    & 1.18$^{+0.04}_{-0.05}$      &0.85    & 0.85$^{+0.01}_{-0.01}$ &-2.43    &-1.18   &10.08   &10.90  & 3 & NGC 2808 & 5.55 & 110$\pm$122\\
          & (0.05) &                              & (0.01)  &                             &(0.01)  &                        &(0.17)   & (0.04) &(0.01)  &(0.70) &   & & 1.97  &\\
\cline{14-14}
          &     & 16.72$^{+0.64}_{-0.61}$      &         & 1.49$^{+0.23}_{-0.18}$      &        & 0.84$^{+0.01}_{-0.02}$ &         &-1.29   &        &9.98   & 4 &NGC 6864 &  2.26 & 167$\pm$167 \\  
&     & &  &&        &                        &         &(0.14)  &        &(---)  &   &   & 0.85  &\\
\cline{1-15}
    &     &  7.40$^{+0.03}_{-0.03}$      &      & 3.14$^{+0.08}_{-0.09}$      &      & 0.40$^{+0.01}_{-0.01}$       &      &-2.19       &         & 12.54  & 5 & NGC 4372 &  14.68 & 12$\pm$27\\
    &     &                              &      &                             &      &                              &      &(0.08)      &         & (---)  & &  & 6.16   &\\
\cline{14-14}
    &     & {\bf 9.11$^{+0.41}_{-0.40}$} &      & {\bf 2.82$^{+0.28}_{-0.30}$}&      & {\bf 0.53$^{+0.02}_{-0.02}$} & &{\bf -1.90} & & {\bf 10.10}& 6 &{\bf NGC 5897}&{\bf 31.88} &{\bf 24$\pm$50}\\
    &     &                              &      &                             &      &                              &      &{\bf (0.06)}& &   {\bf (1.10)}& &  &{\bf 17.66} &\\
\cline{14-14}
HD 201891 & 9.16& 7.55$^{+0.40}_{-0.34}$      & 2.84   & 2.64$^{+0.35}_{-0.30}$     &0.53  & 0.48$^{+0.03}_{-0.03}$       &-1.02  &-0.35       &9.97     & 11.35 & 7 & NGC 6356&  24.44 & 95$\pm$72\\
          & (0.01)&                         & (0.01) &                             &(0.01)  &                             &(0.16) & (0.14)     &(0.11)   &  (0.41) &  & &  6.46 &\\
\cline{14-14}
          &     &  {\bf 8.60$^{+0.15}_{-0.10}$}&      & {\bf 2.71$^{+0.04}_{-0.06}$}&      & {\bf 0.48$^{+0.01}_{-0.01}$} & &{\bf -1.70} & &{\bf 12.50}  & 1 & {\bf NGC 6656}&{\bf 17.27} & {\bf 89$\pm$92}\\
          &     &                              &      &                             &      &                              &      &{\bf (0.08)}&  &{\bf (0.50)} & &  & {\bf 5.92}  &\\
\cline{14-14}
          &     & 10.70$^{+0.37}_{-0.16}$      &      & 3.05$^{+0.06}_{-0.10}$      &      & 0.53$^{+0.01}_{-0.01}$       &      &-2.33       &  & 6.98   & 2 & NGC 7078& 22.85 &23$\pm$26\\	       
          &     &                              &      &                             &      &                              &      &(0.02)      & & (1.19)  & &  & 11.07 &\\
\cline{1-15}
          &     & {\bf 8.21$^{+0.38}_{-0.57}$} &      & {\bf 0.65$^{+0.05}_{-0.11}$}&      &{\bf 0.60$^{+0.01}_{-0.01}$}  & &{\bf -1.31} & & {\bf 10.21} & 2 &{\bf NGC 6284}&{\bf 23.13} & {\bf 16$\pm$28}\\
          &     &                              &      &                             &      &                              & &{\bf (0.09)}&  & {\bf(0.12)} & &  & {\bf 11.80}  &\\
\cline{14-14}
          &     & 5.93$^{+0.17}_{-0.16}$       &      & 0.85$^{+0.11}_{-0.13}$      &      &0.75$^{+0.03}_{-0.02}$        &      &-1.29       &  & 10.56  & 2 & NGC 5946 &  0.00  & 151$\pm$152\\
          &     &                              &      &                             &      &                              &      &(0.14)      & &    (1.93)     & & & 0.00 &\\
\cline{14-14}
HD 194598 &8.39   &11.91$^{+0.29}_{-0.13}$     & 0.76   & 0.08$^{+0.05}_{-0.03}$    &0.83    &0.92$^{+0.03}_{-0.02}$ &-1.17   &-1.30    &10.00    & 10.75   & 1 & NGC ~362 & 15.18 & 59$\pm$80\\
          &(0.01) &                            & (0.01) &                           &(0.01)  &                       &(0.18)  & (0.04)  & (0.05)  & (0.25)  & &        & 6.42 &\\
\cline{14-14}
          &     & 7.86$^{+0.15}_{-0.12}$       &      & 0.93$^{+0.02}_{-0.09}$      &      &0.79$^{+0.01}_{-0.01}$        &      &-1.89       & & 12.50   & 1 & NGC 4833&  2.00 &172$\pm$205\\
          &     &                              &      &                             &      &                              &      & (0.05)     & & (0.50)  & &  & 0.50  & \\
\cline{14-14}
          &     & 8.82$^{+1.48}_{-0.46}$       &      & 0.98$^{+0.30}_{-0.58}$      &      &0.74$^{+0.03}_{-0.03}$        &      &-1.90       & & 10.10  & 6 & NGC 5897& 18.85 &83$ \pm$51\\	       
          &     &                              &      &                             &      &                              &      & (0.06)     &  & (1.10)& && 6.94 &\\
\cline{1-15}
          &     & {\bf 11.78$^{+0.21}_{-0.26}$}&      & {\bf 0.67$^{+0.06}_{-0.13}$}&  &{\bf 0.90$^{+0.02}_{-0.01}$}  & &{\bf -1.30} & & {\bf 10.75}   & 1 &{\bf NGC ~362}& {\bf 7.56} & {\bf 171$\pm$154}\\
          &     &                              &      &                             &      &                              &      &{\bf (0.04)} && {\bf (0.25)}  & &  &{\bf 5.46} &\\
\cline{14-14}
HD 3567   &10.78   & 10.61$^{+0.13}_{-0.13}$   & 0.60   & 0.98$^{+0.03}_{-0.00}$    &0.89   &0.83$^{+0.01}_{-0.01}$ &-1.10    &-2.35       &9.46     & 12.75   & 1 & NGC 6341& 5.55 & 223$\pm$187\\
          &(0.04)  &                           & (0.03) &                           & (0.01)&                       & (0.18)  & (0.05)  &(0.15) & (0.25)  & &  & 1.77 & \\
\cline{14-14}
          &     & 12.32$^{+0.24}_{-0.29}$      &      & 0.63$^{+0.08}_{-0.06}$      &      &0.90$^{+0.01}_{-0.01}$        &      &-2.00       & & 12.75   & 1 & NGC 6779 & 5.66 & 103$\pm$132\\
          &     &                              &      &                             &      &                              &      &(0.09)      &  & (0.50)  & & & 2.16 &\\
\cline{1-15}
\cline{14-14}
          &     & {\bf 10.62$^{+0.41}_{-0.36}$}&      & {\bf 0.18$^{+0.10}_{-0.08}$}&      &{\bf 0.76$^{+0.01}_{-0.01}$}  & &{\bf -2.35} & &{\bf 12.75} & 1 & {\bf NGC 6341}& {\bf 12.19} & {\bf 151$\pm$159}\\
          &     &                              &      &                             &      &                              &      &{\bf (0.05)} &&{\bf (0.25)}& & &{\bf 4.36} &\\
\cline{14-14}
          &     & 12.93$^{+0.64}_{-0.42}$      &      & 0.36$^{+0.21}_{-0.21}$      &      &0.89$^{+0.01}_{-0.02}$        &      &-1.32       &  & 11.50   & 1 & NGC ~288 & 14.83 & 89$\pm$133\\
          &     &                              &      &                             &      &                              &      &(0.02)      & & (0.38)  & &  & 4.77 &\\
\cline{14-14}
HD 84937  &12.74 & 11.91$^{+0.29}_{-0.13}$     & 0.23  & 0.08$^{+0.05}_{-0.03}$     &0.96  &0.92$^{+0.03}_{-0.02}$ &-2.40  &-1.30 &9.86    & 10.75   & 1 & NGC ~362 & 3.24 & 265$\pm$160 \\
          &(0.80)&                             & (0.15)&                            &(0.10)&                       &(0.15) &(0.04)&(0.17)  & (0.25)  & & &  1.08 &\\
\cline{14-14}
          &     & 12.11$^{+0.14}_{-0.10}$      &      & 0.36$^{+0.13}_{-0.10}$      &      &0.80$^{+0.01}_{-0.01}$        &      &-2.00       & & 12.75   & 1 & NGC 6779 & 12.14 & 96$\pm$137\\    
          &     &                              &      &                             &      &                              &      &(0.09)      & & (0.50)  & &        & 4.48 &\\
\hline 
\hline
\end{tabular}
\begin{list}{}{}
\item (1) Vandenberg et al. (2013), (2)  Cezario et al. (2013), (3) Massari et al. (2016), (4) Catelan et al. (2002), (5) de Angeli et al. (2005), (6) Salaris \& Weiss (1998), (7) Koleva et al. (2008), (*) Carretta et al. (2009). 
\end{list}
\end{table*}

Given an understanding of the chemical abundances and kinematics of the HPM stars presented in this study, it is worthwhile to consider the origin of these stars. More specifically, how have they been accelerated to such high velocities? Tidal interactions in a globular cluster (GC) or Galaxy interactions could be a possible source for HPM stars. In order for stars to reach the high velocities observed in this study, it is likely they have been part of a complex three or more body interaction. The dense cores of GCs represent one of the most likely sites for such interactions, as they typically have both a relatively high binary fractions\footnote{For GCs, the modest (photometric) binary fractions range from 1-10\%.} (Milone et al. 2012) and encounter rates (Leigh \& Sills 2011). In fact, searching for a dynamic connection between HPM stars and Galactic GCs is now possible in the {\it Gaia} era, as the proper motions and parallaxes of most GCs and nearby  dwarf spheroidal galaxies (dSph) are now known (Helmi et al. 2018, Vasiliev et al. 2019). 
 
The ages and metalicities of the HPM stars suggest they are all old and metal-poor, also consistent with once being members of a GC population. While its entirely possible the stars were ejected from a cluster that has since reached dissolution, it is interesting to note that the metallicity of the six HPM are comparable to several Galactic GC within uncertainty. Harris (1996; 2010 edition) reported distances, velocities, metallicities, luminosities as well as dynamical parameters for 157 globular clusters in the Milky Way. The up-to-date orbital parameters reported for the common GCs from Helmi et al. (2018) with the help of these metallicities from Harris (1996) were used to estimate origin of six HPM stars in this study. For this aim, we compiled the Galactic orbital parameters (i.e. $R_{\rm a}$, $R_{\rm p}$, $e_{\rm p}$), the metallicities, ages of each HPM star and cluster and listed them in Table 6. In this table, we only included GC candidates that have similar Galactic orbital parameters, metallicities and ages compared to those obtained in this study for the program stars. As it can be inspected from Table 6, even for such a crude comparison, the similarities in Galactic orbital parameters, metallicities and ages are intriguing. In Table 6, the best matching GCs are indicated in bold-type face for clarity. This table also provides encounter probabilities that happens to provide an alternative diagnostic when it comes to testing dynamical origin of the program stars.

The ages of the GCs listed in Table 6 were compiled from seven different studies (Massari et al. 2016, Vandenberg et al. 2013, Cezario et al. 2013, Koleva et al. 2008, de Angeli et al.  2005, Catelan et al. 2002, and Salaris \& Weiss 1998). For these studies, the most homogeneous study of the age of the GCs belongs to Vandenberg et al. (2013). VandenBerg et al. (2013) derived the ages for 55 GCs in the Milky Way and nicely presented their variations with the metallicity. Their computations of mean ages for the GCs as derived by the grouping of [Fe/H] $<$ -1.7 dex and [Fe/H] $>$ -1 dex provided respective ages of 12.5 and 11 Gyr. They estimated errors in age to be 0.25 Gyr, mainly caused by applications of the isochrones to observed data. Also, the errors due to uncertainties in distances and cluster metal abundances ranged from 1.5 to 2 Gyr. The mean ages of the globular clusters with the reported errors in their ages by VandenBerg et al. (2013) show agreement with the mean ages (10 Gyr) and their errors obtained in this study for the program stars. 

In order to test the dynamical origin of the program stars, we followed the same procedure as in Pereira et al. (2017). The orbital evolution of the HPM stars were simulated over 12 Gyr into the past with the orbits of 16 candidate GCs from the literature. As mentioned in Section 4, {\it MilkywayPotential2014} (Bovy 2015) was used to calculate the Galactic orbits of stars and globular clusters. The initially determined orbital parameters of the HPM stars and the GCs are randomly varied. The simulations for each program star were repeated 5000 times considering different orbital parameters for the stars and the clusters within their observational uncertainties. During these orbital simulations, errors in their proper motions, heliocentric distances, and  radial velocity in the equatorial coordinates are considered for HPM stars (see Table 4). The parameters and errors used in the calculation of the orbital parameters of the GCs are also taken from Helmi et al. (2018). For each simulation, the probability for a close encounter between star and the GC was calculated at a certain distance. For a close encounter, the distance to any cluster was assumed to be smaller than 5 tidal radii. The tidal radii for the GCs were obtained from Moreno, Pichardo Velazquez (2014). Table 6 summarizes the encounter probabilities for each HPM star and a candidate GC for distances that corresponds to 5 and 1.5 tidal radii. Moreover, we also computed the average encounter velocities for HPM program stars and they are also reported in Table 6. Orbital simulations have shown that HPM stars have a 5\% to 30\% for 5 tidal radii and 4\% to 18\% probability of encountering with selected GCs for 1.5 tidal radii. The larger values of the listed encounter probabilities are likely to be related to a GC origin.

To further constrain the list of possible GC progenitors, we also compiled abundances of the GC candidates to check whether they also show similar abundances. Figure 9 (also Table A11) presents abundances of HPM star/GC candidate. The discrepant abundances are shown in bold type-face in Table A11. There is a fair agreement for several elements between abundances of HPM stars and assigned GC candidates. Such comparison may also provide valuable information on identifying elements and/or certain lines to trace a GC progenitor for an HPM star.

\begin{figure*}%[!ht]
\centering \includegraphics*[width=17cm,height=21.72cm,angle=0]{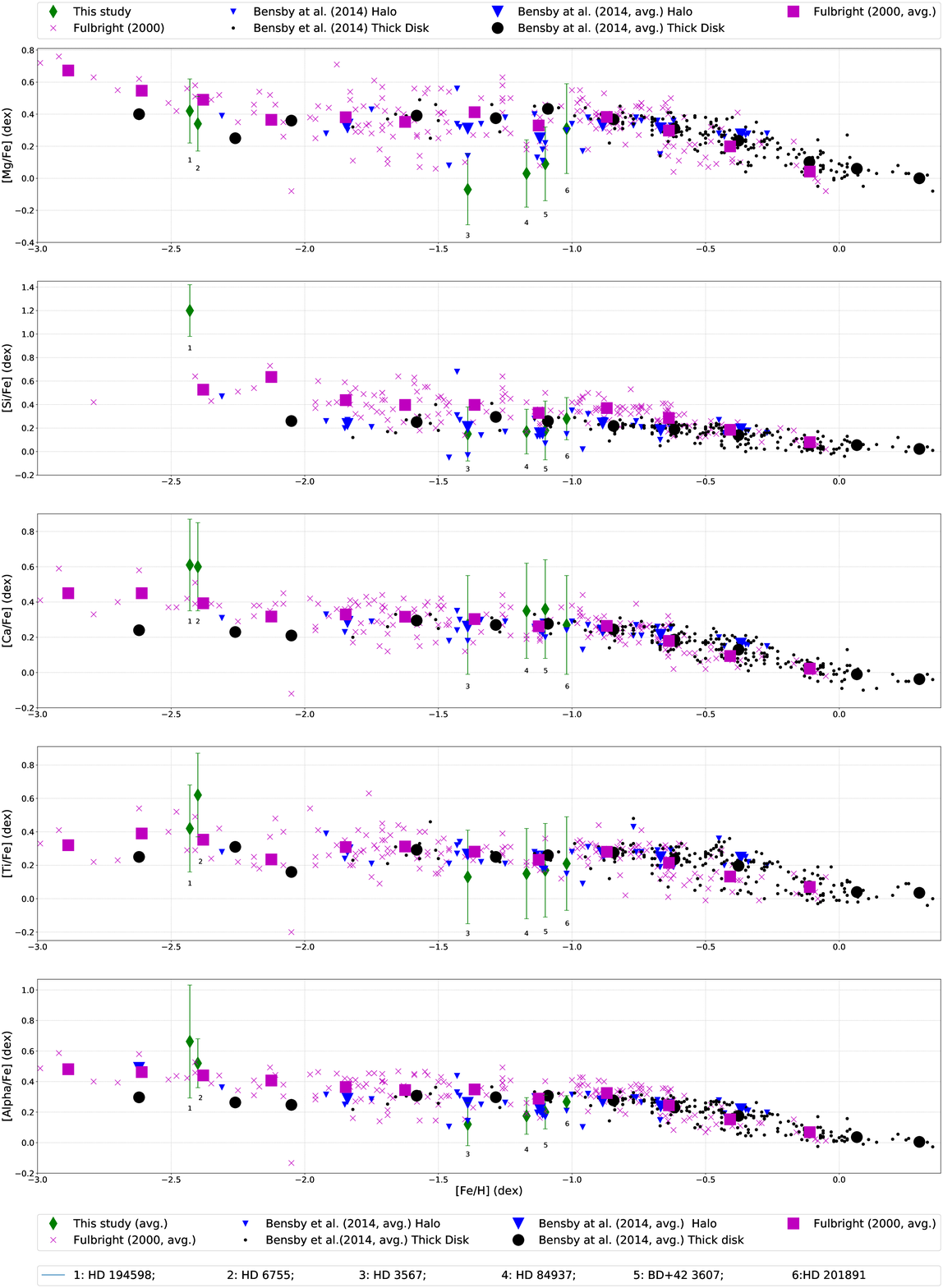}
\caption{Elemental abundances for the four $\alpha$-elements relative to Fe. A representative sample of stars from Bensby et al. (2014) for the Galactic thick disk and halo are also marked by small circles and downward triangles, respectively, and the abundances of the individual elements for the metal poor dwarfs from Fulbright (2000) and the HPM stars in this study by cross and diamond symbols, respectively. Individual error bars in [X/Fe] axis for the HPM stars are also indicated. The large black circles and large downward triangles are average abundances in 0.25 dex intervals of [Fe/H] for thick disk and halo stars, respectively. The large squares show the mean values for the metal poor dwarfs from Fulbright (2000) as indicated in the legend at the bottom. The bottom panel shows the mean values of the four elements for both sample stars and the HPM stars.}
\label{ab_gc_vs_star}
\end{figure*}

Further on the Galactic population classification of the HPM stars, in Figure 10, by considering the abundances determined in this study for magnesium, silicon, calcium, and titanium, we also provided [$\alpha$/Fe] for each of the HPM stars for which the mean values of four elements were used as estimates of [$\alpha$/Fe] (see bottom panel in Figure 10). The [$\alpha$/Fe] values for a representative sample of the Galactic thick disk and halo stars from Bensby et al. (2014) as well as halo and disk stars (including metal-poor dwarf stars) from Fulbright (2000) are shown for comparison of $\alpha$-element abundances. It is apparent from Figure 10 that all HPM stars are rich in $\alpha$-elements. The over all agreement is satisfactory in silicon, calcium, and titanium abundances. Between among the $\alpha$-elements, Ca is generally considered to be true representative of the $\alpha$-elements in the literature since O, Si, and Mg are often thought to be altered due to the recycling of the products of an earlier generation of stars during subsequent star formation inside GCs (Kraft et al. 1997, Gratton et al. 2004, Carretta et al. 2010, Gratton et al. 2012). We also measured the oxygen abundance in the spectra of the six stars by using the triplet oxygen lines at 6156 \AA. To be more specific, we employed O\,{\sc i} line at 6156 \AA~for HD\,194598 and HD\,84937; the OI line at 6158 \AA~ for HD\,6755; the OI lines at 6156 \AA~ and 6158 \AA~ for HD\,3567, however, we did not include the oxygen lines in the current analysis because of the fact that they were weak and contributed by Ca and Fe in the wings of those lines although the contribution of Ca and Fe were removed with success via spectrum synthesis technique. The triplet lines were not observable in the spectrum of BD+42 3607. The oxygen abundances were used to compile [Na/Fe] vs [O/Fe] plots for which the summary of the findings for some selected program stars are as follows: HD 194598 follows the same abundance pattern as the stars of NGC\,362, where anti-correlation is observed with [Na/Fe]: it increases as the [O/Fe] ratio decreases. Similar case was also seen for HD\,194598 and NGC\,4833. HD\,201891 is also seen to show similar abundance pattern as the stars of NGC\,5897. 

{\bf HD\,6755:}
On the basis of its Galactic orbital parameters, HD\,6755 is classified as a halo star. The position of the star in the $Z_{\rm max} \times e_{\rm v}$ plane is also agreed with the membership status of the star as a halo star (i.e. $Z_{\rm max}=$ 13.22$\pm$0.74 kpc).
The Galactic orbital parameters for the star mimics those of NGC\, 5466 with the exception of its apo-galactic distance. The metallicity of the star is far from agreed, for a GC membership for the cluster in the past although a relatively high encounter probability $P~\approx~$16\% for 5 tidal radii was obtained. The encounter probability for the star for 1.5 tidal radii was found to be $P~\approx~$6\%. For HD\,6755, a GC origin from NGC\,5466 can be clearly eliminated, however, a satellite galaxy origin for the star can not be dismissed due retrograde motion of the star, indicated in the Toomre diagram (Figure 5). 

{\bf BD+42\,3607:}
As one of the most metal-poor HPM star in our sample with HD\,84937, its Galactic orbital parameters (and metallicity) indicate a halo membership like HD\,6755 and three GCs are assigned as parent GC candidates, i.e. NGC\,2298, NGC\,2808, and NGC\,6864. The metallicity and age of the star clearly support the assumption of a GC origin, but the encounter probability for NGC\,2298 is found to be relatively low ($P=5\%$ for 1.5 tidal radii). We exclude from consideration the latter two GCs that show $\approx$1.2 dex enrichment in metallicity. Such high metallicities can not be reconciled with the enrichment process between first and second generations in GCs 
(Bastian \& Lardo, 2018) hence, NGC\,2808 and NGC\,6864 can be eliminated from consideration as parent GC candidates for BD+42\,3607.

{\bf HD\,201891:}
The calculated Galactic orbital parameters of the star provided five matches as for GC candidates and indicate a thick disk membership for the star like HD\,3567. The position of the star in the Toomre diagram also corroborates the assessment as the thick disk (Figure 5). The best matching GC candiates between among those listed in Table 6 are NGC\,5897 and NGC\,6656 and they have relatively higher mean metallicities compared to the metallicity of the star (i.e. ${\rm [Fe/H]} = -1.02\pm0.16$ dex). Although an excellent match was otained in orbital parameters, metallicity, and age for NGC\,5897, even preferred candidate polluters, e.g. fast rotating massive stars or the enrichment occurred through equatorial disks of stars between 20 \Msolar\, and 100-120 \Msolar\, (Decressin et al. 2007), or massive AGB stars (Ventura et al. 2001) through slow winds following the hot-bottom burning phase, may not provide the stipulated enrichment ($\approx$1 dex) in the metallicity of the star to support a GC origin. It should be noted that the star has the highest calculated encounter probability for NGC\,5897 with $P\approx18\%$ for 1.5 tidal radii. More onto this, our evaluation for the star's origin from a parent GC candidate solely based on star's element abundances from the high-resolution spectroscopy performed in this study proved to be inconclusive although abundances for HD\,201891 and NGC\,5897 agreed well with the exception of V\,{\sc i}, Cr\,{\sc I}, and Cu\,{\sc I} abundances (see Figure 9). For the second best matching GC candidate, NGC\,6656, the age of the cluster from Vandenberg et al. (2013) is far from being agreed with the star's reported age in this study. Therefore, assigned GC candidates for the star in Table 6 can be excepted for  a GC origin scenario. 

{\bf HD\,194598:}
The position of the star in the Toomre diagram (Figure 5) implies a retrograde motion like HD\,6755 and HD\,3567 hence an alternative origin via tidal distruption of a dwarf galaxy can not be ruled out. Three best matching GC candidates for the star have metallicities that are in accordance with the star's metallicity, NGC\,6284, NGC\,5946, and NGC\,362, but the highest encounter probability was obtained for the former cluster ($P\approx 12\%$). The agreement in the Galactic orbital parameters, metallicity, and age for NGC\,6284 is satisfactory. The outliers are NGC\,4833 and NGC\,5897. They have $\approx$0.7 dex higher metallicities although the age of the latter cluster from Salaris \& Weiss (1998) is in excellent agreement with the age of the star obtained in this study (Table 6). A solely abundance based assignment of a parent GC candidate for the star is again inconclusive for HD\,194598 as in the case of HD\,201891 when the abundances of NGC\,362 and NGC\,4833 along with the abundances of the star were considered (see Figure 9).  Therefore, one may suggest that HD\,194598 is possibly ejected from  NGC\,6284 when consistency at 1$\sigma$ level in Galactic orbital parameters, metallicity, and age for the star are contemplated. However, it is important to note that the metallicity and $Z_{\rm max}$ (Figure 6) of the star are not in accordance with the star's membership status as a thin-disk star suggested by stars Galactic orbital parameters. This with its retrograde motion, we can not rule out the possibility that the star may have originated from a tidally disrupted dwarf galaxy. 

{\bf HD\,3567:}
When the star's position in $Z_{\rm max} \times e_{\rm v}$ plane was inspected, HD\,3567 was seen to be located in a region where thin-disk star are expected to reside. This finding clearly contradicts with the assigned status for the star as a thick disk star from its metallicity and age. Although the match in its Galactic orbital parameters with three parent GC candidates listed in Table 6 is satisfactory, the best matching GC candidate for the star on the basis of its metallicity and age is seen to be NGC\,362. Figure 9 presents abundances of the star along with the abundances of the cluster. The abundances of several species (e.g. Na\,{\sc i}, Mg\,{\sc i}, Si\,{\sc i}, Ca\,{\sc i}, Ti\,{\sc iii}, Cr\,{\sc i}, Co\,{\sc i}, Ni\,{\sc i}, Y\,{\sc ii}, Zr\,{\sc ii}, and Ba\,{\sc ii}) are agreed well with those of the cluster. On the other hand, for the abundances of Sc\,{\sc ii}, V\,{\sc i}, and Mn\,{\sc i} the discrepancies up to $\approx$ 0.1 dex  should be noted. Again, as in the case of HD\,194598, a GC origin can not be easily ruled out in spite of a low encounter probability obtained for NGC\,362, also when compared to HD\,194598. However, the stars retrograde motion seen from Figure 5 leaves open the origin of HD\,3567 like HD\,194598. 

{\bf HD\,84937:}
The Galactic orbital parameters of this {\it Gaia} benchmark star suggest a thin-disk membership status for the star. This is also in accordance with the position of the star in the $Z_{\rm max} \times e_{\rm v}$ plane where the star has $Z_{\rm max}=0.44\pm0.25$ kpc. However, on the basis of its metallicity and age, the star is a member of halo. The best matching GC candidate for the star, NGC\,6341, has a metallicity that is in accordance with the star's metallicity. Also, within the error limits the agreement in orbital parameters is satisfactory. On the basis of its orbital parameters, metallicity, age, and the the low encounter probability calculated for the star, the possibility for a GC origin is the less probable case. 

To conclude, their orbital parameters, metallicities, ages, and also computed encounter probabilities allowed us to determine the origin of the HPM stars, e.g. probable candidate GCs as parent clusters. Tidal disruption from a dwarf galaxy was also considered as an alternative origin. However, this should not be evaluated as a strict assignment. HD\,6755,
HD\,194598, and HD\,3567 with their retrograde orbital motions are likely candidate stars for a dwarf galaxy origin. Despite the fact that a small encounter probability ($P \approx 4\%$) obtained for HD\,84937 for a GC origin, the best candidate in the view of its metallicity would be NGC\,6341. For HD\,201891, the HPM star with the largest encounter probability (i.e. $P\approx 18\%$), the agreement in the Galactic orbital
parameters and age for NGC\,5897 is significant. HD\,194598 presents an interesting case: with the second largest encounter probability between among
the HPM sample stars, it has a retrograde orbital motion. However, in the less probable case that the star originated from a GC, the most likely candidate in the view of its metallicity would be NGC\,6284. Similarly, it is NGC\,0362 for HD\,3567. The age of BD +42 3607 is agreed with NGC\,2298. The dynamical analysis of this star reveals an ambiguous origin.

For such alternative origins to be accepted as a tool to produce HPM star in the Galaxy, the number of HPM sample stars has certainly to be increased. We plan to extend our
study to HPM stars with slightly higher metallicities and G type HPM stars in order to convey spectral type dependence for the HPM nature of a star upon its chemistry in a greater detail, to reveal any possible correlation between their kinematics and Galactic orbits and, also, to test two alternative origins for the HPM stars.

\section*{Acknowledgments}
We would like to thank the anonymous referee for his/her numerous contribution towards improving the paper. We also thank to Eugenio Carretta for fruitful discussion on GC membership and commenting on the content of the paper. We thank to Gizay YOLALAN and Olcay PLEVNE. This research has made use of NASA's (National Aeronautics and Space Administration) Astrophysics Data System and the SIMBAD Astronomical Database, operated at CDS, Strasbourg, France and NASA/IPAC Infrared Science Archive, which is operated by the Jet Propulsion Laboratory, California Institute of Technology, under contract with the National Aeronautics and Space Administration. This work has made use of data from  the European Space Agency (ESA) mission {\it Gaia} (\mbox{https://www.cosmos.esa.int/gaia}), processed by the {\it Gaia} Data Processing and Analysis Consortium (DPAC, \mbox{https://www.cosmos.esa.int/web/gaia/dpac/consortium}). Funding for the DPAC has been provided by national institutions, in particular the institutions participating in the {\it Gaia} Multilateral Agreement.

\software{LIME (\c{S}ahin 2017), SPECTRE (Sneden 1973), MOOG (Sneden 1973), INSPECT program v1.0 (Lind et al. 2012), galpy (Bovy 2015), MilkywayPotential2014 (Bovy 2015).}

\begin{appendix}

\setcounter{table}{0}
\renewcommand{\thetable}{A\arabic{table}}

\setcounter{figure}{0}
\renewcommand{\thefigure}{A\arabic{figure}}

\startlongtable
\begin{deluxetable*}{l|c|c@{}|c@{}|rl@{}|rl@{}|rl@{}|ll@{}|cc@{}|cc|l}
\tabletypesize{\footnotesize}

\tablecaption{Fe\,{\sc i} lines used in the analysis of ELODIE spectra. The equivalent widths are those obtained with the LIME code.}
\tablehead{
\hline
       &           &       &         &\multicolumn{2}{c}{\mbox{HD 194598}}&\multicolumn{2}{c}{\mbox{HD 201891}}&\multicolumn{2}{c}{\mbox{HD 6755}}&\multicolumn{2}{c}{\mbox{HD 3567}}&\multicolumn{2}{c}{\mbox{HD 84937}}&\multicolumn{2}{c|}{\mbox{BD+ 423607}}\\
\cline{5-16} 
Species& $\lambda$ & L.E.P & log(gf) & EW &log$\epsilon$(X)&EW  &log$\epsilon$(X)&EW &log$\epsilon$(X) & EW&log$\epsilon(X)$&EW &log$\epsilon$(X)&EW &log$\epsilon$(X)&ref \\
\cline{2-3}
\cline{5-16}
       &(\AA) & (eV)&  &(m\AA)&(dex) &(m\AA)&(dex) &(m\AA) &(dex) & (m\AA) &(dex)&(m\AA) &(dex)& (m\AA) &(dex)&      \\
\hline
}
\startdata
Fe\,{\sc i}  & 4007.28 & 2.76 & -1.28 &  ...  &  ...  & 59.0  & 6.57 &  ...  &  ... & ...  & ...  & ...  & ...  &...  & ...  &F \\
Fe\,{\sc i}  & 4032.64 & 1.48 & -2.38 &  ...  &  ...  &  ...  &  ... &  ...  &  ... & ...  & ...  & ...  & ...  &24.0 & 5.04 &F \\
Fe\,{\sc i}  & 4059.72 & 3.55 & -1.37 &  ...  &  ...  & 24.0  & 6.44 &  ...  &  ... & ...  & ...  & ...  & ...  &...  & ...  &F \\
Fe\,{\sc i}  & 4062.45 & 2.84 & -0.86 &  ...  &  ...  & 76.0  & 6.59 & 80.4  & 6.08 & ...  & ...  &18.0  & 5.08 &...  & ...  &F \\
Fe\,{\sc i}  & 4065.39 & 3.43 & -1.37 & 26.0  & 6.37  &  ...  &  ... &  ...  &  ... & ...  & ...  & ...  & ...  &...  & ...  &F \\
Fe\,{\sc i}  & 4067.99 & 3.21 & -0.47 & 78.0  & 6.41  & 92.0  & 6.57 &  ...  &  ... & ...  & ...  & ...  & ...  &...  & ...  &F \\
Fe\,{\sc i}  & 4073.77 & 3.26 & -0.90 & 44.0  & 6.17  & 63.0  & 6.54 &  ...  &  ... & ...  & ...  & ...  & ...  &...  & ...  &N \\
Fe\,{\sc i}  & 4079.84 & 2.86 & -1.36 &  ...  &  ...  &  ...  &  ... & 68.7  & 6.26 & ...  & ...  & ...  & ...  &...  & ...  &F \\
\enddata
\tablecomments{References for the adopted $gf$ values: {\bf F:} Fuhr \& Wiese (2006); {\bf N:} NIST Atomic Spectra Database {(http://physics.nist.gov/PhysRefData/ASD)}}
\end{deluxetable*}

\startlongtable
\begin{deluxetable*}{l|c|c@{}|c@{}|rl@{}|rl@{}|rl@{}|ll@{}|cc@{}|cc|l}
\tabletypesize{\scriptsize}
\tablecaption{Fe\,{\sc i} lines used in the analysis of ELODIE spectra. The equivalenth widths are those obtained with the LIME code.}

\tablehead{
\hline
       &           &       &         &\multicolumn{2}{c}{\mbox{HD 194598}}&\multicolumn{2}{c}{\mbox{HD 201891}}&\multicolumn{2}{c}{\mbox{HD 6755}}&\multicolumn{2}{c}{\mbox{HD 3567}}&\multicolumn{2}{c}{\mbox{HD 84937}}&\multicolumn{2}{c|}{\mbox{BD+ 423607}}\\
\cline{5-16} 
Species& $\lambda$ & L.E.P & log(gf) & EW &log$\epsilon$(X)&EW  &log$\epsilon$(X)&EW &log$\epsilon$(X) & EW&log$\epsilon(X)$&EW &log$\epsilon$(X)&EW &log$\epsilon$(X)&ref \\
\cline{2-3}
\cline{5-16}
       &(\AA) & (eV)&  &(m\AA)&(dex) &(m\AA)&(dex) &(m\AA) &(dex) & (m\AA) &(dex)&(m\AA) &(dex)& (m\AA) &(dex)&      \\
\hline
}
\startdata
Fe\,{\sc i} & 4889.00 & 2.20 & -2.55 & ...  & ...  &  ...  & ...  & 59.5 & 6.37 & ...  & ...  & ...  & ...  & ...  & ...  &F \\
Fe\,{\sc i} & 4890.76 & 2.87 & -0.39 & ...  & ...  & 111.0 & 6.26 & ...  & ...  & 88.5 & 6.34 & 40.0 & 5.09 & 54.0 & 5.03 &F \\
Fe\,{\sc i} & 4891.50 & 2.85 & -0.11 & ...  & ...  & 142.0 & 6.24 & ...  & ...  & 99.7 & 6.23 & 48.0 & 4.94 & 59.0 & 4.82 &F \\
Fe\,{\sc i} & 4903.32 & 2.88 & -0.93 & ...  & ...  &  ...  & ...  & ...  & ...  & ...  & ...  & 12.0 & 4.92 & ...  & ...  &F \\
Fe\,{\sc i} & 4917.23 & 4.19 & -1.16 & 17.0 & 6.57 & 20.0  & 6.64 & ...  & ...  & ...  & ...  & ...  & ...  & ...  & ...  &F \\
Fe\,{\sc i} & 4919.00 & 2.86 & -0.34 & ...  & ...  &  ...  & ...  & ...  & ...  & 82.1 & 6.16 & 35.0 & 4.92 & 52.0 & 4.93 &F \\
Fe\,{\sc i} & 4924.78 & 2.28 & -2.11 & 35.0 & 6.20 & 50.0  & 6.53 & 65.0 & 6.13 & 32.0 & 6.22 & ...  & ...  & ...  & ...  &F \\
Fe\,{\sc i} & 4927.43 & 3.57 & -2.07 & ...  & ...  & 11.0  & 6.66 & ...  & ...  & ...  & ...  & ...  & ...  & ...  & ...  &F \\
\enddata
\tablecomments{References for the adopted $gf$ values: {\bf F:} Fuhr \& Wiese (2006); {\bf N:} NIST Atomic Spectra Database {(http://physics.nist.gov/PhysRefData/ASD)}.}
\end{deluxetable*}

\startlongtable
\begin{deluxetable*}{l|c|c@{}|c@{}|rl@{}|rl@{}|rl@{}|ll@{}|cc@{}|cc|l}
\tabletypesize{\scriptsize}
\tablecaption{Lines used in the analysis of ELODIE spectra. The EWs reported for individual lines are those obtained via the LIME code.}
\tablehead{
\hline
       &           &       &         &\multicolumn{2}{c}{\mbox{HD 194598}}&\multicolumn{2}{c}{\mbox{HD 201891}}&\multicolumn{2}{c}{\mbox{HD 6755}}&\multicolumn{2}{c}{\mbox{HD 3567}}&\multicolumn{2}{c}{\mbox{HD 84937}}&\multicolumn{2}{c|}{\mbox{BD+ 423607}}\\
\cline{5-16} 
Species& $\lambda$ & L.E.P & log(gf) & EW &log$\epsilon$(X)&EW  &log$\epsilon$(X)&EW &log$\epsilon$(X) & EW&log$\epsilon(X)$&EW &log$\epsilon$(X)&EW &log$\epsilon$(X)&ref \\
\cline{2-3}
\cline{5-16}
       &(\AA) & (eV)&  &(m\AA)&(dex) &(m\AA)&(dex) &(m\AA) &(dex) & (m\AA) &(dex)&(m\AA) &(dex)& (m\AA) &(dex)&      \\
\hline
}
\startdata
Fe\,{\sc i} & 5633.95 & 4.99 & -0.32 &  ...  &  ... & 18.0  & 6.49 & 16.1  & 6.08 & ...  & ...  & ...  & ...  & ...  & ...  &F \\
Fe\,{\sc i} & 5638.27 & 4.22 & -0.84 &  ...  &  ... & 24.0  & 6.44 & 26.6  & 6.07 & ...  & ...  & ...  & ...  & ...  & ...  &F \\
Fe\,{\sc i} & 5641.45 & 4.26 & -1.15 & 11.0  & 6.37 &  ...  &  ... &  ...  &  ... & ...  & ...  & ...  & ...  & ...  & ...  &F \\
Fe\,{\sc i} & 5662.52 & 4.18 & -0.57 & 32.0  & 6.36 & 36.0  & 6.45 &  ...  &  ... & ...  & ...  & ...  & ...  & ...  & ...  &F \\
Fe\,{\sc i} & 5679.03 & 4.65 & -0.90 & 13.0  & 6.57 & 14.0  & 6.59 & 12.3  & 6.16 & ...  & ...  & ...  & ...  & ...  & ...  &F \\
Fe\,{\sc i} & 5686.54 & 4.55 & -0.45 & 19.0  & 6.24 &  ...  &  ... &  ...  &  ... &27.0  & 6.53 & ...  & ...  & ...  & ...  &F \\
Fe\,{\sc i} & 5701.56 & 2.56 & -2.22 &  ...  &  ... &  ...  &  ... & 44.8  & 6.04 & ...  & ...  & ...  & ...  & ...  & ...  &F \\
Fe\,{\sc i} & 5705.47 & 4.30 & -1.35 &  ...  &  ... &  ...  &  ... & 10.1  & 6.13 &9.5   & 6.59 & ...  & ...  & ...  & ...  &F \\
\enddata

\tablecomments{
References for the adopted $gf$ values: {\bf B:} Kelleher \& Podobedova (2008a); {\bf K:} Kurucz Atomic Spectra Database (http://www.pmp.uni-hannover.de/projekte/kurucz/),
hfs-included; {\bf LG:} Lawler et al. (2013); {\bf P:} Kelleher \& Podobedova (2008b); {\bf N:} NIST Atomic Spectra Database
{(http://physics.nist.gov/PhysRefData/ASD)}; {\bf Ast:} Astrophysical (computed). }

\end{deluxetable*}

%longtable olacak

\startlongtable
\begin{deluxetable*}{l|c|c@{}|c@{}|rl@{}|rl@{}|rl@{}|ll@{}|cc@{}|cc|l}
\tabletypesize{\small}
\tablecaption{Lines used in the analysis of ELODIE spectra. The EWs reported for individual lines are those obtained via LIME code.}
\tablehead{
\hline
       &           &       &         &\multicolumn{2}{c}{\mbox{HD 194598}}&\multicolumn{2}{c}{\mbox{HD 201891}}&\multicolumn{2}{c}{\mbox{HD 6755}}&\multicolumn{2}{c}{\mbox{HD 3567}}&\multicolumn{2}{c}{\mbox{HD 84937}}&\multicolumn{2}{c|}{\mbox{BD+ 423607}}\\
\cline{5-16} 
Species& $\lambda$ & L.E.P & log(gf) & EW &log$\epsilon$(X)&EW  &log$\epsilon$(X)&EW &log$\epsilon$(X) & EW&log$\epsilon(X)$&EW &log$\epsilon$(X)&EW &log$\epsilon$(X)&ref \\
\cline{2-3}
\cline{5-16}
       &(\AA) & (eV)&  &(m\AA)&(dex) &(m\AA)&(dex) &(m\AA) &(dex) & (m\AA) &(dex)&(m\AA) &(dex)& (m\AA) &(dex)&      \\
\hline
}
\startdata
Ti\,{\sc i}  & 6258.71  & 1.46 & -0.24 & 13.0  & 3.95 &  21.0 & 4.20 & 22.7  & 3.57 & 13.0  & 4.14 & ...  & ...  & ...  &  ... &N \\
Ti\,{\sc ii} & 4028.35  & 1.89 & -0.92 &  ...  & ...  &  ...  & ...  & 80.6  & 4.04 & ...   &  ... & 29.1 & 3.09 & ...  &  ... &L \\
Ti\,{\sc ii} & 4290.23  & 1.16 & -0.87 &  ...  & ...  &  ...  & ...  &  ...  & ...  & ...   & ...  & 63.0 & 2.99 & 62.5 & 3.09&LW \\
Ti\,{\sc ii} & 4300.05  & 1.18 & -0.46 &  ...  & ...  &  ...  & ...  &  ...  & ...  & ...   & ...  & 76.0 & 2.89 & 69.7 & 2.90&LW \\
Ti\,{\sc ii} & 4316.80  & 2.05 & -1.62 & 25.0  & 4.11 &  ...  & ...  &  ...  & ...  & 31.2  & 4.09 & ...  &  ... & ...  &  ... &LW \\
Ti\,{\sc ii} & 4394.07  & 1.22 & -1.77 & 54.0  & 4.19 &  ...  & ...  &  ...  & ...  & syn   & 4.10 & ...  &  ... & 16.3 & 2.81 &LW \\
Ti\,{\sc ii} & 4395.04  & 1.08 & -0.54 &  ...  & ...  &  ...  & ...  &  ...  & ...  &  ...  &  ... & 82.0 & 3.02 & ...  &  ... &LW \\
Ti\,{\sc ii} & 4395.85  & 1.24 & -1.93 &  ...  & ...  &  47.0 & 4.22 &  ...  & ...  & ...   & ...  & 17.1 & 3.14 & ...  &  ... &LW \\
\enddata
\tablecomments{References for the adopted $gf$ values -- {\bf LW:} Wood, Lawler, Sneden, \& Cowan (2013); {\bf N:} NIST
Atomic Spectra Database {(http://physics.nist.gov/PhysRefData/ASD)}; {\bf G:} Lawler et al. (2015); {\bf J:} Sobeck, Lawler, Sneden
(2007); {\bf K:} Kurucz Atomic Spectra Database (http://www.pmp.uni-hannover.de/projekte/kurucz/),
hfs-included; {\bf BW:} Blackwell-Whitehead \&
Bergemann (2007); {\bf LWD:} Lawler et al. (2014), hfs-included for 4379 \AA\, but no hfs for
6090 \AA\, V\,{\sc i} line; {\bf W:} Wood, Lawler, Sneden, \& Cowan (2014).}
\end{deluxetable*}

\begin{table*}
\tiny
\caption{Lines used in the analysis of ELODIE spectra. The EWs reported for individual lines are those obtained via LIME code. }
\centering
\begin{tabular}{l|c|c@{}|c@{}|rl@{}|rl@{}|rl@{}|ll@{}|cc@{}|cc|l}
\hline
       &           &       &         &\multicolumn{2}{c}{\mbox{HD 194598}}&\multicolumn{2}{c}{\mbox{HD 201891}}&\multicolumn{2}{c}{\mbox{HD 6755}}&\multicolumn{2}{c}{\mbox{HD 3567}}&\multicolumn{2}{c}{\mbox{HD 84937}}&\multicolumn{2}{c|}{\mbox{BD+ 423607}}\\
\cline{5-16} 
Species& $\lambda$ & L.E.P & log(gf) & EW &log$\epsilon$(X)&EW  &log$\epsilon$(X)&EW &log$\epsilon$(X) & EW&log$\epsilon(X)$&EW &log$\epsilon$(X)&EW &log$\epsilon$(X)&ref \\
\cline{2-3}
\cline{5-16}
       &(\AA) & (eV)&  &(m\AA)&(dex) &(m\AA)&(dex) &(m\AA) &(dex) & (m\AA) &(dex)&(m\AA) &(dex)& (m\AA) &(dex)&      \\
\hline
Ni\,{\sc i}  & 4953.21 & 3.74 & -0.58 &  11.0 & 5.01  &  ...  &  ...  &  ...  &  ...  & 12.6  & 5.11  &   ...  & ...  & ...   & ...  &W  \\
Ni\,{\sc i}  & 4980.18 & 3.61 & 0.07  &  37.0 & 4.98  & 45.0  & 5.13  &  ...  &  ...  & 30.0  & 5.08  &   ...  & ...  &  ...  & ...  &W  \\
Ni\,{\sc i}  & 4998.23 & 3.61 & -0.70 &  12.0 & 5.05  & 13.0  & 5.08  &  ...  &  ...  &  ...  &  ...  &   ...  & ...  &  ...  & ...  &W  \\
Ni\,{\sc i}  & 5000.35 & 3.63 & -0.43 &  21.0 & 5.10  & 25.0  & 5.20  & 30.3  & 4.85  & 18.5  & 5.08  &   ...  & ...  &  ...  & ...  &N  \\
Ni\,{\sc i}  & 5010.94 & 3.63 & -0.87 &  ...  &  ...  & 14.0  & 5.31  & 13.6  & 4.80  &  ...  &  ...  &   ...  & ...  &  ...  & ...  &N  \\
\hline 
\end{tabular}
\begin{list}{}{}
\item References for the adopted $gf$ values -- {\bf A:} Allen \& Porto de Mello (2011); {\bf W:} Wood, Lawler, Sneden, \& Cowan (2014); {\bf N:} NIST Atomic Spectra Database
{(http://physics.nist.gov/PhysRefData/ASD)}; {\bf K:} Kurucz Atomic Spectra Database (http://www.pmp.uni-hannover.de/projekte/kurucz/),
hfs-included; {\bf D:} Ljung, Nilsson, Asplund, \& Johansson (2006); {\bf H:} Hannaford et al. (1982); {\bf LW:} Lawler et al. (2006); {\bf LS:} Lawler et al. (2009);  {\bf DH:}
Dan Hartog, Lawler, Sneden, \& Cowan (2003).
\end{list}
\end{table*}

\begin{table*}
\scriptsize
\caption{Abundances of observed species for HD\,194598, HD\,201891, HD\,6755, and the Sun. Solar abundances were obtained by employing the
Solar atmosphere
from Castelli \& Kurucz (2004).}
\centering
\begin{tabular}{l|ccc|ccc|ccc|cc}
\hline
\hline
        &\multicolumn{3}{c}{HD\,194598}&	\multicolumn{3}{c}{HD\,201891} &\multicolumn{3}{c}{HD\,6755}&	\multicolumn{2}{c}{Sun}\\
\cline{2-12}
 Element&log$\epsilon$(X)  & [X/Fe]  & n & log$\epsilon$(X)  &[X/Fe]  & n &log$\epsilon$(X)  &[X/Fe]& n &log$\epsilon$(X)  & n\\
\hline
%Li\,{\sc i}   &                  &                   &     &                &                  &     &                 &                  &      &                 &    \\             
C\,{\sc i}    & 7.71  $\pm$ 0.00 &  0.52  $\pm$ 0.18 & 1   &  7.63$\pm$0.00 & 0.29  $\pm$ 0.18 & 1   & 7.37 $\pm$0.00  &  0.40 $\pm$ 0.18 &  1   & 8.36 $\pm$0.00  & 1  \\
Na\,{\sc i}   & 5.04  $\pm$ 0.01 &  -0.02 $\pm$ 0.18 & 2   &  5.32$\pm$0.11 & 0.11  $\pm$ 0.19 & 2   & 4.69 $\pm$0.15  & -0.15 $\pm$ 0.23 &  2   & 6.23 $\pm$0.01  & 2  \\
Mg\,{\sc i}   & 6.51  $\pm$ 0.11 &  0.03  $\pm$ 0.21 & 5   &  6.94$\pm$0.09 & 0.31  $\pm$ 0.19 & 2   & 6.19 $\pm$0.12  & -0.07 $\pm$ 0.22 &  5   & 7.65 $\pm$0.04  & 3  \\
Si\,{\sc i}   & 6.60  $\pm$ 0.00 &  0.17  $\pm$ 0.19 & 2   &  6.86$\pm$0.06 & 0.28  $\pm$ 0.18 & 4   & 6.36 $\pm$0.13  & 0.15  $\pm$ 0.23 &  6   & 7.60 $\pm$0.06  & 5  \\
Si\,{\sc ii}  & 6.70  $\pm$ 0.00 &  0.27  $\pm$ 0.20 & 1   &  6.87$\pm$0.07 & 0.29  $\pm$ 0.20 & 2   & 6.41 $\pm$0.21  & 0.2   $\pm$ 0.29 &  2   & 7.60 $\pm$0.09  & 2  \\
Ca\,{\sc i}   & 5.39  $\pm$ 0.13 &  0.35  $\pm$ 0.27 & 17  &  5.46$\pm$0.16 & 0.27  $\pm$ 0.28 & 19  & 5.09 $\pm$0.15  & 0.27  $\pm$ 0.28 &  20  & 6.21 $\pm$0.16  & 14 \\
Sc\,{\sc i}   & 2.15  $\pm$ 0.00 &  0.21  $\pm$ 0.18 & 1   &  2.07$\pm$0.00 & -0.02 $\pm$ 0.16 & 1   & 1.80 $\pm$0.00  & 0.08  $\pm$ 0.18 &  1   & 3.11 $\pm$0.00  & 1  \\
Sc\,{\sc ii}  & 2.17  $\pm$ 0.10 &  0.17  $\pm$ 0.23 & 6   &  2.37$\pm$0.08 & 0.22  $\pm$ 0.20 & 5   & 1.83 $\pm$0.07  & 0.05  $\pm$ 0.22 &  6   & 3.17 $\pm$0.10  & 5  \\
Ti\,{\sc i}   & 3.91  $\pm$ 0.05 &  0.15  $\pm$ 0.22 & 19  &  4.12$\pm$0.07 & 0.21  $\pm$ 0.21 & 19  & 3.67 $\pm$0.10  & 0.13  $\pm$ 0.24 &  21  & 4.93 $\pm$0.12  & 35 \\
Ti\,{\sc ii}  & 4.07  $\pm$ 0.10 &  0.14  $\pm$ 0.25 & 15  &  4.24$\pm$0.08 & 0.16  $\pm$ 0.23 & 17  & 3.84 $\pm$0.11  & 0.13  $\pm$ 0.25 &  12  & 5.10 $\pm$0.14  & 18 \\
V\,{\sc i}    & 2.75  $\pm$ 0.05 &  0.06  $\pm$ 0.21 & 2   &  2.90$\pm$0.04 & 0.08  $\pm$ 0.20 & 2   & 2.24 $\pm$0.04  &-0.22  $\pm$ 0.21 &  2   & 3.83 $\pm$0.11  & 2  \\
Cr\,{\sc i}   & 4.43  $\pm$ 0.07 &  -0.03 $\pm$ 0.22 & 17  &  4.56$\pm$0.10 & -0.05 $\pm$ 0.21 & 14  & 4.14 $\pm$0.09  & -0.1  $\pm$ 0.22 &  13  & 5.63 $\pm$0.10  & 17 \\
Cr\,{\sc ii}  & 4.59  $\pm$ 0.14 &  0.10  $\pm$ 0.24 & 5   &  4.74$\pm$0.09 & 0.10  $\pm$ 0.20 & 4   & 4.32 $\pm$0.04  & 0.05  $\pm$ 0.20 &  3   & 5.66 $\pm$0.07  & 6  \\
Mn\,{\sc i}   & 4.08  $\pm$ 0.05 &  -0.39 $\pm$ 0.25 & 9   &  4.19$\pm$0.09 & -0.43 $\pm$ 0.25 & 8   & 3.62 $\pm$0.10  & -0.63 $\pm$ 0.27 &  9   & 5.64 $\pm$0.17  & 9  \\
Fe\,{\sc i}   & 6.34  $\pm$ 0.13 &  0.03  $\pm$ 0.17 & 149 &  6.47$\pm$0.11 & 0.01  $\pm$ 0.16 & 144 & 6.11 $\pm$0.13  &  0.02 $\pm$ 0.17 &  107 & 7.48 $\pm$0.12  & 160\\
Fe\,{\sc ii}  & 6.31  $\pm$ 0.13 &  0.03  $\pm$ 0.19 & 13  &  6.46$\pm$0.12 & 0.03  $\pm$ 0.19 & 19  & 6.09 $\pm$0.12  &  0.09 $\pm$ 0.19 &  14  & 7.45 $\pm$0.15  & 24 \\
Co\,{\sc i}   & 3.92  $\pm$ 0.00 &   0.13 $\pm$ 0.17 & 2   &  4.10$\pm$0.15 &  0.18 $\pm$ 0.22 & 3   & 3.83 $\pm$0.00  &  0.27 $\pm$ 0.03 &  1   & 4.93 $\pm$0.03  & 3  \\
Ni\,{\sc i}   & 5.03  $\pm$ 0.06 &  -0.05 $\pm$ 0.23 & 26  &  5.22$\pm$0.09 & -0.01 $\pm$ 0.22 & 36  & 4.68 $\pm$0.11  & -0.15 $\pm$ 0.25 &  26  & 6.22 $\pm$0.13  & 42 \\
Cu\,{\sc i}   & 2.71  $\pm$0.15  & -0.31  $\pm$ 0.23 & 2   &  3.05$\pm$0.07 & -0.11 $\pm$ 0.18 & 2   & 1.87 $\pm$0.00  & -0.93 $\pm$ 0.18 &  1   & 4.19 $\pm$0.03           & 2\\
Sr\,{\sc i}   &       ...   	 &	   ...       &     &  1.72$\pm$0.00 & -0.19 $\pm$ 0.16 & 1   & 1.11 $\pm$0.00  &  -0.42$\pm$ 0.18 &  1   & 2.90 $\pm$0.00  & 1  \\
Sr\,{\sc ii}  & 1.75  $\pm$ 0.00 &  -0.01 $\pm$ 0.18 & 1   &  1.97$\pm$0.00 & 0.06  $\pm$ 0.16 & 1   & 1.55 $\pm$0.00  &  0.02 $\pm$ 0.18 &  1   & 2.90 $\pm$0.05  & 2  \\
Y\,{\sc ii}   & 0.86  $\pm$ 0.03 &  -0.24 $\pm$ 0.19 & 2   &  1.04$\pm$0.03 & -0.21 $\pm$ 0.17 & 2   & 0.58 $\pm$0.05  &  -0.29$\pm$ 0.20 &  3   & 2.24 $\pm$0.06  & 4  \\
Zr\,{\sc i}   & 2.10  $\pm$ 0.00 &   0.47 $\pm$ 0.18 & 1   &  2.07$\pm$0.00 & 0.29  $\pm$ 0.16 & 1   & 1.36 $\pm$0.00  & -0.05 $\pm$ 0.18 &  1   & 2.80 $\pm$0.00  & 1  \\
Zr\,{\sc ii}  & 1.60  $\pm$ 0.00 &   0.01 $\pm$ 0.18 & 1   &  1.87$\pm$0.21 & 0.10  $\pm$ 0.26 & 2   & 1.49 $\pm$0.13  &  0.10 $\pm$ 0.22 &  3   & 2.76 $\pm$0.01  & 2  \\
Ba\,{\sc ii}  & 0.76  $\pm$ 0.28 &  -0.16 $\pm$ 0.34 & 3   &  0.81$\pm$0.26 & -0.24 $\pm$ 0.31 & 3   & 0.44 $\pm$0.31  & -0.25 $\pm$ 0.36 &  3   & 2.06 $\pm$0.06  & 3  \\
Ce\,{\sc ii}  & 0.60  $\pm$ 0.00 &   0.17 $\pm$ 0.18 & 1   &  ...	    &  ...	       &     & 0.21 $\pm$0.00  &  0.00 $\pm$ 0.18 &  1   & 1.60 $\pm$0.00  & 1  \\
Nd\,{\sc ii}  &       ...   	 &	   ...       &...  &  ...	    &  ...	       &...  & 0.66 $\pm$0.07  &  0.37 $\pm$ 0.24 &  2   & 1.66 $\pm$0.14  & 3  \\
Sm\,{\sc ii}  & 0.50  $\pm$ 0.00 &   0.49 $\pm$ 0.18 & 1   &  0.45$\pm$0.00 & 0.29  $\pm$ 0.16 & 1   & 0.01 $\pm$0.00  &  0.22 $\pm$ 0.18 &  1   & 1.18 $\pm$0.00  & 1  \\
\hline 
\end{tabular}
\end{table*}

\begin{table*}
\scriptsize
\caption{Abundances of observed species for HD\,3567, HD\,84937, BD\,+42\,3607, and the Sun. Solar abundances were obtained by employing the Solar atmosphere
from Castelli \& Kurucz (2004).}
\centering
\begin{tabular}{l|ccc|ccc|ccc|cc}
\hline
\hline
        &\multicolumn{3}{c}{HD\,3567}&	\multicolumn{3}{c}{HD\,84937} &\multicolumn{3}{c}{BD\,+42\,3607}&\multicolumn{2}{c}{Sun}\\
\cline{2-12}
 Element&log$\epsilon$(X)  & [X/Fe]  & n & log$\epsilon$(X)  &[X/Fe]  & n &log$\epsilon$(X)  &[X/Fe]& n &log$\epsilon$(X)  & n\\
\hline
%Li\,{\sc i}  &                  &                 &      &                  &                 &      &                  &                  &      &                 &    \\             
C\,{\sc i}   &  8.62 $\pm$0.00  & 1.36  $\pm$0.18 &  1   &	     ...    &   ...           & ...  &  	  ...   &   ...            &...   & 8.36 $\pm$0.00  &  1  \\
Na\,{\sc i}  &  4.93 $\pm$0.00  & -0.20 $\pm$0.18 &  1   &   	 ...        & ...	      & ...  &  5.05 $\pm$0.15  &  1.25  $\pm$0.23 &  2   & 6.23 $\pm$0.01  &  2  \\
Mg\,{\sc i}  &  6.64 $\pm$0.14  & 0.09  $\pm$0.23 &  6   &  5.59 $\pm$0.08  & 0.34  $\pm$0.17 &  5   &  5.64 $\pm$0.10  &  0.42  $\pm$0.20 &  6   & 7.65 $\pm$0.04  &  3  \\
Si\,{\sc i}  &  6.68 $\pm$0.16  & 0.18  $\pm$0.25 &  6   &      ...         &       ...       &  ... &  6.37 $\pm$0.13  &  1.20  $\pm$0.22 &  3   & 7.60 $\pm$0.06  &  5  \\
Si\,{\sc ii} &  6.49 $\pm$0.12  & -0.01 $\pm$0.23 &  2   &   	 ...        & 	   ...	      & ...  &  6.32 $\pm$0.00  &  1.15  $\pm$0.19 &  1   & 7.60 $\pm$0.09  &  2  \\
Ca\,{\sc i}  &  5.47 $\pm$0.15  & 0.36  $\pm$0.28 &  21  &  4.41 $\pm$0.12  & 0.60  $\pm$0.25 &  15  &  4.39 $\pm$0.12  &  0.61  $\pm$0.26 &  19  & 6.21 $\pm$0.16  &  14 \\
Sc\,{\sc i}  & ...   	        & ...        	  & ...  &   	   ...      & 	   ...	      & ...  &      ...         &  	   ...	   &...   & 3.11 $\pm$0.00  &  1  \\
Sc\,{\sc ii} &  1.87 $\pm$0.15  & -0.20 $\pm$0.25 &  6   &  0.86 $\pm$0.15  & 0.09  $\pm$0.23 &  3   &  1.02 $\pm$0.13  &  0.28  $\pm$0.24 &  3   & 3.17 $\pm$0.10  &  5  \\
Ti\,{\sc i}  &  4.00 $\pm$0.10  & 0.17  $\pm$0.24 &  25  &  3.15 $\pm$0.12  & 0.62  $\pm$0.23 &  12  &  2.92 $\pm$0.07  &  0.42  $\pm$0.22 &  7   & 4.93 $\pm$0.12  &  35 \\
Ti\,{\sc ii} &  4.02 $\pm$0.16  & 0.02  $\pm$0.28 &  17  &  3.07 $\pm$0.14  & 0.37  $\pm$0.25 &  18  &  2.92 $\pm$0.11  &  0.25  $\pm$0.25 &  12  & 5.10 $\pm$0.14  &  18 \\
V\,{\sc i}   &  2.65 $\pm$0.00  & -0.04 $\pm$0.21 &  1   &  1.70 $\pm$0.00  & 0.29  $\pm$0.19 &  1   &      ...         &  	   ...	   &...   & 3.83 $\pm$0.11  &  2 \\
Cr\,{\sc i}  &  4.45 $\pm$0.13  & -0.08 $\pm$0.24 &  19  &  3.20 $\pm$0.13  & -0.03 $\pm$0.22 &  9   &  3.20 $\pm$0.12  &  0.00	 $\pm$0.23 &   9  & 5.63 $\pm$0.10  &  17 \\
Cr\,{\sc ii} &  4.47 $\pm$0.11  & -0.09 $\pm$0.22 &  4   &  3.57 $\pm$0.09  & 0.31  $\pm$0.19 &  3   &  3.58 $\pm$0.00  &  0.35  $\pm$0.18 & 2    & 5.66 $\pm$0.07  &  6  \\
Mn\,{\sc i}  &  3.95 $\pm$0.08  & -0.59 $\pm$0.26 &  5   &  3.05 $\pm$0.12  & -0.19 $\pm$0.26 &  4   &  3.07 $\pm$0.02  &  -0.14 $\pm$0.24 &  3   & 5.64 $\pm$0.17  &  9  \\
Fe\,{\sc i}  &  6.34 $\pm$0.13  & -0.04 $\pm$0.17 &  83  &  5.06 $\pm$0.09  & -0.02 $\pm$0.15 &  57  &  5.07 $\pm$0.12  &  0.02  $\pm$0.17 &  47  & 7.48 $\pm$0.12  &  160\\
Fe\,{\sc ii} &  6.34 $\pm$0.06  & -0.01 $\pm$0.16 &  10  &  5.06 $\pm$0.07  & 0.01  $\pm$0.16 &  8   &  5.01 $\pm$0.16  &  -0.01 $\pm$0.22 &  5   & 7.45 $\pm$0.15  &  24 \\
Co\,{\sc i}  &  3.57 $\pm$0.00  & -0.22 $\pm$0.18 &  1   &  2.67 $\pm$0.00  & 0.16  $\pm$0.15 &  1   &  2.74 $\pm$0.00  &  0.22  $\pm$0.17 &  1   & 4.93 $\pm$0.03  &  3  \\
Ni\,{\sc i}  &  5.05 $\pm$0.10  & -0.03 $\pm$0.24 &  6   &  3.98 $\pm$0.10  & 0.19  $\pm$0.22 &  15  &  3.98 $\pm$0.15  &  0.16  $\pm$0.26 &  10  & 6.22 $\pm$0.13  &  42 \\
Sr\,{\sc i}  &  1.75 $\pm$0.00  & -0.04 $\pm$0.18 &  1   &  1.00 $\pm$0.00  & 0.53  $\pm$0.15 &  1   &  	  ...	&	   ...	   &...   & 2.90 $\pm$0.00  &  1  \\
Sr\,{\sc ii} &  2.15 $\pm$0.07  & 0.36  $\pm$0.19 &  2   &  1.00 $\pm$0.28  & 0.53  $\pm$0.32 &  2   &  0.95 $\pm$0.03  &  0.45  $\pm$0.17 &  2   & 2.90 $\pm$0.05  &  2  \\
Y\,{\sc ii}  &  0.97 $\pm$0.05  & -0.13 $\pm$0.20 &  3   &  0.13 $\pm$0.07  & 0.32  $\pm$0.18 &  2   &  0.41 $\pm$0.00  &  0.57  $\pm$0.18 &  2   & 2.24 $\pm$0.06  &  4  \\
Zr\,{\sc i}  & ...  	        &  ...       	  & ...  &	     ...    &	...	      & ...  &  ...		&	 ...  	   &...   & 2.80 $\pm$0.00  &  1   \\
Zr\,{\sc ii} &  1.70 $\pm$0.00  & 0.08	$\pm$0.18 &  1   &  0.95 $\pm$0.00  & 0.62  $\pm$0.15 &  1   &  0.77 $\pm$0.00  &  0.41  $\pm$0.17 &  1   & 2.76 $\pm$0.01  &  2  \\
Ba\,{\sc ii} &  0.86 $\pm$0.53  &-0.06  $\pm$0.56 &  2   &  -0.62$\pm$0.18  & -0.30 $\pm$0.24 &  3   &  -0.57$\pm$0.07  &  -0.22 $\pm$0.19 &  2   & 2.06 $\pm$0.06  &  3  \\
Ce\,{\sc ii} & ...  	        & ...        	  & ...  &	     ...    &	...	      & ...  &  	  ...	&	   ...	   &...   & 1.60 $\pm$0.00  &  1  \\
Nd\,{\sc ii} &  1.00 $\pm$0.00  & 0.48  $\pm$0.23 &  1   &	     ...    &	...	      & ...  &  0.07 $\pm$0.00  &  0.82  $\pm$0.22 &  1   & 1.66 $\pm$0.14  &  3  \\
Sm\,{\sc ii} &  0.50 $\pm$0.00  & 0.46	$\pm$0.18 &  1   &	     ...    &	...	      & ...  &  	  ...	&	   ...	   &...   & 1.18 $\pm$0.00  &  1  \\\hline 

\end{tabular}
\end{table*}

%\newpage
% \startlongtable
\begin{longtable*}{l|l|c|c|c|l}
% \tabletypesize{\small}
\caption{Atmospheric parameters of HPM sample stars from this study -I.}\\

\hline
 Stars 	&\Teff	  & \logg   &   [Fe/H]    & $\xi$              & Notes	\\
            &(\kelvin) &   (cgs)  &    (dex)    & (km s$^{-1}$) &               \\    
\hline 
\hline

\endfirsthead

\caption* {Atmospheric parameters of HPM sample stars from this study -I.}\\

\hline
 Stars 	&\Teff	  & \logg   &   [Fe/H]    & $\xi$              & Notes	\\
            &(\kelvin) &   (cgs)  &    (dex)    & (km s$^{-1}$) &               \\    
\hline 
\hline
\endhead

\hline
\endfoot

\hline
\endlastfoot

HD\,84937   &   6000$\pm$140  & 3.50$\pm$0.18   &  -2.40$\pm$0.09  & 1.60   &This study \\
            &   6264          & 3.97            &  -2.15           & --     &Prugniel (2007) \\
            &   6090$\pm$50   & 4.00$\pm$0.15   &  -2.34$\pm$0.07  & 1.80   &Nissen, Gustafsson, Edvardsson et al. (1994)\\
            &   6150$\pm$90   & 3.70$\pm$0.17   &  -2.23$\pm$0.03  & 1.60   &Carney, Wright, Sneden et al. (1997)\\
            &   6210$\pm$120  & 4.00$\pm$0.10   &  -2.25$\pm$0.15  & --     &Israelian, Lopez \& Rebolo (1998)\\
            &   6350$\pm$80   & 4.03$\pm$0.10   &  -2.07$\pm$0.05  & 1.70   &Mashonkina \& Gehren (2000)\\
            &   6353$\pm$80   & 4.03$\pm$0.10   &  -2.07$\pm$0.09  & --     &Fuhrmann(2000)\\
            &   6375$\pm$40   & 4.10$\pm$0.06   &  -2.08$\pm$0.04  & 0.80   &Fulbright (2000)\\
            &   6250$\pm$100  & 3.80$\pm$0.30   &  -2.00$\pm$0.15  & 1.50   &Mishenina \& Kovtyukh (2001)\\
            &   6290          & 4.02            &  -2.18           & 1.25   & Gratton, Carretta, Claudi et al. (2003)\\
            &   6346$\pm$100  & 4.00$\pm$0.08   &  -2.16           & 1.80   &Bergemann \& Gehren (2008)\\
            &   6365          & 4.00$\pm$0.12   &  -2.15$\pm$0.04  & 1.60   &  Mashonkina, Zhao, Gehren et al. (2008)\\
            &   6365          & 4.00            &  -2.15           & 1.60   & Shi, Gehren, Mashonkina et al. (2009)\\
            &   6431$\pm$100  & 4.08$\pm$0.30   &  -2.14$\pm$0.08  & 1.20   &Ishigaki, Chiba \& Aoki (2012)\\
            &   6377$\pm$36   & 4.15$\pm$0.06   &  -2.02$\pm$0.08  & 1.79   &Ramirez, Allende Prieto \& Lambert (2013)\\
            &   6275$\pm$97   & 4.11$\pm$0.06   &  -2.03$\pm$0.02  & 1.50   &Jofre, Heiter, Soubiran et al. (2014)\\
            &   6541$\pm$145  & 4.23$\pm$0.15   &  -1.92$\pm$0.12  & 0.10   &Bensby, Feltzing \& Oey (2014)\\
            &   6242$\pm$70   & 3.93$\pm$0.08   &  -2.11$\pm$0.14  & 1.00   &Schonrich \& Bergemann (2014)\\
            &   6350$\pm$85   & 4.09$\pm$0.08   &  -2.12$\pm$0.07  & 1.70   &Sitnova, Zhao, Mashonkina et al. (2015)\\
            &   6556          & 4.52            &  -2.00           &  --    & Boeche \& Grebel (2016)\\
            &     6300        &  4.0            &    -2.25         & 1.30   &Spite et al. (2017)\\
\hline
HD\,3567   &	6000$\pm$175  & 3.58$\pm$0.27   &  -1.10$\pm$0.13  & 0.93   &This study  \\
	   &   5991           & 3.96            &  -1.25           &  --    &Prugniel (2007)  \\
	   &   5950$\pm$40    & 3.90$\pm$0.06   &  -1.32$\pm$0.04  &1.10    &Fulbright (2000)\\
	   &   6041$\pm$70    & 4.01$\pm$0.10   &  -1.20$\pm$0.10  & --     &Nissen, Chen, Schuster et al. (2000)\\
	   &   5950$\pm$100   & 3.80$\pm$0.30   &  -1.20$\pm$0.15  &0.50    &Mishenina \& Kovtyukh (2001)\\
	   &   6000$\pm$50    & 4.07$\pm$0.15   &  -1.16$\pm$0.08  &1.50    &Nissen, Primas, Asplund et al. (2002)\\
	   &   6087$\pm$50    & 4.16$\pm$0.10   &  -1.19$\pm$0.10  & 0.98   &Gratton, Carretta, Claudi et al. (2003)\\
	   &   6022$\pm$100   & 4.12$\pm$0.20   &  -1.05$\pm$0.10  &1.15    &Zhang \& Zhao (2005)\\
	   &   6060$\pm$?     & 4.47            &  -1.04           & --     &Reddy, Lambert \& Allende Prieto (2006)\\
	   &   6177$\pm$50    & 4.14$\pm$0.01   &  -1.17$\pm$0.03  & --     &Melendez, Casagrande, Ramirez et al. (2010)\\
	   &   6051$\pm$30    & 4.02$\pm$0.05   &  -1.16$\pm$0.04  &1.50    &Nissen \& Schuster (2010)\\
          &    6035          &  4.08           &    -1.33         &1.5     &Hansen et al. (2012)    \\
	   &   6319$\pm$62    & 4.29$\pm$0.10   &  -0.94$\pm$0.08  &1.53    &Ramirez, Allende Prieto \& Lambert (2013)\\
	   &   6157$\pm$92    & 4.19$\pm$0.14   &  -1.13$\pm$0.06  &1.70$\pm$0.29&Bensby, Feltzing \& Oey (2014)\\
	   &    6073          & 3.80            &  -1.21           & --     &Boeche \& Grebel (2016)\\
          &   6051           & 4.02            &  -1.16           &1.5     &Yan, Shi, Nissen and Zhao (2016)\\           
\hline
HD\,194598 &	5875$\pm$165  & 4.10$\pm$0.33   &  -1.17$\pm$0.13  & 0.83   &This study  \\
	   &   5981           &4.26             &  -1.15           &--	    &Prugniel (2007)  \\
	   &   6050$\pm$100   &4.27$\pm$0.10	&  -1.11$\pm$0.10  & 1.31   &Fuhrmann, Pfeiffer, Frank et al. (1997)\\
	   &   5960$\pm$120   &4.15$\pm$0.10	&  -1.40$\pm$0.15  & --     &Israelian, Lopez \& Rebolo (1998)\\
	   &   6058$\pm$80    &4.27$\pm$0.10	&  -1.12$\pm$0.07  & 1.45   &Fuhrmann (1998)\\
	   &   5920$\pm$20    &4.20$\pm$0.20	&  -1.13$\pm$0.03  &1.00    &Jehin, Magain, Neuforge et al. (1999)\\
	   &   5875$\pm$40    &4.20$\pm$0.06	&  -1.23$\pm$0.04  &1.40    &Fulbright (2000)\\
	   &   6058$\pm$80    &4.33$\pm$0.10	&  -1.12$\pm$0.07  &1.45    &Zhao \& Gehren (2000)\\
	   &   6060$\pm$80    &4.27$\pm$0.10	&  -1.12$\pm$0.05  &1.40    &Mashonkina \& Gehren (2000)\\
	   &   5890$\pm$100   &4.00$\pm$0.30	&  -1.16$\pm$0.15  &1.50    &Mishenina \& Kovtyukh (2001)\\
	   &   6023$\pm$50    &4.31$\pm$0.10    &  -1.12$\pm$0.10  &1.26    &Gratton, Carretta, Claudi et al. (2003)\\
	   &   5890$\pm$20    &4.00$\pm$0.20	&  -1.21$\pm$0.10  &1.50    &Mishenina, Soubiran, Kovtyukh et al. (2004)\\
	   &   5980           &4.27	        &   -1.16          &1.60    &Gehren, Shi, Zhang et al. (2006)\\
	   &   6118           &4.37	        &   -1.13          &--	    &Melendez, Casagrande, Ramirez et al. (2010)\\
	   &   5926$\pm$30    &4.32$\pm$0.05	&  -1.08$\pm$0.04  &1.40    &Nissen \& Schuster (2010)\\
	   &   5900           &4.00	        &   -1.20          & 1.10   &Peterson (2013)\\
	   &   6002$\pm$80    &4.31$\pm$0.10	&  -1.11$\pm$0.06  &1.31    &Bensby, Feltzing \& Oey (2014)\\
	   &   5876           &4.08	        &   -1.16          &--	    &Boeche \& Grebel (2016)\\
	   &   5942           &4.33	        &   -1.09          &1.40    &Yan, Shi, Nissen \& Zhao (2016)\\ 
	   &   5942           &4.33	        &   -1.09          & 1.40   &Fishlock, Yong, Karakas et al. (2017)\\
\hline
HD\,201891 &   5850$\pm$160  & 4.20$\pm$0.33  &  -1.02$\pm$0.11 & 0.85  &This study\\
	   &   5904          & 4.34           & -1.04           & --	&Prugniel (2007)\\
          &   5948$\pm$100  & 4.19$\pm$0.10  & -1.05$\pm$0.10  &1.15   &Fuhrmann, Pfeiffer, Frank et al. (1997)\\
          &   5943$\pm$80   & 4.24$\pm$0.10  & -1.05$\pm$0.08  &1.18   &Fuhrmann (1998)\\
          &   5870$\pm$100  & 4.00$\pm$0.10  & -1.20$\pm$0.20  & --	&Israelian, Lopez \& Rebolo (1998)\\
          &   5940$\pm$80   & 4.24$\pm$0.10  & -1.05$\pm$0.05  &1.20   &Mashonkina \& Gehren (2000)\\
          &   5943$\pm$80   & 4.28$\pm$0.10  & -1.05$\pm$0.07  &1.18   &Zhao \& Gehren (2000)\\
          &   5825$\pm$40   & 4.30$\pm$0.06  & -1.12$\pm$0.04  & 1.00  &Fulbright (2000)\\
          &   5827$\pm$70   & 4.43$\pm$0.10  & -1.04$\pm$0.10  & --    &Chen, Nissen, Zhao et al. (2000)\\
          &   5850$\pm$100  & 4.45$\pm$0.30  & -0.99$\pm$0.15  &1.00   &Mishenina \& Kovtyukh (2001)\\
	   &   5917$\pm$50   & 4.28$\pm$0.10  & -1.08$\pm$0.10  &1.27   &Gratton, Carretta, Claudi et al. (2003)\\
	   &   5688$\pm$44   & 4.39$\pm$0.06  & -1.12$\pm$0.03  & 0.85  &Valenti \& Fischer (2005)\\
	   &   5730          & 4.34           & -1.07           & --    &Reddy, Lambert \& Allende Prieto (2006)\\
	   &   5900          & 4.22           & -1.07           &1.20   &Gehren, Shi, Zhang et al. (2006)\\
	   &   5947          & 4.31           & -1.05           & --    &Melendez, Casagrande, Ramirez et al. (2010)\\
          &   5850          & 4.40           & -0.96           &  --   &Mishenina, Gorbeneva, Basak et al. (2011)\\
          &   5985$\pm$37   & 4.32$\pm$0.03  & -1.01$\pm$0.08  & 1.47  &Ramirez, Allende Prieto \& Lambert (2013)\\
          &   5840$\pm$41   & 4.10$\pm$0.32  & -1.20$\pm$0.10  &1.10   &Roederer, Preston, Thompson et al. (2014)\\
          &   5973$\pm$69   & 4.35$\pm$0.10  & -1.08$\pm$0.06  &1.33   &Bensby, Feltzing \& Oey (2014)\\
	   &   5913          & 4.26           & -1.06           & --    &Boeche \& Grebel (2016)\\
\hline
BD+423607 &   5480$\pm$130   & 3.72$\pm$0.45  &  -2.43$\pm$0.12  & 0.90  &This study  \\
	  &   5893           & 4.31           & -2.03            & --	 &Prugniel (2007)  \\
	  &   5650$\pm$60    & 5.00$\pm$0.15  & -2.25$\pm$0.04   & 1.20  &Carney, Wright, Sneden et al. (1997)\\
          &   5850$\pm$100   & 4.00$\pm$0.30  & -1.97$\pm$0.15   &0.90   &Mishenina \& Kovtyukh (2001)\\
          &   5965$\pm$50    & 4.58$\pm$0.10  & -2.06$\pm$0.10   &0.52   & Gratton, Carretta, Claudi et al. (2003)\\
          &   5710$\pm$100   & 4.31$\pm$0.20  & -1.99$\pm$0.10   &1.50   &  Zhang \& Zhao (2005)\\
          &   5655           & 3.81           & -2.29            &1.43   & Boesgard, Rich, Levesque et al. (2011)\\
          &   5900           & 4.40           & -2.10            &1.10   & Peterson (2013)\\
          &   5462           & 4.25           & -2.37            & --    & Boeche \& Grebel (2016))\\
\hline
HD\,6755   &   5175$\pm$140  & 2.90$\pm$0.33  & -1.39$\pm$0.13   & 1.30  &This study  \\
	   &   5133          & 2.69           & -1.50            & --	 &Prugniel (2007)  \\
	   &   5150          & 2.70           & -1.50            &1.40   & Plachowski, Sneden \& Kraft (1996)\\
          &   5150$\pm$40   & 2.80$\pm$0.06  & -1.58$\pm$0.04   &  1.50 &  Fulbright (2000)\\
	   &   5150          & 2.70           & -1.50            & 1.40  & Burris, Plachowski, Armandroff et al. (2000)\\
	   &   5100$\pm$100  & 2.70$\pm$0.30  & -1.47$\pm$0.15   & 1.20  &Mishenina \& Kovtyukh (2001)\\
	   &   5128          & 2.70           & -1.60            &  --   & Boeche \& Grebel (2016)\\
\hline
\end{longtable*}

\begin{table*}
\tiny
\centering
\caption{Sensitivity of the derived abundances to the uncertainties of
$\Delta\Teff$, $\Delta\logg$, and $\Delta\xi$ in the model atmosphere
parameters for HPM stars.}
\begin{tabular}{l|cccc|cccc|cccc}
\hline\hline
\multicolumn {13}{c}{$\Delta$ log$\epsilon$}\\
\hline
         & \multicolumn{4}{c|}{HD201891} &\multicolumn{4}{c|}{HD194598} &\multicolumn{4}{c|}{HD6755} \\
\cline{2-13}
Species   & $\Delta$Teff  & $\Delta$logg  & $\Delta\xi$(-)  & $\Delta\xi$(+)  & $\Delta$Teff  & $\Delta$logg  & $\Delta\xi$(-)  & $\Delta\xi$(+)  & $\Delta$Teff  & $\Delta$logg  & $\Delta\xi$(-)  & $\Delta\xi$(+)  \\
& (+160) & (+0.33) & (-0.50) & (+0.50)  & (+165) & (+0.33) & (-0.50) & (+0.50) & (+140) & (+0.33) & (-0.50) & (+0.50) \\
& (\kelvin)& (cgs) & (km s$^{\rm -1}$) & (km s$^{\rm -1}$)  & (\kelvin)& (cgs) & (km s$^{\rm -1}$) & (km s$^{\rm -1}$) & (\kelvin)& (cgs)  & (km s$^{\rm -1}$) & (km s$^{\rm -1}$) \\
\hline\hline
C\,{\sc i}   &  0.05  & -0.15  & -0.05  & -0.05  &  0.10  &  0.00  &  0.00  &  0.10  &  0.00  & -0.05  &  0.00  &  0.00  \\
Na\,{\sc i}  & -0.07  &  0.00  & -0.01  &  0.01  & -0.07  &  0.00  & -0.01  &  0.00  &  0.00  &  0.05  &  0.13  &  0.10  \\
Mg\,{\sc i}  & -0.12  &  0.02  & -0.06  &  0.08  & -0.12  &  0.07  & -0.03  &  0.04  & -0.10  &  0.07  & -0.05  &  0.06  \\
Si\,{\sc i}  & -0.05  & -0.02  & -0.07  & -0.02  & -0.05  &  0.00  &  0.00  & -0.02  & -0.03  &  0.02  &  0.07  &  0.12  \\
Si\,{\sc ii} &  0.07  & -0.07  &  0.10  &  0.03  &  0.00  & -0.10  &  0.00  &  0.10  &  0.03  & -0.15  &  0.00  &  0.02  \\
Ca\,{\sc i}  & -0.10  &  0.06  & -0.05  &  0.07  & -0.11  &  0.06  & -0.05  &  0.07  & -0.11  &  0.05  & -0.11  &  0.11  \\
Sc\,{\sc i}  & -0.15  &  0.00  &  0.00  &  0.00  & -0.10  &  0.00  &  0.00  &  0.00  & -0.10  &  0.00  &  0.05  &  0.10  \\
Sc\,{\sc ii} & -0.05  & -0.09  & -0.05  &  0.08  & -0.06  & -0.10  & -0.04  &  0.06  &  0.02  & -0.03  &  0.05  &  0.12  \\
Ti\,{\sc i}  & -0.13  &  0.01  & -0.04  &  0.06  & -0.14  &  0.01  & -0.04  &  0.05  & -0.17  &  0.02  & -0.11  &  0.08  \\
Ti\,{\sc ii} & -0.04  & -0.10  & -0.09  &  0.10  & -0.04  & -0.11  & -0.06  &  0.07  & -0.03  & -0.12  & -0.19  &  0.16  \\
V\,{\sc i}   & -0.10  &  0.00  &  0.00  &  0.04  & -0.10  & -0.10  &  0.00  &  0.03  & -0.15  & -0.03  & -0.03  & -0.03  \\
Cr\,{\sc i}  & -0.14  &  0.02  & -0.07  &  0.08  & -0.12  &  0.01  & -0.05  &  0.05  & -0.15  &  0.02  & -0.17  &  0.13  \\
Cr\,{\sc ii} &  0.02  & -0.11  & -0.04  &  0.05  &  0.01  & -0.13  & -0.03  &  0.03  &  0.03  & -0.12  & -0.13  &  0.10  \\
Mn\,{\sc i}  & -0.05  &  0.05  & -0.02  &  0.13  & -0.10  &  0.00  & -0.01  &  0.02  & -0.12  & -0.10  &  0.00  &  0.03  \\
Fe\,{\sc i}  & -0.12  &  0.04  & -0.05  &  0.08  & -0.13  &  0.03  & -0.07  &  0.08  & -0.14  &  0.04  & -0.14  &  0.13  \\
Fe\,{\sc ii} &  0.00  & -0.10  & -0.08  &  0.09  & -0.01  & -0.11  & -0.07  &  0.08  &  0.01  & -0.12  & -0.14  &  0.12  \\
Co\,{\sc i}  & -0.05  &  0.05  &  0.10  &  0.10  & -0.15  &  0.00  & -0.10  &  0.00  & -0.05  & -0.05  & -0.20  &  0.20  \\
Ni\,{\sc i}  & -0.10  &  0.00  & -0.03  &  0.04  & -0.10  &  0.00  & -0.03  &  0.04  & -0.11  &  0.00  & -0.06  &  0.05  \\
Cu\,{\sc i}  & -0.13  &  0.00  &  0.00  &  0.00  & -0.10  & -0.05  & -0.02  & -0.05  & -0.15  & -0.10  & -0.10  & -0.05  \\
Sr\,{\sc i}  &  ....  &  ....  &  ....  &  ....  &  ....  &  ....  &  ....  &  ....  & -0.10  &  0.00  &  0.00  &  0.05  \\
Sr\,{\sc ii} &  0.00  &  0.10  &  0.00  &  0.10  & -0.10  &  0.00  & -0.05  & -0.05  & -0.01  &  0.04  & -0.11  & -0.01  \\
Y\,{\sc ii}  &  0.00  & -0.08  & -0.05  &  0.07  &  0.00  & -0.10  & -0.03  &  0.05  & -0.03  & -0.08  & -0.10  &  0.10  \\
Zr\,{\sc i}  & -0.15  &  0.00  &  0.00  &  0.00  & -0.10  &  0.00  & -0.05  & -0.05  & -0.05  &  0.15  &  0.15  &  0.15  \\
Zr\,{\sc ii} & -0.07  & -0.12  & -0.07  &  0.00  & -0.10  & -0.15  & -0.05  &  0.00  & -0.10  & -0.17  & -0.20  & -0.02  \\
Ba\,{\sc ii} & -0.12  & -0.09  & -0.17  &  0.05  &  0.00  &  0.00  & -0.10  &  0.03  & -0.06  & -0.08  & -0.16  &  0.12  \\
Ce\,{\sc ii} &  ....  &  ....  &  ....  &  ....  &  0.00  & -0.05  &  0.00  &  0.05  &  0.00  & -0.15  &  0.00  &  0.00  \\
Nd\,{\sc ii} &  ....  &  ....  &  ....  &  ....  &  ....  &  ....  &  ....  &  ....  &  0.00  & -0.12  & -0.02  &  0.05  \\
Sm\,{\sc ii} & -0.12  & -0.02  & -0.12  & -0.12  & -0.05  & -0.05  &  0.00  &  0.05  &  0.00  & -0.10  &  0.00  &  0.05  \\
	&&&&&&&&&&&&\\
\hline\hline
         &\multicolumn{4}{c|}{HD3567} &\multicolumn{4}{c|}{BD+423607} &\multicolumn{4}{c|}{HD84937} \\
\cline{2-13}
Species   & $\Delta$Teff  & $\Delta$logg  & $\Delta\xi$(-)  & $\Delta\xi$(+)  & $\Delta$Teff  & $\Delta$logg  &  $\Delta\xi$(-)  & $\Delta\xi$(+)  &  $\Delta$Teff  &  $\Delta$logg  & $\Delta\xi$(-)  & $\Delta\xi$(+)  \\
 & (+175) & (+0.27) & (-0.50) & (+0.50) & (+130) & (+0.45) & (-0.50) & (+0.50) & (+140) & (+0.18) & (-0.50) & (+0.50) \\
& (\kelvin)& (cgs) & (km s$^{\rm -1}$) & (km s$^{\rm -1}$)  & (\kelvin)& (cgs) & (km s$^{\rm -1}$) & (km s$^{\rm -1}$) & (\kelvin)& (cgs)  & (km s$^{\rm -1}$) & (km s$^{\rm -1}$) \\
 \hline\hline
C\,{\sc i}   &  ....  &  ....  &  ....  &  ....  &  ....  &  ....  &  ....  &  ....  &  ....  &  ....  &  ....  &  ....  \\
Na\,{\sc i}  & -0.05  &  0.00  &  0.05  &  0.00  & -0.10  & -0.07  & -0.10  &  0.00  &  ....  &  ....  &  ....  &  ....  \\
Mg\,{\sc i}  & -0.06  &  0.16  &  0.11  &  0.21  & -0.09  &  0.02  &  0.00  & -0.01  & -0.06  &  0.01  & -0.05  &  0.03  \\
Si\,{\sc i}  & -0.06  & -0.01  & -0.01  &  0.00  &  0.02  &  0.00  &  0.02  &  0.09  &  ....  &  ....  &  ....  &  ....  \\
Si\,{\sc ii} &  0.10  & -0.11  & -0.02  &  0.02  &  0.05  & -0.15  &  0.13  & -0.20  &  ....  &  ....  &  ....  &  ....  \\
Ca\,{\sc i}  & -0.11  &  0.04  & -0.08  &  0.11  & -0.08  &  0.03  & -0.04  &  0.04  & -0.08  &  0.00  & -0.04  &  0.02  \\
Sc\,{\sc i}  &  ....  &  ....  &  ....  &  ....  &  ....  &  ....  &  ....  &  ....  &  ....  &  ....  &  ....  &  ....  \\
Sc\,{\sc ii} & -0.04  & -0.12  & -0.07  & -0.04  & -0.04  & -0.14  & -0.09  & -0.01  & -0.09  & -0.14  & -0.06  & -0.06  \\
Ti\,{\sc i}  & -0.15  &  0.00  & -0.05  &  0.05  & -0.13  &  0.00  & -0.03  &  0.04  & -0.11  &  0.01  & -0.02  &  0.02  \\
Ti\,{\sc ii} & -0.05  & -0.09  & -0.10  &  0.10  & -0.07  & -0.13  & -0.13  &  0.14  & -0.06  & -0.06  & -0.14  &  0.09  \\
V\,{\sc i}   & -0.10  &  0.05  &  0.05  &  0.05  &  ....  &  ....  &  ....  &  ....  &  0.00  & -0.10  & -0.05  &  0.00  \\
Cr\,{\sc i}  & -0.14  &  0.00  & -0.06  &  0.05  & -0.14  &  0.05  & -0.07  &  0.10  & -0.12  &  0.00  & -0.07  &  0.04  \\
Cr\,{\sc ii} &  0.00  & -0.09  & -0.07  &  0.08  &  0.00  & -0.15  &  0.00  &  0.05  &  0.00  & -0.02  & -0.02  &  0.03  \\
Mn\,{\sc i}  &  0.00  &  0.08  &  0.08  &  0.05  & -0.05  &  0.00  & -0.15  &  0.00  &  0.00  &  0.00  &  0.00  &  0.05  \\
Fe\,{\sc i}  & -0.13  &  0.03  & -0.10  &  0.11  & -0.12  &  0.02  & -0.07  &  0.06  & -0.10  &  0.00  & -0.08  &  0.05  \\
Fe\,{\sc ii} & -0.01  & -0.09  & -0.09  &  0.09  & -0.02  & -0.14  & -0.06  &  0.08  & -0.02  & -0.07  & -0.09  &  0.06  \\
Co\,{\sc i}  & -0.15  & -0.05  & -0.10  &  0.05  & -0.15  & -0.05  & -0.40  & -0.05  & -0.15  &  0.00  &  0.00  &  0.05  \\
Ni\,{\sc i}  & -0.11  &  0.00  & -0.03  &  0.04  & -0.08  & -0.03  & -0.02  &  0.01  & -0.10  &  0.00  & -0.03  &  0.01  \\
Cu\,{\sc i}  &  ....  &  ....  &  ....  &  ....  &  ....  &  ....  &  ....  &  ....  &  ....  &  ....  &  ....  &  ....  \\
Sr\,{\sc i}  & -0.10  & -0.05  & -0.05  & -0.05  &  ....  &  ....  &  ....  &  ....  & -0.10  & -0.15  & -0.15  & -0.10  \\
Sr\,{\sc ii} &  0.00  &  0.02  &  0.00  &  0.25  & -0.13  & -0.03  & -0.38  &  0.27  & -0.05  &  0.10  & -0.10  &  0.40  \\
Y\,{\sc ii}  &  0.00  &  0.00  &  0.00  &  0.15  &  0.05  & -0.05  & -0.05  &  0.15  &  0.00  &  0.00  &  0.03  &  0.03  \\
Zr\,{\sc i}  &  ....  &  ....  &  ....  &  ....  &  ....  &  ....  &  ....  &  ....  &  ....  &  ....  &  ....  &  ....  \\
Zr\,{\sc ii} &  0.00  & -0.10  & -0.20  &  0.05  & -0.15  & -0.30  & -0.30  & -0.10  & -0.05  & -0.10  &  0.05  & -0.05  \\
Ba\,{\sc ii} &  0.01  &  0.01  & -0.04  &  0.18  & -0.18  & -0.23  & -0.40  &  0.00  & -0.05  &  0.00  & -0.02  &  0.05  \\
Ce\,{\sc ii} &  ....  &  ....  &  ....  &  ....  &  ....  &  ....  &  ....  &  ....  &  ....  &  ....  &  ....  &  ....  \\
Nd\,{\sc ii} &  0.00  &  0.00  &  0.00  &  0.05  &  0.00  & -0.10  & -0.20  &  0.00  &  ....  &  ....  &  ....  &  ....  \\
Sm\,{\sc ii} & -0.15  & -0.20  & -0.10  & -0.10  &  ....  &  ....  &  ....  &  ....  &  ....  &  ....  &  ....  &  ....  \\
\hline\hline
\end{tabular}
\end{table*}

%\newpage

\begin{table*}
\scriptsize
\caption{Abundances of NGC\,5897, NGC\,4833, and NGC\,~362 along with the abundances of HD\,201891, HD\,194598, and HD\,3567. The
reported errors in abundances are the standard errors.}
\centering
\begin{tabular}{l|c c|c|c||c c|c|c||c c|c|c}
\hline
\hline
        & \multicolumn{2}{c|}{NGC\,5897} &\multicolumn{2}{c||}{HD\,201891} & \multicolumn{2}{c|}{NGC\,4833} &\multicolumn{2}{c||}{HD\,194598}&  \multicolumn{2}{c|}{NGC\,~362} &\multicolumn{2}{c}{HD\,3567}   \\
\cline{2-13}
 Element& [X/Fe] & Ref& [X/Fe] & N & [X/Fe] & Ref& [X/Fe] & N &[X/Fe]& Ref & [X/Fe]& N\\
\hline
Na\,{\sc i}   &  0.27$\pm$0.24 &1 &  0.11$\pm$ 0.13     &2    &  0.39$\pm$0.22   &2 & -0.02$\pm$0.13    &2  &-0.04$\pm$0.00  &3 &-0.20$\pm$0.18        &1    \\
Mg\,{\sc i}   &  0.46$\pm$0.08 &1 &  0.31$\pm$ 0.13     &2    &  0.35$\pm$0.24   &2 &  0.03$\pm$0.09    &5  & 0.14$\pm$0.08  &3 & 0.09$\pm$0.09        &6    \\
Si\,{\sc i}   &  0.17$\pm$0.10 &1 &  0.28$\pm$ 0.09     &4    &{\bf 0.74$\pm$0.09}&2&{\bf 0.17$\pm$0.13}&2  & 0.29$\pm$0.08  &3a & 0.18$\pm$0.10        &6    \\
Ca\,{\sc i}   &  0.39$\pm$0.04 &1 &  0.27$\pm$ 0.06     &19   &  0.49$\pm$0.04   &2 &  0.35$\pm$0.07    &17 & 0.27$\pm$0.12  &3 & 0.36$\pm$0.06        &21   \\
Sc\,{\sc ii}  &  0.08$\pm$0.07 &1 &  0.22$\pm$ 0.09     &5    &  0.19$\pm$0.11   &2 &  0.17$\pm$0.09    &6  & 0.06$\pm$0.09  &3a &-0.20$\pm$0.10        &6    \\
Ti\,{\sc i}   &  0.16$\pm$0.04 &1 &  0.21$\pm$ 0.05     &19   &  0.08$\pm$0.06   &2 &  0.15$\pm$0.05    &19 & 0.13$\pm$0.08  &3a & 0.17$\pm$0.05        &25   \\
Ti\,{\sc ii}  &  0.30$\pm$0.07 &1 &  0.16$\pm$ 0.06     &17   &  0.39$\pm$0.08   &2 &  0.14$\pm$0.06    &15 & 0.12$\pm$0.13  &3a & 0.02$\pm$0.07        &17   \\
V\,{\sc i}    & -0.14$\pm$0.07 &1 &  0.08$\pm$ 0.14     &2    &  0.05$\pm$0.19   &2 &  0.06$\pm$0.15    &2  &{\bf -0.14$\pm$0.04}&3a&{\bf-0.50$\pm$0.18}&1    \\
Cr\,{\sc i}   & -0.20$\pm$0.05 &1 & -0.05$\pm$ 0.06     &14   & -0.26$\pm$0.07   &2 & -0.03$\pm$0.05    &17 &-0.13$\pm$0.07  &3a &-0.08$\pm$0.06        &19   \\
Mn\,{\sc i}   & -0.43$\pm$0.05 &1 & -0.43$\pm$ 0.09     &8    & -0.58$\pm$0.11   &2 & -0.39$\pm$0.08    &9  &-0.36$\pm$0.07  &3a &-0.59$\pm$0.12        &5    \\
Co\,{\sc i}   &  0.03$\pm$0.07 &1 &  0.18$\pm$ 0.13     &3    & -0.39$\pm$0.14   &2 &  0.13$\pm$0.12    &2  & 0.11$\pm$0.05  &3a & 0.09$\pm$0.18        &1    \\
Ni\,{\sc i}   & -0.07$\pm$0.02 &1 & -0.01$\pm$ 0.04     &36   & -0.06$\pm$0.06   &2 & -0.05$\pm$0.05    &26 &-0.13$\pm$0.03  &3a &-0.03$\pm$0.10        &6    \\
Cu\,{\sc i}&{\bf -0.70$\pm$0.08}&1&{\bf -0.11$\pm$ 0.09}&2    &{\bf -0.65$\pm$0.14}&2&{\bf -0.31$\pm$0.16}&2&-0.56$\pm$0.01  &3a & ...                  & ... \\
Y\,{\sc ii}   &-0.29$\pm$0.11  &1 & -0.21$\pm$ 0.12     &2    &-0.22$\pm$0.11    &2 & -0.24$\pm$0.13    &2  &-0.01$\pm$0.07  &3b &-0.13$\pm$0.12        &3    \\
Zr\,{\sc ii}  &    ...         &... &  0.10$\pm$0.18      &2    &  0.25$\pm$0.11   &2 &  0.01$\pm$0.18    &1  & 0.17$\pm$0.11  &3a & 0.08$\pm$0.18        &1    \\
Ba\,{\sc ii}  & -0.03$\pm$0.08 &1 & -0.24$\pm$ 0.18     &3    & -0.02$\pm$0.06   &2 & -0.16$\pm$0.20    &3  & 0.25$\pm$0.07  &3a &-0.02$\pm$0.18        &2    \\
Sm\,{\sc ii}  &  0.31$\pm$0.06 &1 &  0.29$\pm$ 0.16     &1    & 0.29$\pm$0.11    &2 &  0.49$\pm$0.18    &1  &...             &... &...   & ...   \\
\hline 
\end{tabular}
\begin{list}{}{}
\item 1: Koch \& McWilliam (2014); 2: Roederer \& Thompson (2015); 3: Larsen et al. (2017); 3a: Colucci et al. (2017); 3b: Worley \& Cottrell (2010)
\end{list}
\end{table*}

\begin{figure*}[h]
\centering
\includegraphics*[width=8.57cm,height=7.3cm,angle=0]{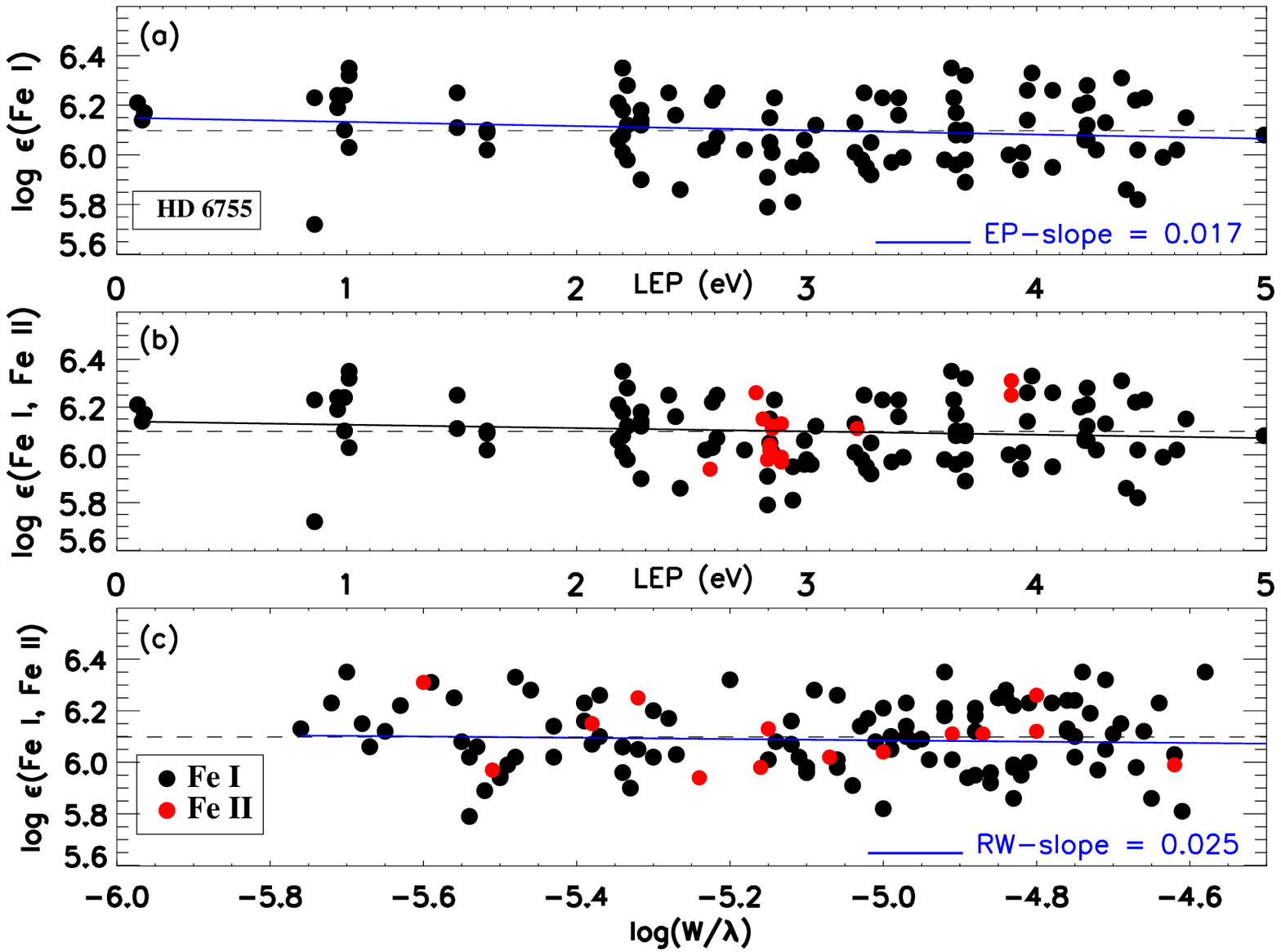}
\includegraphics*[width=8.57cm,height=7.3cm,angle=0]{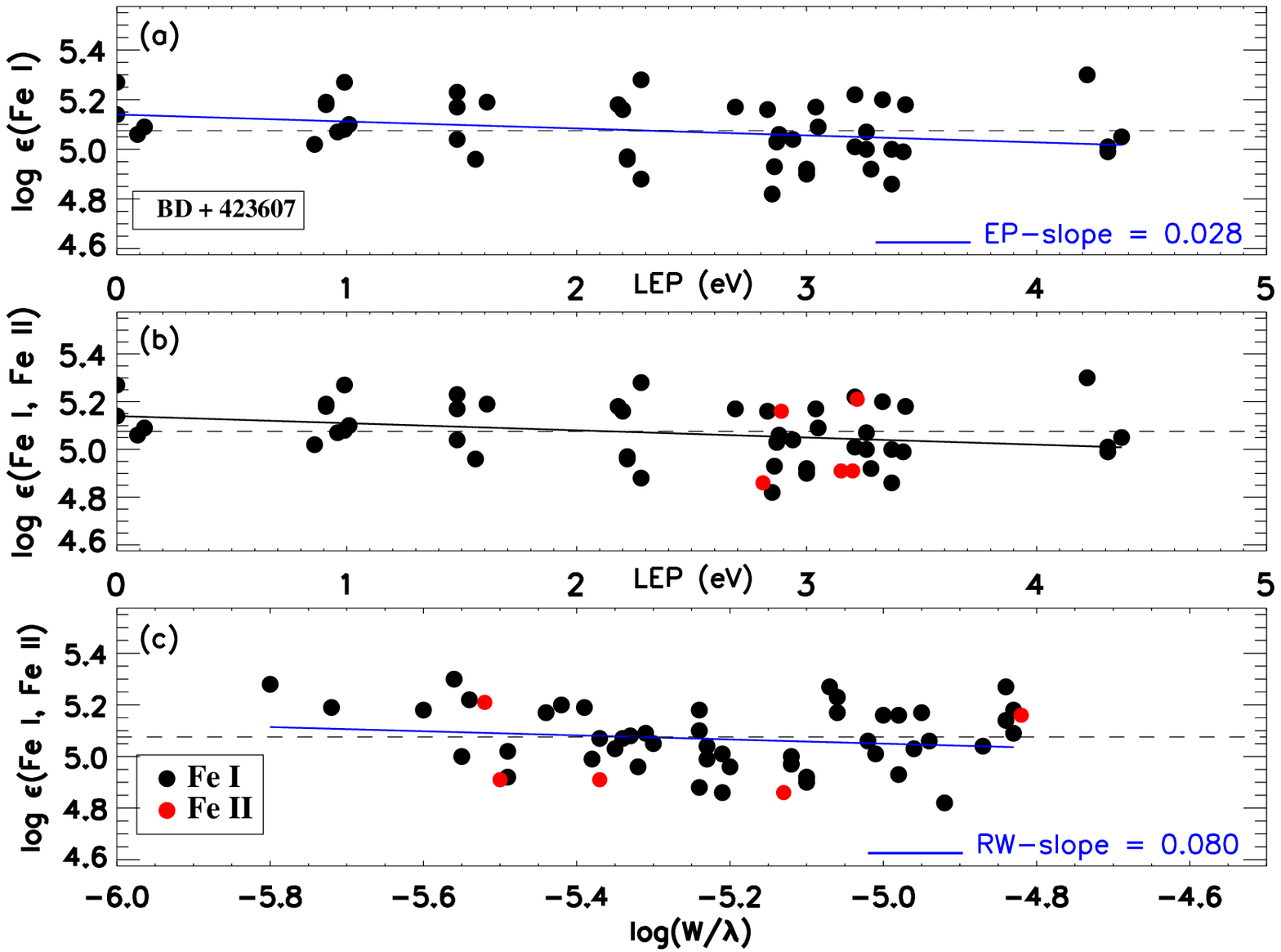}
\includegraphics*[width=8.57cm,height=7.3cm,angle=0]{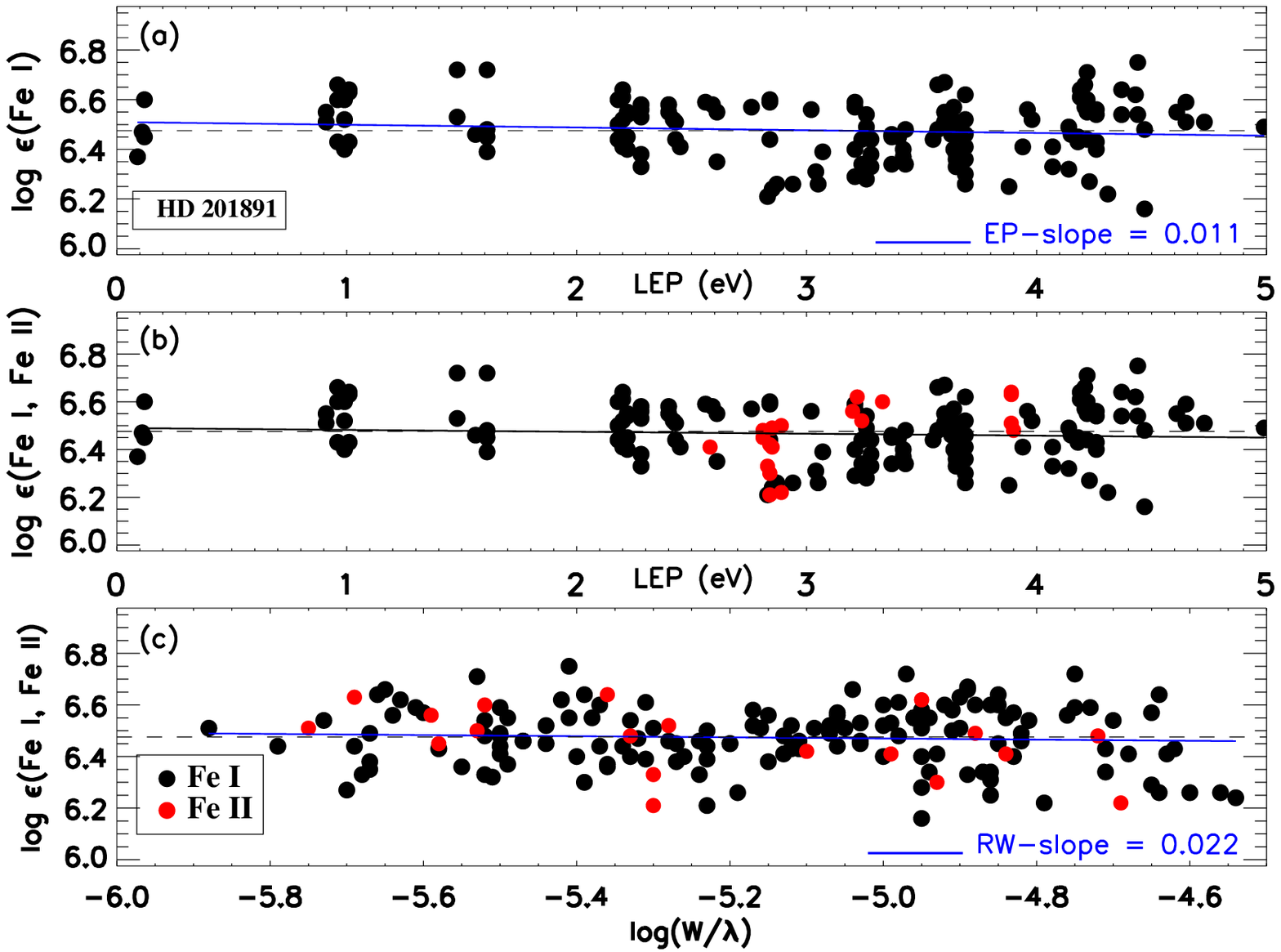}
\includegraphics*[width=8.57cm,height=7.3cm,angle=0]{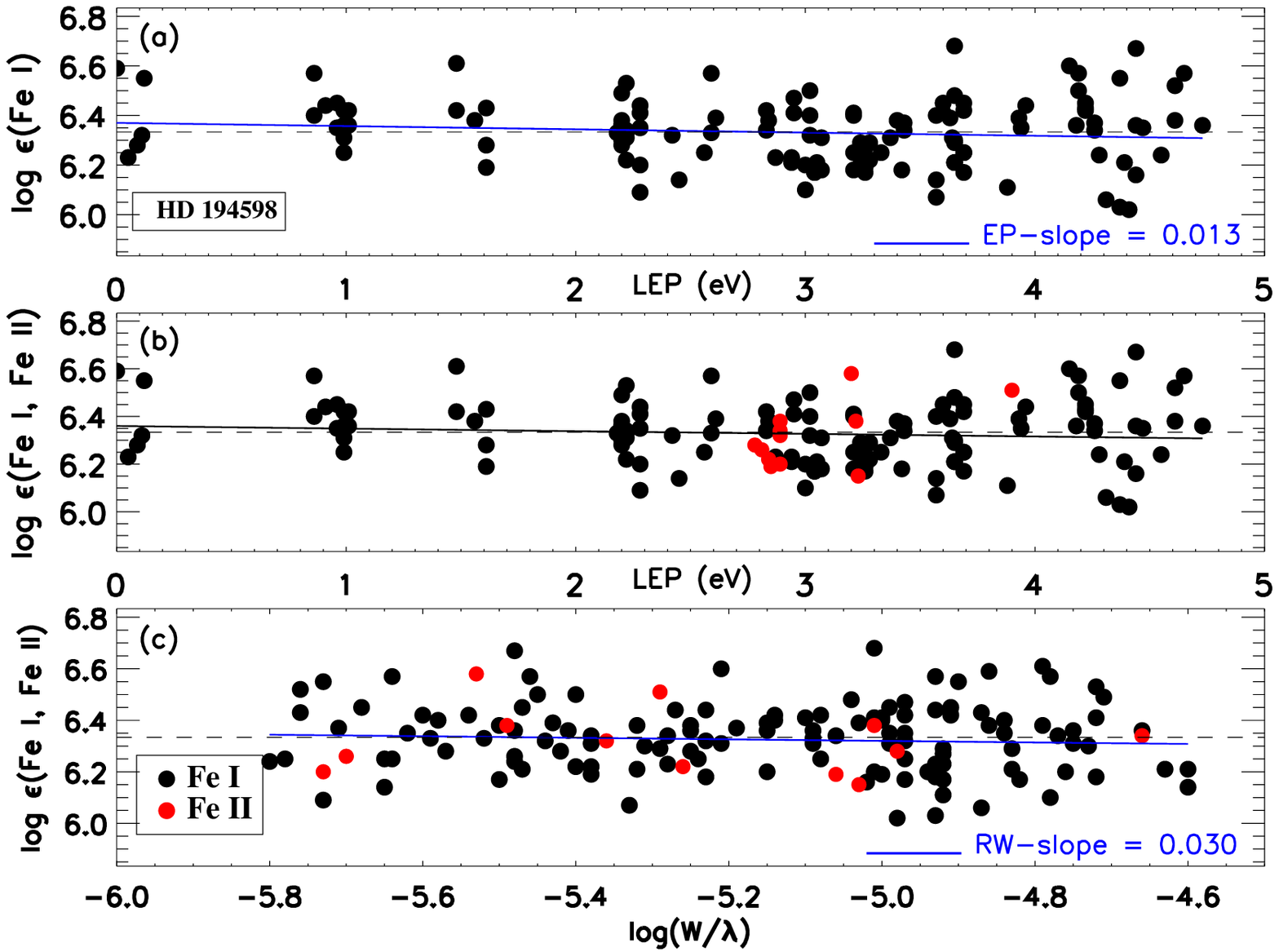}
\includegraphics*[width=8.57cm,height=7.3cm,angle=0]{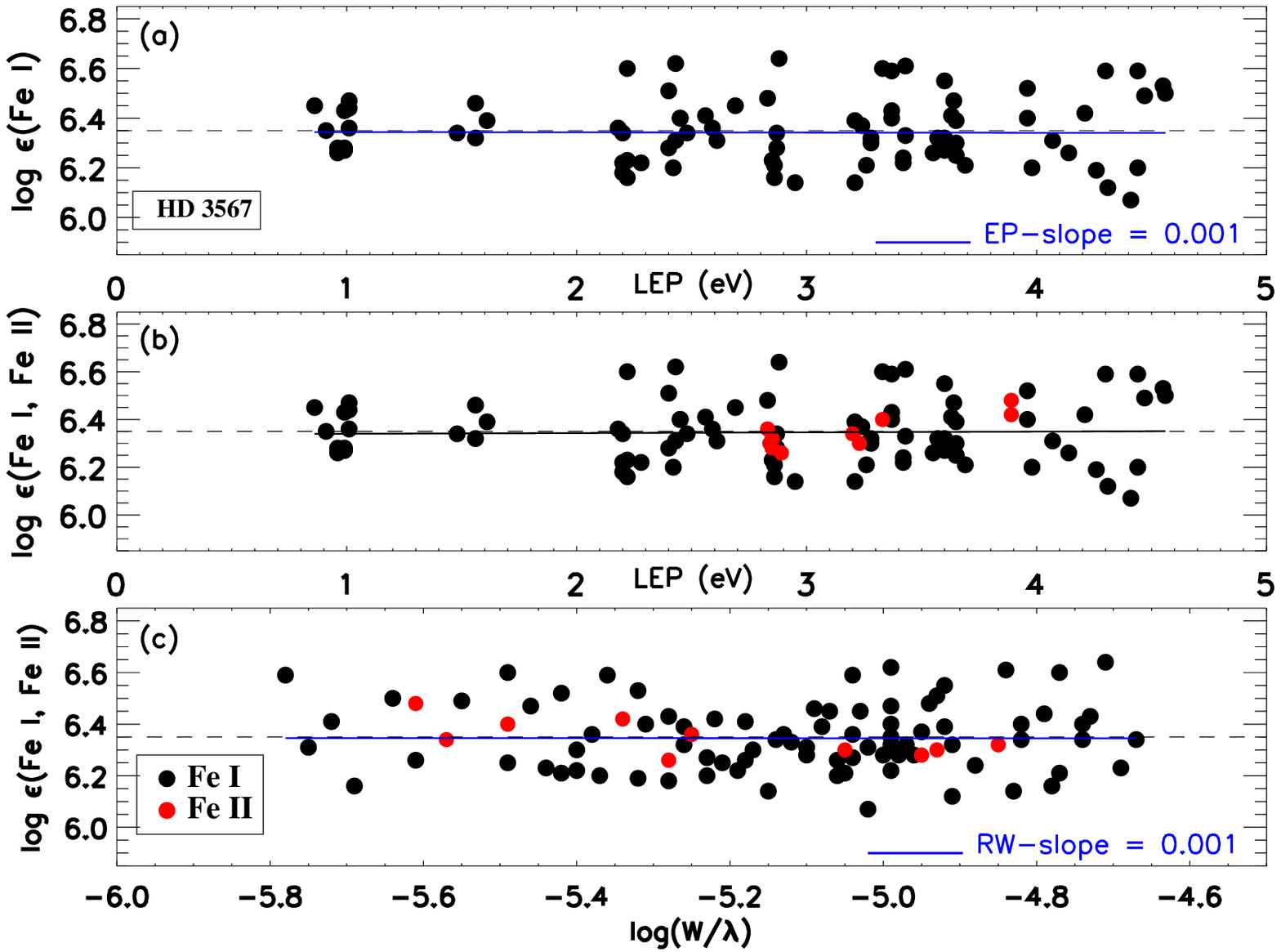}
\includegraphics*[width=8.57cm,height=7.3cm,angle=0]{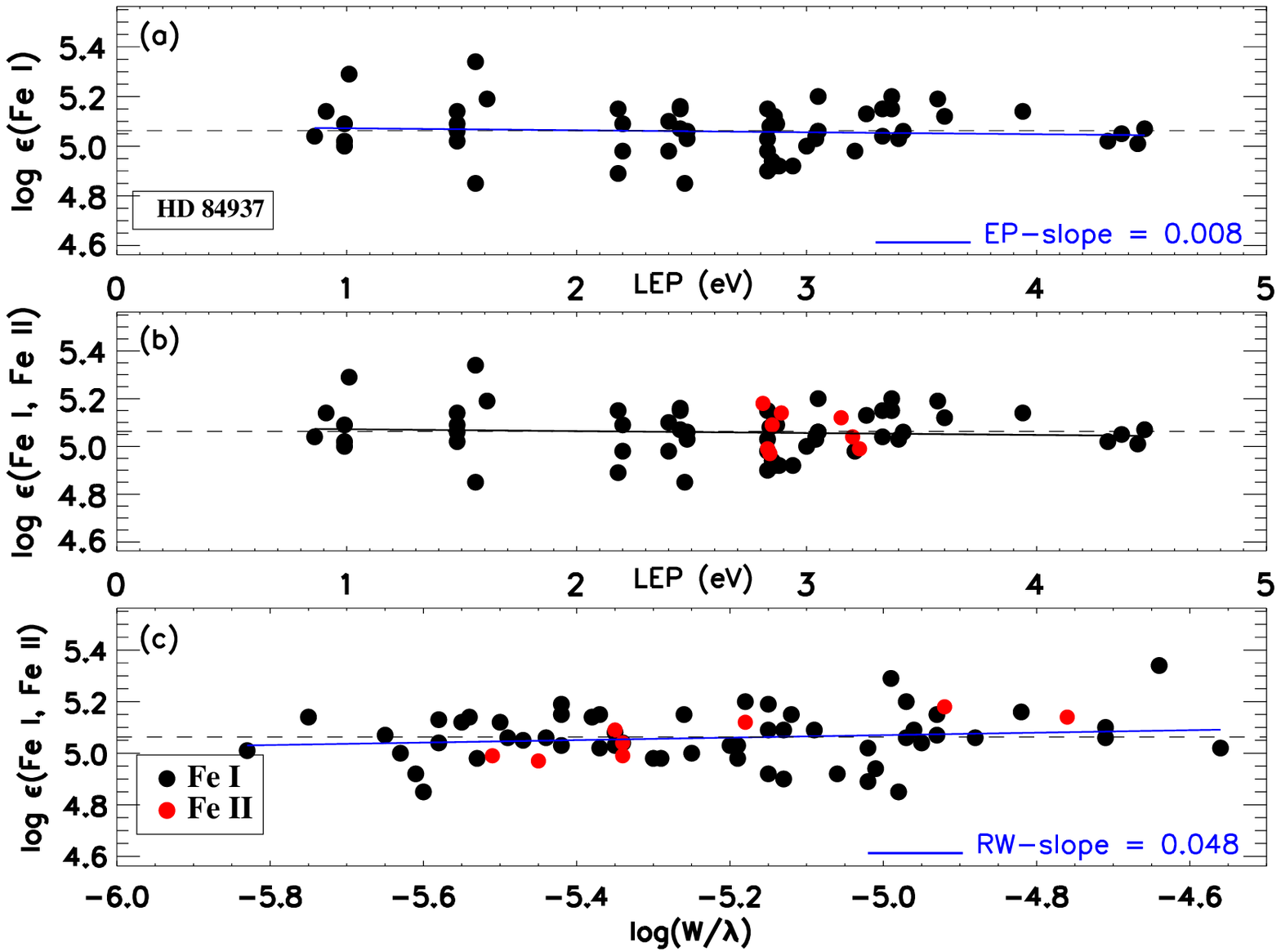}
\caption{An example for the determination of atmospheric parameters $T_{\rm
    eff}$ and $\xi$ using abundance (log$\epsilon$) as a function of both lower
  level excitation potential (LEP, upper and middle panels) and reduced EW (REW; log
  (EW/$\lambda$), bottom panel) for HD\,6755, BD\,+42\,3607, HD\,201891, HD\,194598, HD\,3567, HD\,84937. 
  The solid line in the all panels is the
  least-square fit to the data.}
\label{e_spec_HD84937}
\end{figure*}

\begin{figure*}
\centering 
\includegraphics*[width=8.66cm,height=7.0cm,angle=0]{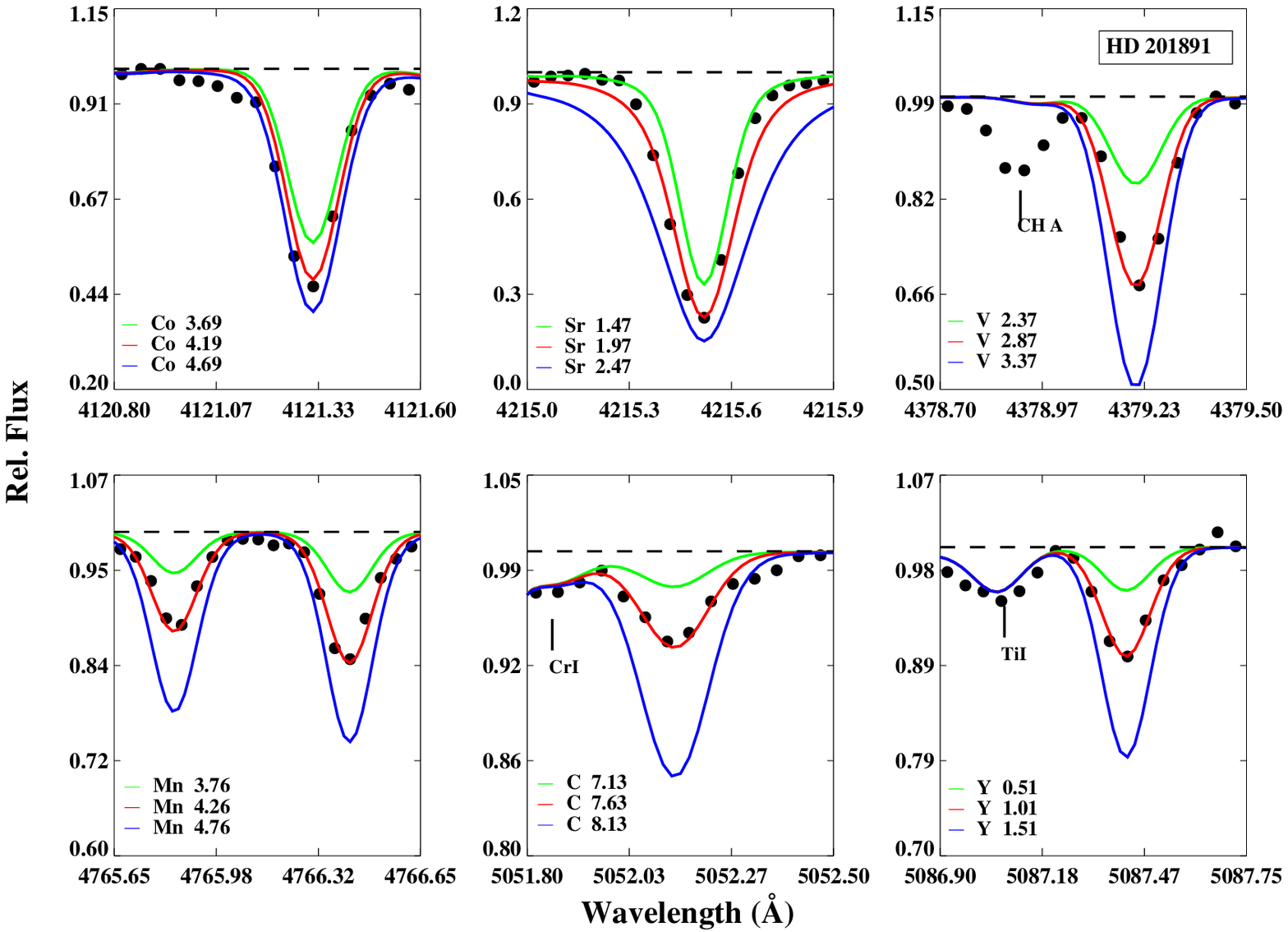}
\includegraphics*[width=8.66cm,height=7.0cm,angle=0]{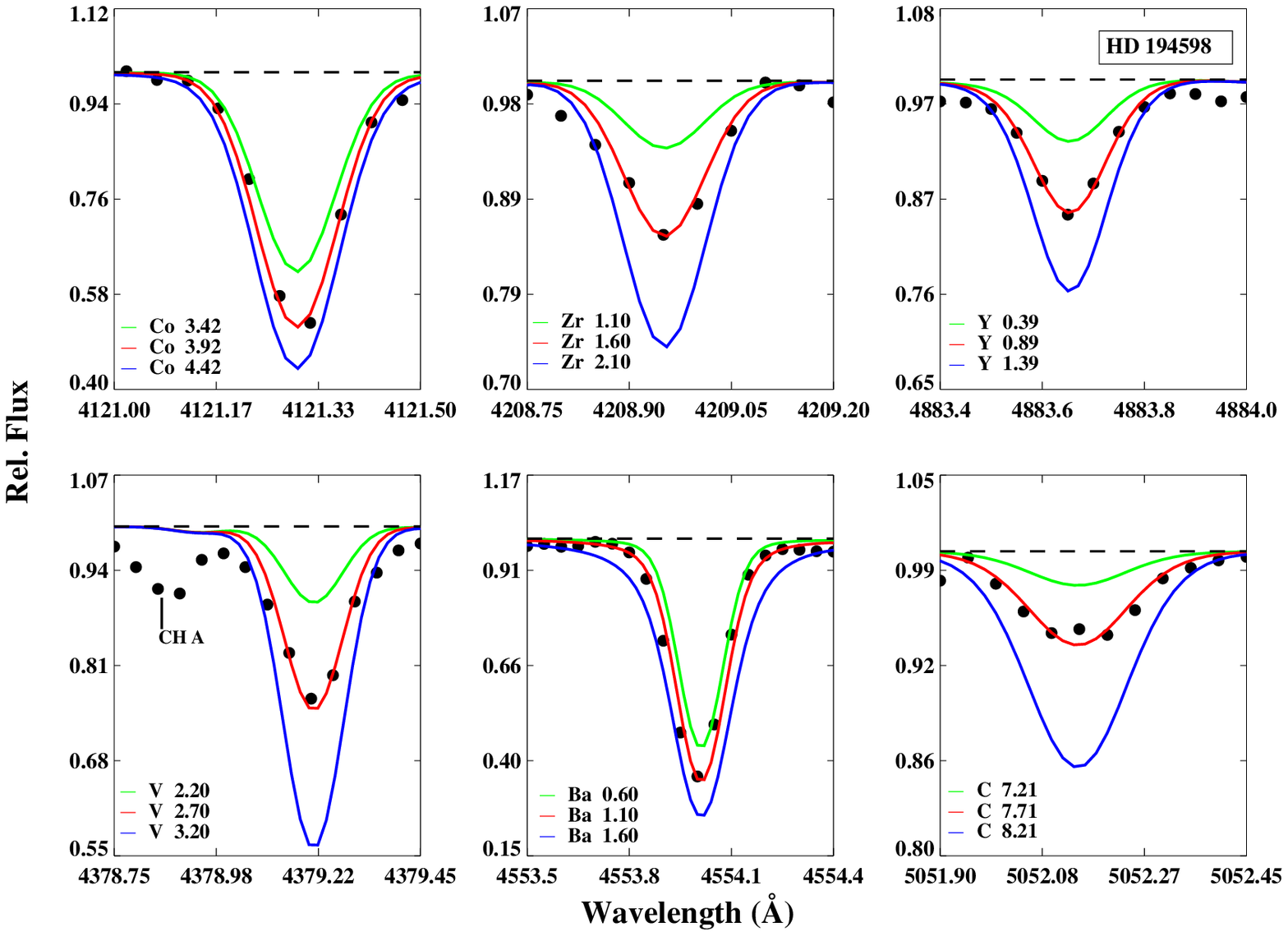}
\includegraphics*[width=8.66cm,height=7.0cm,angle=0]{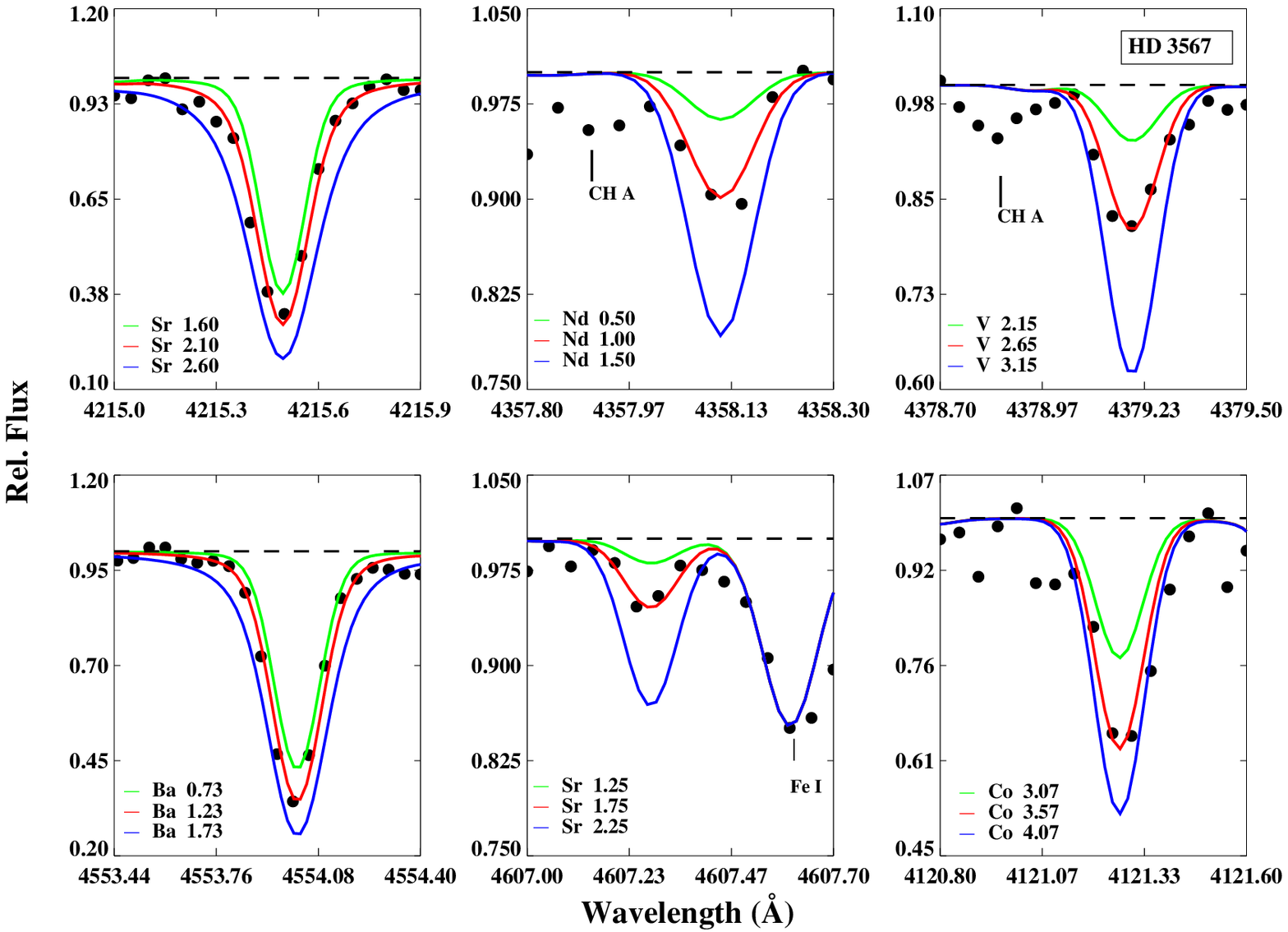}
\includegraphics*[width=8.66cm,height=7.0cm,angle=0]{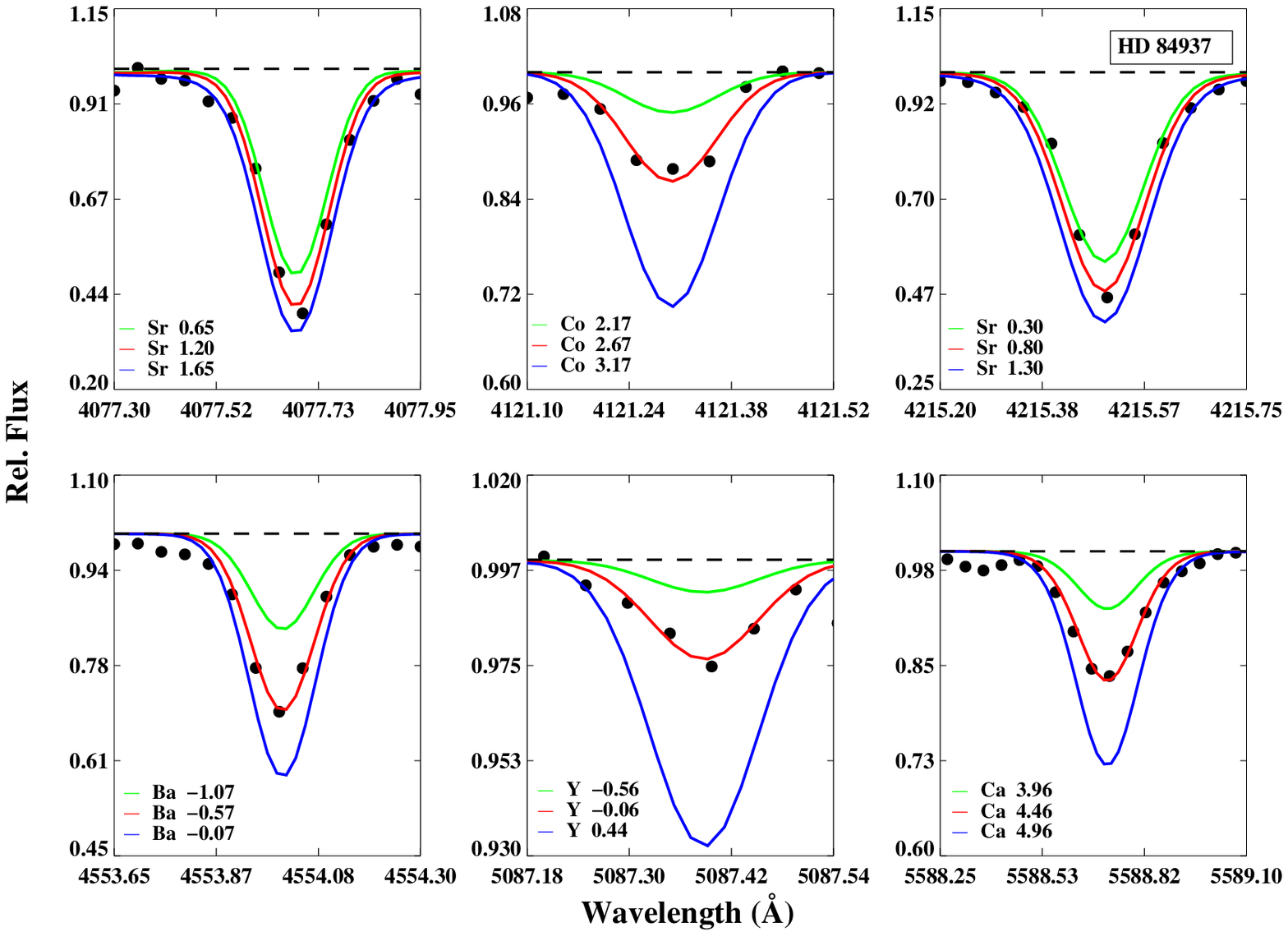}
\includegraphics*[width=8.66cm,height=7.0cm,angle=0]{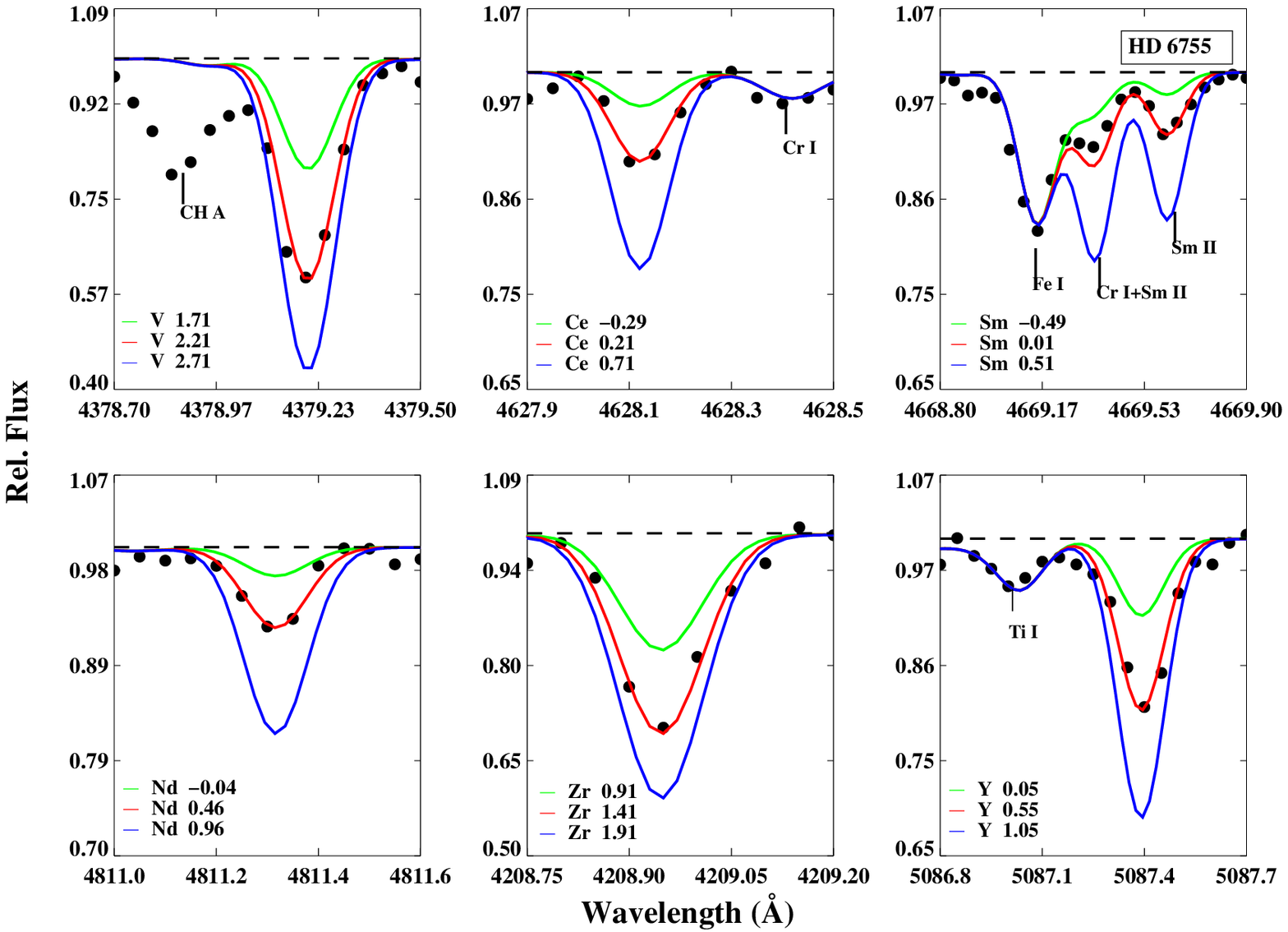}
\includegraphics*[width=8.66cm,height=7.0cm,angle=0]{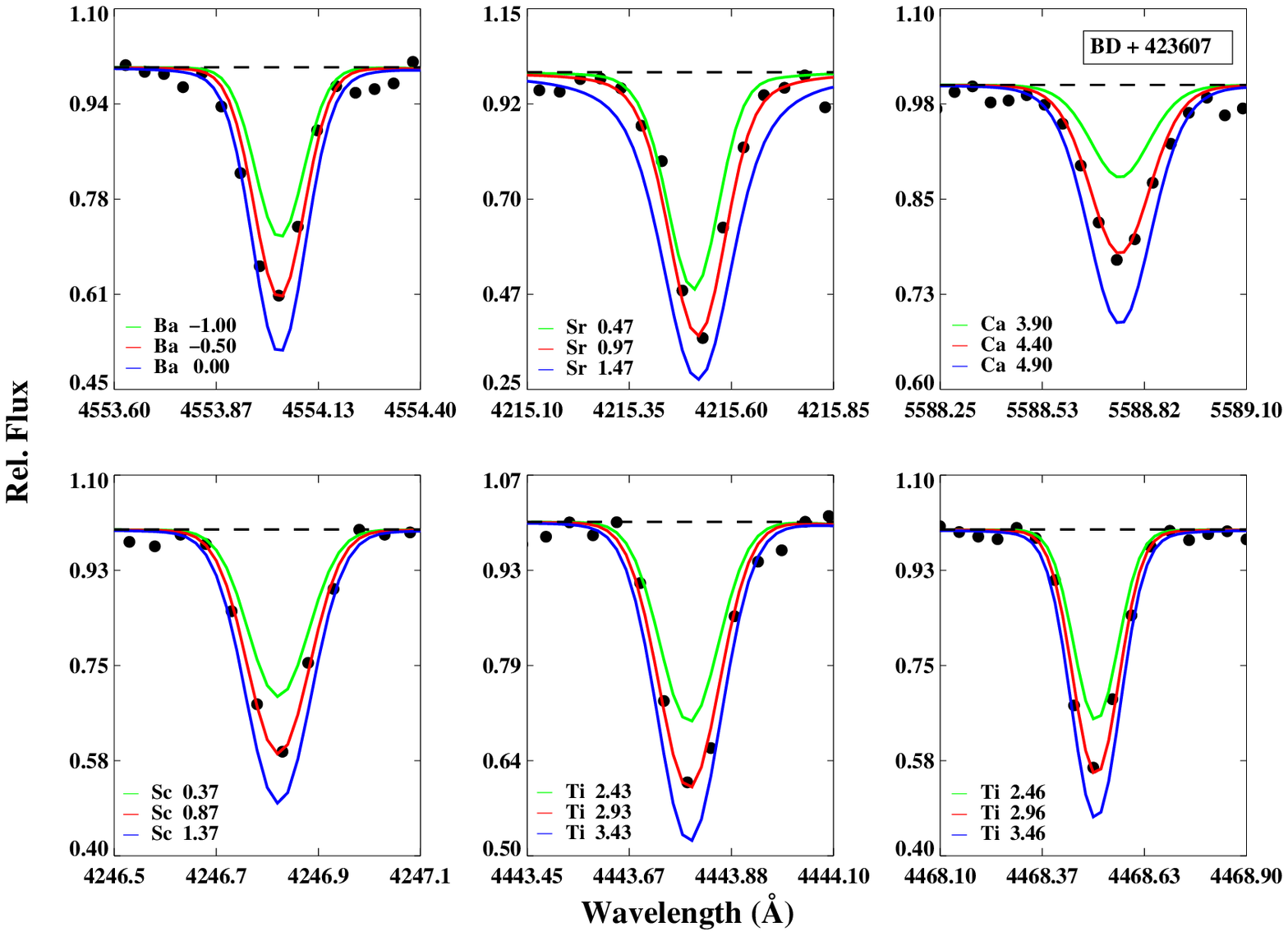}
\caption{The observed (filled circles) and computed (full red line) line profiles for some of the neutral metal lines used in the
analysis of all program stars. The computed profiles show synthetic spectra for the abundances reported in Table A4-A6.}
\label{feh_w}
\end{figure*}

\section{Notes on abundances analysis of HD\,6755}
From KPNO spectra of the star, the reported neutral silicon and calcium abundances by Pilachowski, Sneden, Kraft (1996) are in excellent agreement with
our measurement of those elements (i.e. $\Delta$([X/Fe]$_{\rm TS}$-[X/Fe]$_{\rm PSK}$) = 0.00 dex and 0.09 dex, respectively). The nickel abundance is
also agreed within 0.1 dex. The largest scatter was observed for abundances of neutral sodium ($\Delta$([Na/Fe]$_{\rm TS}$-[Na/Fe]$_{\rm PSK}$ = 0.25
dex), magnesium (-0.24 dex), and single ionized scandium (-0.21 dex). They reported sodium abundance via spectrum synthesis of 5682 \AA\, and 5688
\AA\, Na\,{\sc i} lines. The abundance for magnesium by Pilachowski, Sneden \& Kraft (1996) comes from 5711 \AA\, Mg\,{\sc i} line. Our magnesium
abundance is provided by five magnesium lines reported in Table A4\footnote{Table A4 is available
online.} with log$gf$-values from Kelleher \& Podobedova (2008). The difference in reported
log$gf$-values for this common magnesium line is 0.11 dex. The scandium abundance by Pilachowski et al. (1996) is
determined from 5669 \AA\, Sc\,{\sc ii} line. The mean difference in the log$gf$-values of the scandium lines ($\Delta$(log$gf_{\rm PSK}$ - log$gf_{\rm
TS}$)) is 0.23 dex, hence the difference in scandium abundance is probably due to the difference in the adopted log$gf$ values for this element. 

The star was listed in the {\sc SIMBAD} database as a spectroscopic binary. We did not see any indication for the binarity in its ELODIE
spectrum.\\
Several studies in the literature reported a sub-giant status for the star (i.e. Pilachowski et al. 1996; Burris et al. 2000). However, Carretta
et al. (2000), with no given concrete evidence to support, noted that the star was evolving to a giant stage in the H-R diagram.

The overall agreement between reported abundances by Fulbright (2000) and in this study for all common elements are satisfactory. For
instance, for neutral sodium, silicon, and single ionized barium abundances, the difference is 0.2 dex. The agreement in abundances for
calcium, titanium, chromium, nickel, yttrium, and zirconium is also excellent (i.e. $\Delta$([X/Fe]$_{\rm TS}$-[X/Fe]$_{\rm F}$) = 0.00 dex, -0.11 dex,
0.06 dex, -0.06 dex, -0.07 dex, and 0.02 dex, respectively). Magnesium abundance from the listed neutral magnesium lines in Table A4 provided
$\approx$0.3 dex lower magnesium abundance. There are six magnesium lines that are common to both studies. The 4703, 4730, 5528, and 5711
\AA\, Mg\,{\sc i} lines have comparable equivalent widths in both studies (see Table A4). Their log$gf$-values are also agreed well except the
4703 \AA\, Mg\,{\sc i} line,i.e. the difference is $\approx$0.1 dex. The difference in abundances reported in two studies
($\Delta$([X/Fe]$_{\rm TS}$-[X/Fe]$_{\rm F}$)) is $\approx$ -0.17 dex for vanadium. Our determination of neutral vanadium abundance is from 4379
\AA\, and 6090 \AA\, V\,{\sc i} lines. The lines were not included in the analysis by Fulbright (2000).

Burris et al. (2000) via spectrum synthesis technique from KPNO spectrum of the star ($R\approx$ 20\,000) reported 
$\approx$ 0.3 dex higher Sr\,{\sc ii} and Ba\,{\sc ii} abundances. Their strontium abundance is provided by 4077 \AA\, and 4215 \AA\, Sr\,{\sc ii}
lines. In this study, the reported strontium abundance is from Sr\,{\sc ii} 4215 \AA\, line. The log$gf$-values in both studies are adequate. For Y\,{\sc ii},
Zr\,{\sc ii}, and Nd\,{\sc ii} abundances, the difference is $<$0.1 dex (i.e.$\Delta$([X/Fe]$_{\rm TS}$-[X/Fe]$_{\rm F}$)= -0.03 dex, 0.03 dex, and -0.08
dex, respectively). We also searched for La\,{\sc ii}, Eu\,{\sc ii} and Dy\,{\sc ii} lines in the ELODIE spectrum of the star. The line profile for La\,{\sc ii} line at 4333 \AA\, is seen to have contribution
in its red wing. Inspection of line profiles for Eu\,{\sc ii} lines at 4129 \AA\, and 4205 \AA\, via spectrum synthesis
indicated blending for those lines in the ELODIE spectrum. Also, the red wing of the Dy\,{\sc ii} line at 4077 \AA\, is found to be highly contributed by Sr\,{\sc
ii}.   

The reported abundances by Mishenina \& Kovtyukh (2001) for Mg\,{\sc i} and Sr\,{\sc i} are $\approx$0.3 dex higher. However, the agreement for
the Ca\,{\sc i} abundance is excellent (i.e. $\Delta$([Ca/Fe]$_{\rm TS}$-[Ca/Fe]$_{\rm MK}$) = 0.02 dex). The Si\,{\sc i},
Ba\,{\sc ii}, and Nd\,{\sc ii} abundances present $\approx$0.1 dex differences. The strontium abundance by Mishenina \& Kovtyukh (2001) is provided by neutral strontium line at 4607 \AA. The
difference (MK-TS) in oscillator strengths between two studies is -0.11 dex.

The abundances by Boeche \& Grebel (2016) from medium resolution spectrum show good agreement with the abundances in this study, with the
exceptions of two elements. The exceptions are for vanadium and cobalt. The [V/Fe] ratios between two studies showed 0.3 dex discrepancy. The
difference is $\approx$0.2 dex for cobalt. The differences in [element/Fe] ratios for silicon, scandium, titanium, chromium, and nickel is
less than 0.1 dex. Their reported calcium abundance (i.e.[Ca/Fe]) is $\approx$ 0.2 dex lower. Boeche \& Grebel (2016) did not provide atomic
data for their analysis.

\section{Notes on abundances analysis of BD\,+42\,3607}
Carney, Wright, Sneden et al. (1997) obtained abundances of Li, Mg, Ca, Ti, and Ba abundances for BD\,+42\,3607. No silicon abundance was
reported. Their reported titanium ([Ti\,{\sc ii}/Fe]=0.66 dex) and magnesium abundances ([Mg/Fe]=0.25 dex) are agreed with our abundances of
those elements in this study within error limits. The abundances for calcium and barium are somewhat discrepant with our reported abundances
for those elements. For instance, the difference in element over iron ratio for the calcium is $\approx$0.3 dex and it may partially be
ascribed to adopted log$gf$ values for the calcium lines employed in the analysis. For instance, when the atomic data for the common calcium lines at 5581 \AA,
5588 \AA, and 5590 \AA\, are scrutinized, their log$gf$ values are seed to differ $\approx$ 0.2 dex. The difference in barium abundance is $\approx$ 0.2 dex. The barium
abundances in this study was provided by Ba\,{\sc ii} lines at 4554 \AA\, and 5853 \AA. The lines were synthesized for abundance
determination and the hfs were also
considered for the lines.

The nickel and calcium abundances reported by Gratton, Carretta, Claudi et al. (2003) are agreed with the abundances reported for the star in
this study. The agreement in nickel abundance in two studies is excellent, i.e. the difference in abundance is 0.04 dex. However, calcium
abundance by Gratton, Carretta, Claudi et al. (2003) is 0.35 dex lower. The difference can be explained via differences in the adopted log$gf$ values in
two studies. There are eleven common calcium lines to both studies. The common calcium lines at 5261 \AA, 5588 \AA, 6166 \AA, and 6169 \AA\,
show $\approx$0.2 dex discrepancies in their log$gf$-values. The difference in log$gf$ for common calcium lines at 5590 \AA, 6439 \AA, and 6717
\AA\, is at 0.1 dex level.   

Zhang \& Zhao (2005) analyzed spectra of 32 metal poor stars. Some discrepancies are present for the abundances of Si\,{\sc i/ii} and Ca\,{\sc i}. 

Boesgard et al. (2011) used HIRES spectrum and their reported abundances as for [Mg/Fe] and [Ti/Fe] show good agreement
with our abundances reported for these elements. The difference in [Mg\,{\sc i}/Fe] ratios in both study is 0.04 dex. Their reported titanium
abundance ([Ti/Fe]) is in excellent agreement with [Ti\,{\sc ii}/Fe] ratio obtained in this study. 

Peterson (2013) used ultraviolet part of the HIRES spectrum. The reported abundances by Peterson (2013)
for common elements in both studies as for Ca, Mn, Sr, Y, Zr, Nd show fair agreement for manganese and zirconium. The exceptions are for
calcium, strontium, and yttrium abundances. The abundances for those elements in the ultraviolet present somewhat discrepant
abundances. For instance, the abundance difference for yttrium is 0.57 dex while the difference in element over iron ratios between Peterson
(2013) and this study for strontium is 0.45 dex.

Boeche \& Grebel (2016) determined model parameters of the star from a moderate resolution spectrum. They did not report silicon
abundances. Their reported calcium abundance of [Ca/Fe]=0.20 differs 0.4 dex from our measurement in this study. The element over iron ratios
for Mg, Sc, Ti, and Cr show excellent agreement (i.e. the differences are within 0.1 dex). They reported $\approx$ 0.2 dex lower nickel
abundance for the star. 

\section{Notes on abundances analysis of HD\,201891}
In a search for possible non-LTE affects on magnesium abundances from magnesium lines at 4703 \AA, 5528 \AA\,, and 5711 \AA\, in the FOCES
spectrum of metal poor dwarf stars, Zhao \& Gehren (2000) reported LTE-non-LTE differences in abundances of these magnesium lines to be
$\approx$ -0.1 dex. In the current study, the reported magnesium abundance for HD\,201891 in Table A7 are provided by the Mg\,{\sc i} lines at
4571 \AA\, and 5711 \AA. The magnesium over iron ratios for those lines from the ELODIE spectrum are 0.37 dex and 0.24 dex, respectively. The
abundance for the common magnesium line at 5711 \AA\, show excellent agreement with that reported by Zhao \& Gehren
(2000) for this line (i.e. 0.24 dex vs. 0.29 dex). 

The abundances (e.g. [Element/Fe]) by Fulbright (2000)\footnote{The tables for the summary of the model parameters and mean abundances by
Fulbright (2000) listed two different [Fe/H]  values for HD\,201891 as -1.0 dex and -1.12 dex, respectively.} for Na\,{\sc i}, Mg\,{\sc i},
Si\,{\sc i}, Ca\,{\sc i}, Ti\,{\sc i}, Cr\,{\sc i}, V\,{\sc i} and Ni\,{\sc i} are in excellent agreement with the reported abundances of
those elements in this study. The difference is $\le$0.1 dex. For single ionized, yttrium, zirconium, and barium abundances, the difference in
abundances between Fulbright (2000) and this study is ranging from $\approx$0.1 dex for [Zr\,{\sc ii}/Fe] to $\approx$0.2 dex for [Ba\,{\sc
ii}/Fe]. 

Mishenina \& Kovtyukh (2001) also used the ELODIE spectrum for their spectroscopic analysis but reported somewhat discrepant abundances for
Mg\,{\sc i} and Y\,{\sc ii}. The differences for those elements between two studies are at $\approx$0.3 dex level. Mishenina \& Kovtyukh (2001)
did not provide their list of magnesium lines. We have two common yttrium lines in the ELODIE spectrum. They are 4883 \AA\, and 5087 \AA\,
Y\,{\sc ii} lines. Their adopted log$gf$-values by Mishenina \& Kovtyukh (2001) are slightly different. For
instance, the log$gf$ for the former line agreed within 0.04 dex, however, the difference in log$gf$ for 5087 \AA\, line is 0.12. The [Ba\,{\sc ii}/Fe] ratio
by Mishenina \& Kovtyukh (2001) also differed 0.2 dex from the barium abundance in this study. The element over iron ratios for Si\,{\sc i},
Ca\,{\sc i}, and Sr\,{\sc i} are agreed within 0.1 dex. 

The reported abundances by Gratton et al. (2003) for Na\,{\sc i}, Si\,{\sc i}, Ca\,{\sc i}, Ti\,{\sc i}, Ti\,{\sc ii},
V\,{\sc i}, Sc\,{\sc ii}, and Ni\,{\sc i} show excellent agreement with the abundance reported in this study for those elements (i.e.
$\Delta$[X/Fe]$_{\rm GCC}$ - [X/Fe]$_{\rm TS}$ $\le$0.1 dex). The exception is for Cr\,{\sc i} abundance: the difference is 0.08 dex. The difference
in element over iron ratios for magnesium, single ionized chromium, and manganese abundances is $\approx$0.2 dex.  

Mishenina et al. (2003) used model atmosphere parameters from Mishenina \& Kovtyukh (2001) and reported non-LTE Na abundance for star as [Na/Fe]=0.07
dex from
relatively weak Na\,{\sc i} lines at 5682 \AA\,(of 29.0 m\AA), 6154 \AA\,(of 5.0 m\AA), and 6160 \AA\,(of 13.0 m\AA). It is in excellent
agreement with our measurement of sodium abundance, i.e. the difference is 0.04 dex.

The abundances reported by Reddy, Lambert \& Allende Prieto (2006) via high resolution spectroscopy for Na\,{\sc i}, Mg\,{\sc i}, Si\,{\sc
i}, Ca\,{\sc i}, Sc\,{\sc ii}, Ti\,{\sc i}, V\,{\sc i}, Cr\,{\sc i}, Ni\,{\sc i}, Y\,{\sc ii}, and Ba\,{\sc ii} is agreed well within 0.1 dex. The [C/Fe]
differed $\approx$0.2 dex. Also, the manganese abundance showed only $\approx$0.1 dex difference. Conversely, the difference for the cobalt
abundances was $\approx$0.3 dex. No common lines of cobalt were available. 

Gehren et al. (2006) used high resolution ($R\approx$ 40\,000) FOCES spectrum of the star and reported LTE and non-LTE abundances
of Mg, Na, and Al. They're as follows: ([Mg/Fe]$_{\rm LTE}$,[Mg/Fe]$_{\rm NLTE}$) = (0.32, 0.43), ([Na/Fe]$_{\rm LTE}$,[Na/Fe]$_{\rm NLTE}$) = (0.21, 0.07), and
([Al/Fe]$_{\rm LTE}$,[Al/Fe]$_{\rm NLTE}$)= (0.24, 0.59). The reported LTE abundance of magnesium by Gehren et al. (2006) showed excellent
agreement (i.e. 0.01 dex different) with the reported magnesium abundance in this study. On the other hand, they reported 0.1 dex higher sodium
abundance, however, they did not provide the atomic data for the lines of those elements that are included in their analysis. 

Mishenina et al. (2011) reported abundances for Na, Al, Cu, and Zn. They reported [Na/Fe]
= -0.02 dex. We did not dedect lines of aluminum, copper, and zinc in the ELODIE spectrum. The sodium abundance reported for the star in Table
A7 is agreed with the sodium abundance reported by Mishenina et al. (2011) within 0.1 dex. 

The $\alpha$-element abundances by Bensby, Feltzing and Oey (2014) shows excellent agreement with the $\alpha$-element abundances reported for
the star in this study (i.e. $\Delta$[X/Fe]$_{\rm BFO}$ - [X/Fe]$_{\rm TS}$ $\le$0.1 dex). The [Mn\,{\sc i}/Fe] ratio in this study differs 0.17 dex
from that of Bensby, Feltzing and Oey (2014). The differences in element over iron ratios between two studies for Sc\,{\sc ii}, Ti\,{\sc i},
Cr\,{\sc i}, Ni\,{\sc i}, Y\,{\sc ii} and Ba\,{\sc ii} is ranging between 0.02 and 0.07 dex. 

Battisini \& Bensby (2015) adopted model parameters from Bensby, Feltzing and Oey (2014) and reported abundances for the star for Sc\,{\sc
ii}, and Mn\,{\sc i} as [Sc\,{\sc ii}/Fe]= 0.17 dex and [Mn\,{\sc i}/Fe]= -0.26 dex, respectively. Their reported manganese abundance is from
6016 \AA\, Mn\,{\sc i} line. The logarithmic abundance reported for the line by Battisini \& Bensby (2015) is 4.12 dex. This is in excellent
agreement with the logarithmic abundance reported for this line in this study (see Table A7). The listed manganese abundance for the star in this study
is from eight Mn\,{\sc i} lines in Table A7\footnote{Table A7 is available online.}. The scandium abundance is
also agreed well, i.e. the difference in scandium abundances is 0.05 dex.     
For vanadium, the reported abundance by Boeche \& Grebel (2016) is agreed with our determination. The difference is negligible, i.e. 0.03 dex.
It is $\approx$ 0.3 dex for cobalt. They employed astrophysically calibrated log$gf$-values for their spectroscopic analysis. 

The reported zirconium abundance by  Battisini \& Bensby (2015) via model parameters from Bensby, Feltzing and Oey (2014) for the Zr\,{\sc ii} line at 4208 \AA\, is
in excellent agreement with the calculated abundance for this line in this study. 

On the basis of its computed abundances in this study, the spectrum of this candidate benchmark star for the {\it Gaia} (Hawkins et al. 2016) indicates a halo
like chemisty.

\section{Notes on abundances analysis of HD\,194598}
Zao \& Gehren (2000) using FOCES spectrum, reported LTE and non-LTE magnesium abundances of the star over four neutral lines of magnesium: 4571 \AA\, (LTE/non-LTE; 0.25/0.33), 4703 \AA\, (0.20/0.32), 5528 \AA\, (0.24/0.32) ve 5711
\AA\, (0.19/0.31) with the model parameters reported in Table A9\footnote{Table A9 is available
online.}.  

The abundances (i.e. [element/Fe]) obtained by Fulbright (2000)\footnote{The tables for the summary of the model parameters and mean
abundances by Fulbright (2000) present two different [Fe/H] values for HD\,194598 as -1.1 dex and -1.23 dex, respectively.} with the model
atmosphere parameters reported in
this study (Table A9) are in very good agreement (within 0.1 dex) for Na\,{\sc i}, Si\,{\sc i}, Ca\,{\sc i}, Ti\,{\sc i}, V\,{\sc i}, Cr\,{\sc i},
Ni\,{\sc i}, and Y\,{\sc ii}. The vanadium lines employed by Fulbright (2000) at 4389 \AA, 6090 \AA, and 6216 \AA\, have measured equivalent widths of 9.0 m\AA, 0.3 m\AA,
and 0.2 m\AA, respectively. The lines are too weak to be measured in the ELODIE spectrum. Their reported single ionized zirconium and barium abundances are only $\approx$0.2 dex higher. The difference
in [Mg\,{\sc i}/Fe] is 0.25 dex.    

The abundances obtained from the ELODIE spectrum by Mishenina \& Kovtyukh (2001) is in very good agreement with our measurements, i.e. the
differences in [element/Fe] ratios are less than 0.1 dex for Mg\,{\sc i}, Si\,{\sc i}, Ca\,{\sc i}, Y\,{\sc ii}, and Ba\,{\sc ii}. The single
ionized cerium abundance (i.e. [Ce\,{\sc ii}/Fe]) shows 0.18 dex difference. Mishenina \& Kovtyukh (2001) reported Ce\,{\sc ii} lines at 4486,
4562, 4572, 4628, and 5274 \AA\, however, it was unclear whether these lines were used for cerium abundance determination by Mishenina \&
Kovtyukh (2001) for HD\,194598. The 4628 \AA\, Ce\,{\sc ii} line, apparently a single common cerium line in both studies, and has a reported
log$gf$-value which is 0.04 dex less than the adopted log$gf$-value for the line in the present study.

The reported abundances of, Na\,{\sc i}, Si\,{\sc i}, Ca\,{\sc i}, Ti\,{\sc i \ ii}, Cr\,{\sc i \ ii}, Mn\,{\sc i}, and Ni\,{\sc i} over
model parameters reported in Table A9 by Gratton et al. (2003) are in agreement with the computed abundances of those
elements in this study within 0.1 dex. The difference for single ionized scandium abundance (in [Sc/Fe]) is 0.2 dex. The highest difference
in abundance was observed for magnesium: it is 0.3 dex. For the three common Mg\,{\sc i} lines to both studies at 4703 \AA, 5528 \AA,
and 5711 \AA, their measured equivalent widths in the ELODIE spectrum as well as adopted log$gf$-values in this study are in accordance with
those reported by Gratton et al. (2003). Vanadium abundance by Gratton et al.
(2003) is in excellent agreement with the vanadium abundance in this study, [V\,{\sc i}/Fe]$_{\rm TS}$ -
[V\,{\sc i}/Fe]$_{\rm GCC}$ $\approx$ 0.03 dex. However, their vanadium abundance is provided by V\,{\sc i} lines at 5727 \AA, 6090 \AA, and 6216 \AA.
and the lines are too weak to provide reliable abundances. Their equivalent widths by Gratton et al.
(2003) are 5.8 m\AA, 2.3 m\AA, and 3.2 m\AA, respectively.

Mishenina et al. (2003) computed non-LTE abundances of weak sodium lines at 6154 \AA\, with equivalent width of 4.2
m\AA\, and at 6160 \AA\, of 5.9 m\AA. They reported [Na/Fe]=0.00 dex which is in excellent agreement with the sodium abundance reported for the
star in Table A9 of this study.

Gehren et al. (2006) used FOCES ($R~$40000) spectrum of the star and reported both LTE and non-LTE abundances of magnesium, sodium, and
aluminum abundances of the star. The reported abundances by Gehren et al. (2006) for those elements over the model parameters reported in
Table A7 are as follows: ([Na/Fe]$_{\rm LTE}$,[Na/Fe]$_{\rm NLTE}$)= (0.00, -0.14) and ([Mg/Fe]$_{\rm LTE}$, [Mg/Fe]$_{\rm NLTE}$)= (0.16, 0.28). The
agreement in abundances for those elements is good within 0.1 dex. 

Nissen \& Schuster (2010) used UVES and FIES spectra of the star for their abundance analysis. Their analysis was based on abundances computed with
astrophysical log$gf$-values. The calculated magnesium abundance in the LTE over MARCS model atmospheres and FIES spectrum slightly differs, i.e.
+0.05 dex. The overall agreement in abundances for Na\,{\sc i}, Mg\,{\sc i}, Si\,{\sc i}, Ca\,{\sc i}, Ti\,{\sc i}, Cr\,{\sc i},  and Ni\,{\sc i} is
noteworthy, i.e. the differences in abundance between two studies is less than 0.1 dex.  

Abundances reported for Mn ([Mn/Fe]=-0.1 dex), Sr ([Sr/Fe]=-0.3 dex), and Zr ([Zr/Fe]=0.2 dex) by Peterson (2013) over HIRES spectrum agreed with
the obtained abundances of those elements in this study. The differences in abundances for cobalt and yttrium is also agreed within $\approx$
0.1 dex. The reported cobalt abundance by Peterson (2013) is obtained from the cobalt lines at 3405 \AA. The line appears to be a blend
of Ti\,{\sc i}. Our cobalt abundance is from 4121 \AA\, Co\,{\sc i} line (Table A5).   

Bensby, Feltzing, and Oey (2014) reported abundances for Na\,{\sc i}, Mg\,{\sc i}, Si\,{\sc i}, Ca\,{\sc i}, Ti\,{\sc i}, Cr\,{\sc i}, Ni\,{\sc i},
Y\,{\sc ii}, and Ba\,{\sc ii}. The over all agreement in abundances for those elements is satisfactory, i.e. the differences are  $<$0.1
dex.

Battisini \& Bensby (2015) adopted model parameters reported by Bensby, Feltzing and Oey (2014) and used the same spectrum for revised abundances of
the star, however, they did not report abundances for Sc\,{\sc ii}, V\,{\sc i}, Mn\,{\sc i} and Co\,{\sc i}. In the ELODIE spectrum, even for a
relatively short wavelength coverage, we were able to report abundances for those elements (Table A7).

The reported element over iron ratios for  Mg\,{\sc i}, Si\,{\sc i}, Ti\,{\sc i}, and Cr\,{\sc i} are in excellent agreement with the
abundances of these elements in this work. The differences are seen to be within 0.1 dex. The [Ca/Fe], [Co/Fe], and [Ni/Fe] by Boeche \& Grebel (2016)
differed $\approx$0.2 dex from those reported in this study. The scandium and vanadium abundances showed $\approx$0.3 dex
differences. 

Yan et al. (2016) used UVES spectrum of the star. With adopted model parameters from Nissen \& Schuster (2010), they reported
both LTE and non-LTE abundances for Cu\,{\sc i} lines at 5105 \AA\, as [Cu\,{\sc i}/Fe] = -0.46 (-0.35) and for 5218 \AA\, as [Cu\,{\sc i}/Fe]
= -0.44 (-0.38). Their atomic data for copper was from the NIST database. 

Fishlock et al. (2017) used MIKE spectrum ($\lambda\lambda$ 3350 - 9000 \AA) of the star to report abundances for Sc, Zr, and Ce.
Their reported abundances are as follows: [Sc/Fe]=-0.07, [Zr/Fe]=-0.11, and [Ce/Fe] = -0.11. The scandium ([Sc\,{\sc ii}/Fe]) and cerium ([Ce\,{\sc ii}/Fe])
abundances show $\approx$ 0.3 dex difference between two studies. Their reported cerium abundances are from cerium lines at  4562 \AA\, and
4628 \AA. The latter line was also available in the ELODIE spectrum and its reported abundance in
Table A6\footnote{Table A6 is available online.} was determined via spectrum
synthesis. The line provided a [Ce\,{\sc ii}/Fe] of 0.17 dex. The zirconium abundance is obtained by 4208 \AA\, Zr\,{\sc ii} line. The line was
observed in the ELODIE spectrum (see Table A6). Its abundance is determined via line synthesis. Fishlock et
al.(2017) determined scandium abundance from 5526 \AA\, Sc\,{\sc ii} line via spectrum synthesis.    

\section{Notes on abundances analysis of HD\,3567}
The abundances (i.e. [element/Fe]) by Fulbright (2000) for neutral magnesium, silicon, and titanium agreed within 0.2 dex. The same
difference was also observed for single ionized yttrium, zirconium, and barium abundances. The calcium abundance of the star in this study is
in excellent agreement with the calcium abundance by Fulbright (2000) (i.e. $\Delta$([Ca/Fe]$_{TS}$-[Ca/Fe]$_{F}$) = +0.03 dex).
Our determination of vanadium abundance is obtained using 4379 \AA\, V\,{\sc i} line. There is no common lines of vanadium in the
spectroscopic analysis by Fulbright (2000), however, the agreement is good, i.e. $\Delta$([V/Fe]$_{TS}$-[V/Fe]$_{F}$) = -0.13 dex. Our chromium abundance from
19 neutral chromium lines shows a good agreement with their result (i.e. $\Delta$([Cr\,{\sc i}/Fe]$_{TS}$-[Cr\,{\sc i}/Fe]$_{F}$) =
-0.07 dex). Nickel abundance also shows very good agreement (i.e. $\Delta$([Ni/Fe]$_{TS}$-[Ni/Fe]$_{F}$) = +0.02 dex). Overall, the comparison is satisfactory. 

Gratton et al. (2003), over UVES and SARG spectra of the star, reported  abundances for neutral sodium, magnesium,
calcium, vanadium, and manganese that are in agreement with elemental abundances in this study within 0.1 dex. A similar difference was also seen for
single ionized chromium (i.e. $\Delta$([Cr\,{\sc ii}/Fe]$_{TS}$-[Cr\,{\sc ii}/Fe]$_{GCC}$)= -0.14 dex). The neutral silicon and neutral
titanium abundances are in excellent agreement with those reported by Gratton et al. (2003). The abundance for single
ionized scandium differed only 0.15 dex and the difference in abundance for single ionized titanium was 0.25 dex. The Ti\,{\sc ii}
line at 4583 \AA\, is common in both studies (Table A5). Its log$gf$ in this study is from Wood et al. (2013) and is agreed with the log$gf$ from Gratton et al. (2003). 

Zhang \& Zhao (2005) via medium resolution ($R \approx$37\,000) spectra over $\lambda\lambda$5500-8700 \AA\, wavelength region. The neutral sodium,
magnesium, titanium, and nickel abundances are agreed within 0.1 dex with those reported in the current study. The silicon abundance from
neutral silicon lines shows excellent agreement (i.e.$\Delta$([Si/Fe]$_{TS}$-[Si/Fe]$_{ZZ}$) = -0.02 dex). The chromium abundance is also
shows a very good agreement (i.e. $\Delta$(ours-Zhang) = -0.01 dex). As it comes to abundances for neutral calcium and single ionized barium,
the difference in abundance is $\approx$0.2 dex at its maximum. The highest discrepancies were observed for single ionized scandium, and
neutral vanadium and manganese abundances. The log$gf$ values for the scandium are taken from the NIST database (Table A4). Our log$gf$
values for manganese lines are from Blackwell-Whitehead \& Bergemann (2007). The hfs is only included for 6016 \AA\, Mn\,{\sc i} line. The
vanadium abundance in this study is provided from the V\,{\sc i} 4379 \AA\, line and the listed log$gf$-value for the line in Table A5 is the
astrophysical log$gf$. Figure A2 shows the best-fitting synthetic spectrum for the 4379 \AA\, V\,{\sc i} line. 

Reddy, Lambert and Allende Prieto (2006) used a high resolution spectra ($R\approx$55000, $\lambda\lambda$3500-5600 \AA) to obtain abundances
of several species. The differences (i.e.$\Delta$([X/Fe]$_{TS}$-[X/Fe]$_{RLP}$)) for chromium, magnesium, silicon, and manganese abundances
are ranging from -0.07 dex for Mn to +0.13 dex for Cr. The differences for sodium, calcium, and scandium abundances are at 0.3 dex level.
For cobalt and yttrium abundances, it was -0.5 dex and -0.19 dex, respectively. The only common cobalt line with Reddy, Lambert and Allende
Prieto (2006) is the 4792 \AA\, Co\,{\sc i} line. We adopted
astrophysical log$gf$-value for the line from Reddy, Lambert and Allende
Prieto (2003). It was obtained via inverting solar and stellar spectra. 

Nissen \& Schuster (2010), in a study for spectroscopic analysis of 94 dwarf stars used UVES spectrum of the star to obtain abundances for
Na\,{\sc i}, Mg\,{\sc i}, Si\,{\sc i}, Ca\,{\sc i}, Ti\,{\sc i}, Cr\,{\sc i}, and Ni\,{\sc i}. Their reported abundances for those elements
are in excellent agreement with our measurements (less than 0.1 dex). 

The study by Hansen et al. (2012) over a mixture of giant and dwarf stars including HD\,3567 reported abundances for common elements. These
included Sr, Y, Zr, and Ba abundances for the star. The element over iron ratios by Hansen et al. (2012) for Sr and Y is in excellent
agreement with our results (the difference is 0.06 dex). The abundance ratio for Zr showed $\approx$0.2 dex difference while the difference
for Ba was $\approx$0.3 dex. The zirconium lines at 4208 \AA\, and 4317 \AA\, are common to both studies. The adopted log$gf$-values for the
lines by Hansen et al. (2012) is slightly lower (0.05 dex for 4208 \AA; 0.07 dex for 4317 \AA) compared to the log$gf$-values reported in
Table A6 in this study. 

The reported abundances by Bensby, Feltzing \& Oey (2014) for Na, Mg, Si, Ca, Ti, Cr, Ni, and Y show excellent agreement ($\le$0.1 dex) with the abundances of
those elements listed in Table A8.

The largest difference in abundances (i.e. differences in [Element/Fe] ratios) reported for the star by Boeche \& Grebel (2016) was for
nickel. It is 0.25 dex. For the silicon, magnesium, scandium, and chromium abundances, the difference in abundance between two study
was less than 0.1 dex and $\approx$0.2 dex for calcium and titanium abundances. The largest difference in abundance is observed for cobalt,
i.e.$\Delta$([Co/Fe]$_{TS}$-[Co/Fe]$_{BG}$ = -0.36 dex. Boeche \& Grebel (2016) did not provide atomic data for their
spectroscopic analysis.

Yan et al. (2016) used UVES spectrum and model atmosphere parameters from Nissen \& Schuster (2010) to obtain both
LTE and non-LTE abundances for Cu for their halo sample stars. The atomic data was taken from NIST database. Using the
Cu line at 5105 \AA\, they reported an LTE abundances of [Cu/Fe]=-0.73 dex. (\Teff=6051, \logg=4.02, [Fe/H]=-1.16 dex
$\xi$ =1.5 km s$^{\rm -1}$). The line is too weak to be measured in the {\sc ELODIE} spectrum.

Fishlock et. al.(2017) used MIKE spectrum of the star ($\lambda\lambda$3350-9000 \AA). For the common elements in both studies,
the agreement in abundances is satisfactory. That included abundances of single ionized scandium, zirconium, and neodymium. The differences in
abundances (i.e.$\Delta$([X/Fe]$_{TS}$-[X/Fe]$_{FYK}$)) for scandium and zirconium are -0.16 dex and +0.03 dex, respectively. The observed
difference in neodymium abundance in both studies is 0.15 dex. The Nd\,{\sc ii} abundance by Fishlock et. al.(2017) was based on 4706
Nd\,{\sc ii} line RMT-3. However, the line has a measured equivalent width of 4 m\AA\, in their MIKE spectrum.

Bergemann et al. (2017) adopted model parameters from Hansen et al. (2012) to report magnesium abundance for the star. The [Mg/Fe]
ratio by Bergemann et al. (2017) is 0.1 dex higher (i.e.$\Delta$([X/Fe]$_{\rm TS}$-[X/Fe]$_{\rm B}$)= -0.1 dex). Our determination of magnesium
abundance is from six neutral magnesium lines (see Table A4). These included the resonance line of Mg\,{\sc i} at 4571 \AA, and optical triplet lines
at 5172 \AA\, and 5183 \AA. All these lines are known to be sensitive to the model atmosphere structure. However, we also included Mg\,{\sc
i} lines at 5528 \AA\, and 5711 \AA. The
calculated logarithmic abundances for latter two are in excellent agreement with the reported mean magnesium abundance over six magnesium
lines in Table A8.   

\section{A {\it Gaia} benchmark star: HD\,84937}
Table A8 gives the results on equivalent width analysis of the spectrum for the following model
parameters: $T_{\rm eff}$ = 6000$\pm$140 \kelvin, $\log\,g =$ 3.50$\pm$0.18 cgs, [Fe/H] = -2.40$\pm$0.15 dex, and
$\xi$ = 1.60 km s$^{\rm -1}$.

Fulbright (2000), over a high resolution spectrum of the star, reported abundances for several species including Na, Mg, Si, Al, Ca, Ti, V,
Cr, Ni, Y, Zr, Ba ve Eu. Their listed [element/Fe] ratios for HD\,84937 slightly differ for some of the species (i.e. they reported
[Y/Fe]=0.03$\pm$0.06 and [Ba/Fe]=0.0$\pm$0.11 dex), mostly due to their preference for atomic data. Magnesium, cromium (Cr\,{\sc i}), and
titanium (Ti\,{\sc ii}) abundances in both studies are in excellent agreement. Our results for the calcium and nickel is consistent with
their results (i.e. within $\approx$0.2 dex).

Based on {\sc ELODIE} spectra of the metal-poor stars, Mishenina \& Kovtyukh (2001) reports [Y\,{\sc ii}/Fe] = -0.02$\pm$0.14 dex. The reported
yttrium abundance in this study is [Y/Fe] = 0.32$\pm$0.18 dex. No further investigation on the difference for yttrium abundance was carried out,
since Mishenina \& Kovtyukh (2001) did not report the yttrium lines employed in their analysis. Their magnesium ([Mg\,{\sc
i}/Fe]=0.36$\pm$0.11 dex) and calcium ([Ca\,{\sc i}/Fe] = 0.34$\pm$0.09 dex) abundances are in good agreement with our
results. Their barium abundance, i.e. [Ba\,{\sc ii}/Fe] = -0.10$\pm$0.29, is also in accordance with the quoted barium abundance obtained in this study
within error limits. Figure A2 show the best-fitting synthetic spectrum for the 4554 \AA\, Ba\,{\sc ii} line for the listed abundance in Table
A6 for the line.

Over high resolution McD, UVES and SARG spectra of the selected metal-poor stars, the abundances reported by Gratton et al.
(2003) for HD\,84937 showed only $\approx$0.1 dex differences for manganese, scandium, and chromium. For magnesium, titanium, and nickel, it was $\approx$0.2
dex. Relatively higher errors reported in abundances by Gratton et al.
(2003) are important to note.

The (LTE) silicon abundance for the star by Shi et al. (2009) was [Si\,{\sc i}/Fe] = 0.18$\pm$ 0.07 dex. They employed model
atmosphere parameters computed by Mashonkina et al. (2008). Non-LTE contribution on the neutral silicon abundance was estimated to be
+0.13 dex. The silicon abundance was provided from Si\,{\sc i} lines at 3905 \AA\, and 4102 \AA. The former was not observed in the {\sc ELODIE}
spectra where the 4102 \AA\, Si\,{\sc i} line is seen to be blended hence the line is not used in the silicon abundance
determination in current study.

Ishigaki et al. (2012, 2013) via high resolution ($R\approx$50000) spectroscopy in the $\lambda\lambda$4000-6800 \AA\, 
wavelength range performed spectroscopic analysis for Na, Si, Ca, Mg, Ti, Sc, V, Cr, Mn, Co, Ni, Zn, Sr,
Zr, Y, La, Cu, Ba, Eu, Nd, Sm for 93 metal-poor stars. Their spectrum for HD\,84937 yielded $\approx$0.1 dex higher abundances when compared to the
abundances of neutral
magnesium, titanium, chromium, and manganese in this study. For single ionized titanium, chromium, strontium, and barium, the difference was at
0.1 dex level. Exceptions are yttrium and
zirconium: the differences in abundances for those elements were at 0.3 dex level.
 
The neutral magnesium abundance by Bensby, Feltzing \& Oey (2014) is in excellent agreement with the reported magnesium abundance in this study ($\Delta$([X/Fe]$_{TS}$-[X/Fe]$_{BFO}$) = +0.06 dex). Ti\,{\sc i}, Ni\,{\sc i},
and Y\,{\sc ii} abundances were agreed within 0.2 dex. Exception is for neutral calcium and single ionized barium abundances: the differences were
+0.27 dex and -0.55 dex, respectively. The log$gf$ values for Ba\,{\sc ii} by Bensby, Feltzing \& Oey (2014) showed 0.2 dex difference
from the log$gf$-values reported in this study for
4554 \AA\, and 5853 \AA\, Ba\,{\sc ii} lines. The difference in log$gf$ is $\approx$0.3 dex for the 6496 \AA\, Ba\,{\sc ii} line. Hence, it is apparent that the difference in
barium abundance between two studies may partially be due to atomic data differences in both studies. For instance, for calcium, over five common Ca\,{\sc i}
lines in both studies, the mean difference in loggf values is -0.05$\pm$0.24 dex. The 5261 \AA, and 6169 \AA\, Ca\,{\sc i} lines
showed $\approx$0.3
dex differences in their reported log$gf$-values by Bensby, Feltzing \& Oey (2014). The log$gf$ values for Ca\,{\sc i} in this study is from the NIST database.

The magnesium abundance by Boeche \& Grebel (2016) showed a fair agreement, of about $\approx$0.2 dex level. Calcium and scandium abundances
showed 0.3 dex and $\approx$0.5 dex differences, respectively. Unfortunately, they did not report atomic data for these elements. The chromium
abundance by Boeche \& Grebel (2016) is in
excellent agreement with our determination of the single ionized chromium abundance.

Zhao et al. (2016) reported both LTE and non-LTE abundances of elements from Li to Eu for a sample of thin disk,
thick disk, and halo stars with metallicities (i.e. [Fe/H]) ranging between -2.6 and 0.2 from the high resolution
($R=$60\,000; $\lambda\lambda$3700-9300 \AA) Hamilton spectrum. The reported Mg\,{\sc i} and Sc\,{\sc ii} abundances for the star were
agreed within 0.1 dex. [Ca\,{\sc i}/Fe] ratio differed by $\approx$0.3 dex (i.e. $\Delta$([Ca\,{\sc i}/Fe]$_{TS}$-[Ca\,{\sc i}/Fe]$_{Z}$) = +0.28 dex). Single ionized titanium and
barium abundances showed excellent agreement (i.e.$\Delta$([X/Fe]$_{TS}$-[X/Fe]$_{Z}$)= 0.0 dex and -0.03 dex, respectively). The difference in abundance for
single ionized zirconium was 0.2 dex. The largest difference in abundance was observed for the strontium, i.e.
$\Delta$([Sr/Fe]$_{TS}$-[Sr/Fe]$_{Z}$)= +0.6 dex.   
Their adopted log$gf$ values for the common Sr\,{\sc ii} lines at 4077 \AA\, and 4215 \AA\, are in excellent agreement with those adopted from the
NIST database\footnote{http://physics.nist.gov/PhysRefData/ASD} in this study.

More recent work by Spite et al. (2017) reported only Mg and Ca abundances as for $\alpha$-element (e.g. Mg, Si, and Ca) for this halo
main-sequence (turnoff) star. In addition to the modest overabundance of $\alpha$-elements, a solar value of [C/Fe] was reported. Our result
for magnesium abundance is in excellent agreement with Spite et al. (2017). The difference in calcium abundance between two studies is only
0.05 dex.

\begin{figure*}%[!ht]
\centering \includegraphics*[width=18cm,height=10.50cm,angle=0]{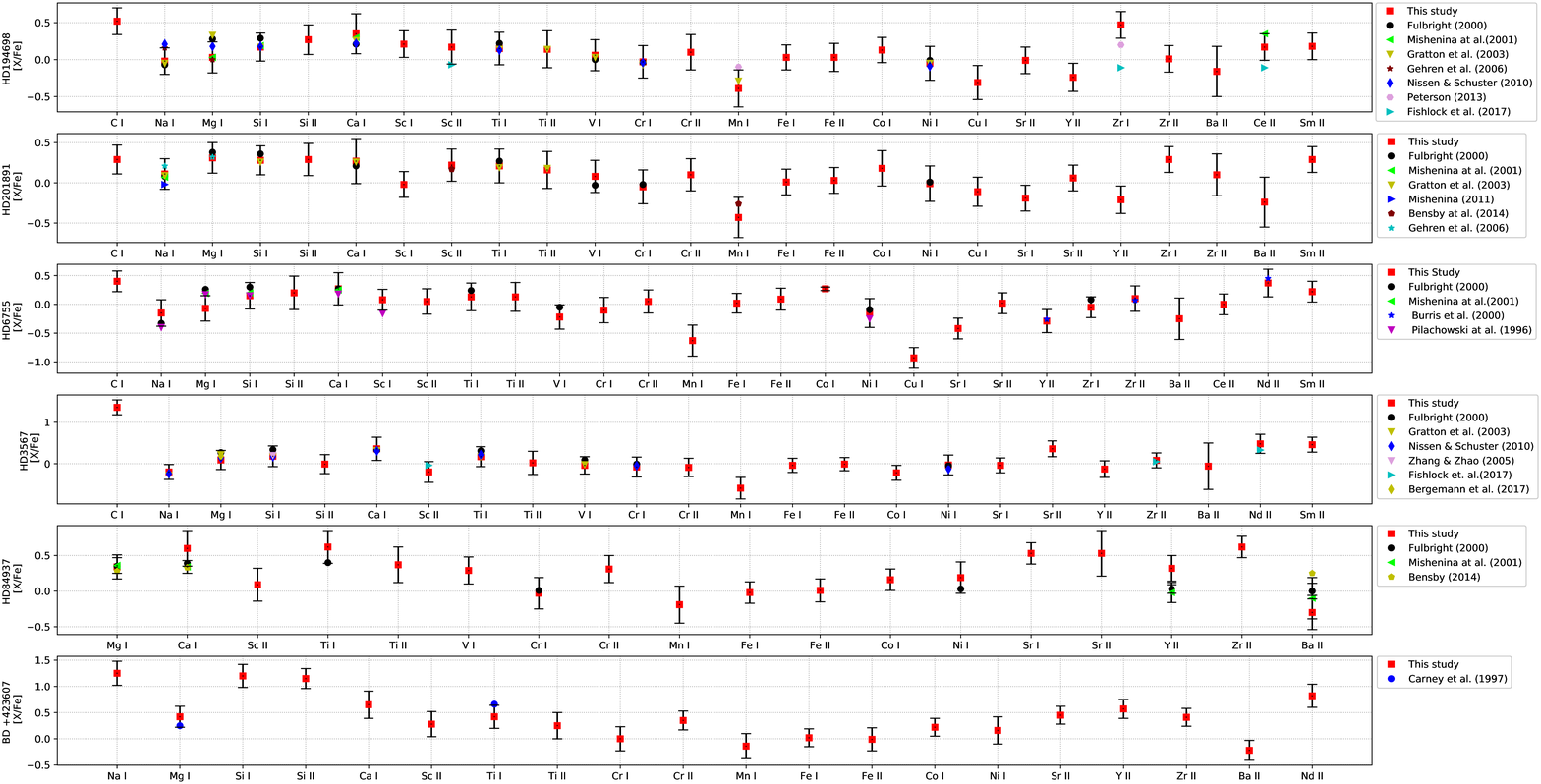}
\caption{Comparison of abundances relative to Fe for the HPM stars. The formal errors on abundances are also included.}
\label{ab_gc_vs_star}
\end{figure*}

\end{appendix}


\begin{thebibliography}{}

\bibitem[\protect\citeauthoryear{Adibekyan et al.}{2012}] {adibekyan2012}
Adibekyan V. Zh., Santos N. C., Sousa S. G. et al., 2012, A\&A, 543, A89

\bibitem[\protect\citeauthoryear{Aguilera et al.}{2018}] {guilera2012}
Aguilera-G{\'o}mez C., Ram{\'\i}rez I., Chanam{\'e} J., 2018, A\&A, 614, A55

\bibitem[\protect\citeauthoryear{Allende Prieto et al.}{2008}] {allende2008}
Allende Prieto C., Majewski S.~R., Schiavon R., C. et al., 2008,
AN, 329, 1018

\bibitem[\protect\citeauthoryear{Allen \&  Porto de Mello}{2011}] {allen2011}
Allen D.~M., Porto de Mello G.~F., 2011, A\&A, 525, A63

\bibitem[\protect\citeauthoryear{Asplund et al.}{2009}] {asplund2009} 
Asplund M., Grevesse N., Sauval A.~J., Scott P., 2009, ARA\&A, 47, 481

\bibitem[\protect\citeauthoryear{Balbinot \& Gieles}{2018}] {balbinot2018} 
Balbinot E., Gieles M., 2018, MNRAS, 474, 2479

\bibitem[\protect\citeauthoryear{Bastian \& Lardo}{2018}]{Bastian2018} 
Bastian N., Lardo C., 2018, ARA\&A, 56, 83

\bibitem[\protect\citeauthoryear{Battisini \& Bensby}{2015}] {battisini2015} 
Battisini C., Bensby T., 2015, A\&A, 577, A9 

\bibitem[\protect\citeauthoryear{Belokurov et al.}{2007}] {belokurov2007}
Belokurov  V. et al., 2007, ApJ, 657, 89

\bibitem[\protect\citeauthoryear{Bennett et al.}{2013}] {bennett2013}
Bennett C. L. et al., 2013, ApJS, 208, 20

\bibitem[\protect\citeauthoryear{Bensby et al.}{2003}] {bensby2003} 
Bensby T., Feltzing S., Lundstr{\"o}m I., 2003, A\&A, 410, 527

\bibitem[\protect\citeauthoryear{Bensby et al.}{2003}] {bensby2003} 
Bensby T., Feltzing S., Lundstr{\"o}m I., Ilyin I., 2005, A\&A, 433, 185

\bibitem[\protect\citeauthoryear{Bensby et al.}{2014}] {bensby2014}
Bensby T., Feltzing S., Oey M.~S., 2014, A\&A, 562, A71

\bibitem[\protect\citeauthoryear{Bergemann \& Gehren}{2008}] {bergemann2008} 
Bergemann M., Gehren T., 2008, A\&A 492, 823--831

\bibitem[\protect\citeauthoryear{Bergemann et al.}{2010}] {bergemann2010}
Bergemann M., Pickering J.~C., Gehren T., 2010, MNRAS, 401, 1334

\bibitem[\protect\citeauthoryear{Bergemann et al.}{2012}] {bergemann2012}
Bergemann M., Lind K., Collet R., Magic Z., Asplund M., 2012, MNRAS, 427, 27

\bibitem[\protect\citeauthoryear{Bergemann et al.}{2017}] {bergemann2017}
Bergemann M., Collet R., Sch\"{o}nrich R., Andrae R., Kovalev M., Ruchti G., Hansen C.~J., Magic Z., 2017, ApJ, 847, 16

\bibitem[\protect\citeauthoryear{Bergemann et al.}{2019}] {bergemann2019}
Bergemann M. et al., 2019, A\&A, 631, A80

\bibitem[\protect\citeauthoryear{Bertelli et al.}{1994}] {bertelli1994}
Bertelli G., Bressan A., Chiosi C., Fagotto F., Nasi E., 1994, A\&AS, 106, 275

\bibitem[\protect\citeauthoryear{Bertin \& Varri}{2008}] {bertin2008}
Bertin G., Varri A.~L., 2008, ApJ, 689, 1005

\bibitem[\protect\citeauthoryear{Bilir et al.}{2006}] {Bilir2006}
 Bilir S., Karaali S., Gilmore G., 2006, MNRAS, 366, 1295
 
\bibitem[\protect\citeauthoryear{Bilir et al.}{2008}] {Bilir2008}
Bilir S., Cabrera-Lavers A., Karaali S., Ak S., Yaz E., L{\'o}pez-Corredoira M., 2008, PASA, 25, 69

\bibitem[\protect\citeauthoryear{Bilir, et al.}{2012}]{Bilir2012} 
Bilir S., Karaali S., Ak, S. et al., 2012, MNRAS, 421, 3362

\bibitem[\protect\citeauthoryear{Binney \& Tremaine}{1987}] {binney1987}
Binney, J., Tremaine, S. 1987, Galactic Dynamics (Princeton, NJ: Princeton Univ. Press)

\bibitem[\protect\citeauthoryear{Blackwell \& Bergemann}{1987}] {blackwell2007}
Blackwell-Whitehead R., Bergemann M., 2007, A\&A, 472, L43

\bibitem[\protect\citeauthoryear{Boeche \&  Grebel}{2016}] {boeche2016}
Boeche C., Grebel E.~K., 2016, A\&A, 587, A2 

\bibitem[\protect\citeauthoryear{Boesgard et al.}{2011}] {boesgard2011} 
Boesgard A.~M., Rich J.~A., Levesque E.~M., Bowler B.~P., 2011, ApJ, 743, 140

\bibitem[\protect\citeauthoryear{Bovy}{2015}] {bovy2015}
Bovy J., 2015, ApJS, 216, 29 

\bibitem[\protect\citeauthoryear{Bressan et al.}{2012}] {bressan2012}
Bressan A., Marigo P., Girardi L., Salasnich B., Dal Cero C., Rubele S., Nanni A., 2012, MNRAS, 427, 127

\bibitem[\protect\citeauthoryear{Burris et al.}{2000}]{burris2000} 
Burris D.~L., Pilachowski C.~A., Armandroff T.~E., Sneden C., Cowan J. J., Roe H., 2000, ApJ, 544, 302

\bibitem[\protect\citeauthoryear{Buser et al.}{1999}] {buser1999}
Buser R., Rong J., Karaali S., 1999, A\&A, 348, 98

\bibitem[\protect\citeauthoryear{Cabrera-Lavers et al.}{2007}] {Cabrera-Lavers07} Cabrera-Lavers A., Bilir S., Ak S., Yaz E., L{\'o}pez-Corredoira M., 2007, A\&A, 464, 565

\bibitem[\protect\citeauthoryear{Carney et al.}{1997}] {carney1997}
Carney B. W., Wright J. S., Sneden C., Laird J.~B., Aguilar L.~A., Latham D.~W., 1997, AJ, 114, 363

\bibitem[\protect\citeauthoryear{Carretta et al.}{2000}]{carretta2000}
Carretta E., Gratton R.~G., Clementini G., Pecci F.~F., 2000, ApJ, 533, 215

\bibitem[\protect\citeauthoryear{Carretta et al.}{2009}]{carretta2009} 
Carretta E., Bragaglia A., Gratton R. et al., 2009, \aap, 508, 695

\bibitem[\protect\citeauthoryear{Casagrande et al.}{2011}]{casagrande2011}
Casagrande L., Sch\"{o}nrich R., Asplund M., Cassisi S., Ramirez I., Melendez J., Bensby T., Feltzing S., 2011, A\&A, 530, A138

\bibitem[\protect\citeauthoryear{Castelli \& Kurucz}{2004}]{cas04} 
Castelli F., Kurucz R.~L., 2004, preprint (arXiv:astro-ph/0405087) 

\bibitem[\protect\citeauthoryear{Catelan et al.}{2002}]{catelan2002} 
Catelan M., Borissova J., Ferraro F. R., Corwin T. M., Smith H. A., Kurtev R., 2002, AJ, 124, 364

\bibitem[\protect\citeauthoryear{Cayrel et al.}{2004}] {caryel2004}
Cayrel R. Depagne E., Spite M. et al., 2004, A\&A, 416, 1117

\bibitem[\protect\citeauthoryear{Cezario et al.}{2013}]{Cezario2013} 
Cezario E., Coelho P.~R.~T., Alves-Brito A., Forbes D.~A., Brodie J.~P., 2013, A\&A, 549, A60

\bibitem[\protect\citeauthoryear{Chen et al.} {2000}] {chen2000}
Chen Y.~Q., Nissen P. E., Zhao G., Zhang H.~W., Benoni T. 2000, A\&AS, 141, 491

\bibitem[\protect\citeauthoryear{Chen et al.}{2001}] {chen2001}
Chen H.~W., Lanzetta K. M., Webb J.~K., Barcons X., 2001, ApJ, 559, 654

\bibitem[\protect\citeauthoryear{Colucci et al.}{2017}] {colucci2017}
Colucci J.~E., Bernstein, R.~A., McWilliam, A., 2017, ApJ, 834, 105

\bibitem[\protect\citeauthoryear{Coskunoglu et al.}{2011}] {coskunoglu2011}
Co\c{s}kuno\u{g}lu B. et al., 2011, MNRAS, 412, 1237

\bibitem[\protect\citeauthoryear{De Angeli, et al.}{2005}]{deangeli2005} 
De Angeli F. et al., 2005, AJ, 130, 116

\bibitem[\protect\citeauthoryear{Decressin et al.}{2007}]{Decressin2007} 
Decressin, T., Meynet, G., Charbonnel, C., et al.\ 2007, \aap, 464, 1029

\bibitem[\protect\citeauthoryear{de la Fuente Marcos \& de la Fuente Marcos}{2019}] {dela2019}
de la Fuente Marcos R., de la Fuente Marcos C., 2019, A\&A, 627, 104

\bibitem[\protect\citeauthoryear{De Marchi \& Pulone}{2007}] {demarchi2007}
De Marchi G., Pulone L., 2007, A\&A, 467, 107

\bibitem[\protect\citeauthoryear{Siva et al.}{2015}] {silva2015}
De Silva G.~M. et al., 2015, MNRAS, 449, 2604

\bibitem[\protect\citeauthoryear{Den Hartog et al.}{2003}] {denhartog2003}
Den Hartog E.~A., Lawler J.~E., Sneden C., Cowan J.~J., 2003, ApJS, 148, 543

\bibitem[\protect\citeauthoryear{Den Hartog et al.}{2011}] {denhartog2011}
Den Hartog E.~A., Lawler J.~E., Sobeck J.~S., Sneden C., Cowan J.~J., 2011, ApJS, 194, 35

\bibitem[\protect\citeauthoryear{Dobrovolskas et al.}{2013}] {dobrovolskas2013}
Dobrovolskas V., Kucinskas A., Steffen M., Ludwig H. G., Prakapavicius D., Klevas J., Caffau E., Bonifacio P., 2013, A\&A, 559, A102

\bibitem[\protect\citeauthoryear{Feltzing et al.}{2001}] {feltzing2001}
Feltzing S., Holmberg J., Hurley J. R., 2001, A\&A, 911, 924

\bibitem[\protect\citeauthoryear{Fishlock et al.}{2017}] {fishlock2017}
Fishlock C.~K., Yong D., Karakas A.~I. et al., 2017, MNRAS, 466, 4672

\bibitem[\protect\citeauthoryear{Fulbright}{2000}] {fulbright2000}
Fulbright J.~P., 2000, AJ, 120, 1841

\bibitem[\protect\citeauthoryear{Fuhr \& Wiese}{2006}] {fuhr2006}
Fuhr J.~R., Wiese W.~L., 2006, J. Phys. Chem. Ref. Data, 35, 4

\bibitem[\protect\citeauthoryear{Fuhrmann et al.}{1997}] {fhurmann1997}
Fuhrmann K., Pfeiffer M., Frank C., Reetz J., Gehren T., 1997, A\&A, 323, 909

\bibitem[\protect\citeauthoryear{Fuhrmann}{1998}] {fhurmann1998}
Fuhrmann K., 1998, A\&A, 338, 161 

\bibitem[\protect\citeauthoryear{Gaia Collaboration}{2018}] {gaia2018}
Gaia Collaboration, Katz D. et al., 2018, A\&A, 616, 11

\bibitem[\protect\citeauthoryear{Ge et al.} {2016}] {ge2016}
Ge Z. S., Bi S. L., Chen Y. Q., Li T. D., Zhao J. K., Liu K., Ferguson J. W., Wu Y. Q., 2016, ApJ, 833, 161 

\bibitem[\protect\citeauthoryear{Gehren et al.} {2006}] {gehren2006}
Gehren T., Shi, J.~R., Zhang H.~W., Zhao G., Korn A.~J., 2006, A\&A, 451, 1065

\bibitem[\protect\citeauthoryear{Gilmore et al.}{2012}] {gilmore2012}
Gilmore et al., 2012, The Messenger, 147, 25

\bibitem[\protect\citeauthoryear{Gratton et al.} {2003}] {gratton2003}
Gratton R.~G., Carretta E., Claudi R., Lucatello S., Barbieri M., 2003, A\&A, 404, 187

\bibitem[\protect\citeauthoryear{Gratton et al.}{2004}]{Gratton2004} 
Gratton, R., Sneden, C., \& Carretta, E.\ 2004, \araa, 42, 385

\bibitem[\protect\citeauthoryear{Gratton et al.}{2012}]{Gratton2012} 
Gratton, R.~G., Carretta, E., \& Bragaglia, A.\ 2012, \aapr, 20, 50

\bibitem[\protect\citeauthoryear{Grillmair}{2006}] {grillmair2006}
Grillmair C.~J., 2006, ApJ, 651, 29

\bibitem[\protect\citeauthoryear{Guctekin et al.}{2019}] {guctekin2019}
G\"{u}\c ctekin S., Bilir S., Karaali S., Plevne O., Ak S., 2019, AdSpR, 63, 1360

\bibitem[\protect\citeauthoryear{Hannaford et al.}{1982}] {hannaford1982}
Hannaford P., Lowe R.~M., Grevesse N., Bi\'{e}mont E., 1982, ApJ, 261, 736

\bibitem[\protect\citeauthoryear{Hansen et al.} {2012}] {hansen2012}
Hansen C.~J., Primas F., Hartman H. et al., 2012, A\&A, 545, A31

\bibitem[\protect\citeauthoryear{Harris} {1996}] {harris1996}
Harris W.~E, 1996, AJ, 112, 1487 

\bibitem[\protect\citeauthoryear{Hawkins et al.}{2016}] {hawkins2016}
Hawkins K., Jofre P., Heiter U. et al., 2016, A\&A, 592, A70

\bibitem[\protect\citeauthoryear{Heijmans et al.}{2012}] {heijmans2012}
Heijmans J., Asplund M., Barden S. et al., 2012, Proceedings of the SPIE, 8446, 84460W

\bibitem[\protect\citeauthoryear{Heiter et al.}{2015}] {heiter2015}
Heiter U., Jofre P., Gustafsson B., Korn A.~J., Soubiran C., Thevenin F., 2015, A\&A, 582, A49

\bibitem[\protect\citeauthoryear{Helmi et al.}{2017}] {helmi2017}
Helmi A. et al., 2017, A\&A, 598, 58

\bibitem[\protect\citeauthoryear{Helmi et al.}{2018}]{helmi2018} Helmi A., van Leeuwen F. et al.\ 2018, \aap, 616, A12

\bibitem[\protect\citeauthoryear{Ibukiyama \& Arimoto}{2002}] {ibukiyama2002}
Ibukiyama A., Arimoto N., 2002, A\&A, 394, 927

\bibitem[\protect\citeauthoryear{Isaacson \& Fischer}{2010}] {isaacson2010}
Isaacson H., Fischer D., 2010, ApJ, 725, 875

\bibitem[\protect\citeauthoryear{Ishigaki et al.}{2012}] {ishigaki2012}
Ishigaki M.~N., Chiba M., Aoki W., 2012, ApJ, 753, 64

\bibitem[\protect\citeauthoryear{Ishigaki et al.}{2013}] {ishigaki2013}
Ishigaki M.~N., Aoki W., Chiba M., 2013, ApJ, 771, 67 

\bibitem[\protect\citeauthoryear{Israelian et al.}{1998}] {israelian1998}
Israelian G., L\'{o}pez R.~J.~G., Rebolo R., 1998, ApJ, 507, 805

\bibitem[\protect\citeauthoryear{Jehin et al}{1999}] {jehin1999}
Jehin E., Magain P., Neuforge C., Noels A., Parmentier G., Thoul A.~A., 1999, A\&A, 341, 241

\bibitem[\protect\citeauthoryear{Jofre et al.}{2014}] {jofre2014}
Jofre P., Heiter U., Soubiran C. et al., 2014, A\&A, 564, A133

\bibitem[\protect\citeauthoryear{Jorgensen \& Lindegren}{2005}] {jorgensen2005}
Jorgensen B.~R., Lindegren L., 2005, A\&A, 436, 127

\bibitem[\protect\citeauthoryear{Johnson \& Soderblom}{1987}] {jhonson1987}
Johnson D.~R.~H., Soderblom D.~R. 1987, AJ, 93, 864

\bibitem[\protect\citeauthoryear{Karaali, et al.}{2003}]{Karaali2003} 
Karaali S., Ak S.~G., Bilir S., Karata{\textcommabelow s} Y., Gilmore G., 2003, MNRAS, 343, 1013

\bibitem[\protect\citeauthoryear{Karaali, Yaz G{\"o}k{\c{c}}e \& Bilir}{2016}]{Karaali2016} 
Karaali S., Yaz G{\"o}k{\c{c}}e E., Bilir S., 2016, Ap\&SS, 361, 354

\bibitem[\protect\citeauthoryear{Karaali et al.}{2019}] {Karaali2019}
Karaali S., Bilir S.,  G\"ok\c ce E. Yaz, Plevne O., 2019, PASA, 36, 40 

\bibitem[\protect\citeauthoryear{Kelleher \& Podobedova}{2008a}]{Kelleher08a}
Kelleher D.~E., Podobedova L.~I., 2008, Journal of Physical and Chemical Reference Data, 37, 267

\bibitem[\protect\citeauthoryear{Kelleher \& Podobedova}{2008b}]{Kelleher08b}
Kelleher D.~E., Podobedova L.~I., 2008, Journal of Physical and Chemical Reference Data, 37, 1285

\bibitem[\protect\citeauthoryear{Kepley et al.}{2007}] {kepley2007}
Kepley A.~A., Morrison H.~L., Helmi A. et al., 2007, AJ, 134, 1579

\bibitem[\protect\citeauthoryear{Klement}{2010}] {klement2010}
Klement R.~J., 2010, A\&ARv, 18, 567

\bibitem[\protect\citeauthoryear{Klose et al.}{2002}] {klose2002}
Klose J.~Z., Fuhr J.~R., Wiese W.~L., 2002, Journal of Physical and Chemical Reference Data, 31, 217

\bibitem[\protect\citeauthoryear{Koch \& McWilliam}{2014}] {McWilliam2014}
Koch A., McWilliam A., 2014, A\&A,565, A23 

\bibitem[\protect\citeauthoryear{Koleva, et al.}{2008}]{Koleva2008}
Koleva M., Prugniel P., Ocvirk P., Le Borgne D., Soubiran C., 2008, MNRAS, 385, 1998

\bibitem[\protect\citeauthoryear{Kraft et al.}{1997}]{kraft997} 
Kraft, R.~P., Sneden, C., Smith, G.~H., et al.\ 1997, \aj, 113, 279

\bibitem[\protect\citeauthoryear{Kurucz et al.}{1984}]{kurucz84}
Kurucz R.~L., Furenlid I., Brault J., Testerman L., 1984, in Kurucz R. L., Furenlid I. Brault J. Testerman L., eds, Solar Flux Atlas from 296
to 1300 nm. National Solar Observatory, Sunspot, NM

\bibitem[\protect\citeauthoryear{Kruijssen \& Mieske}{2009}] {kruijssen2009}
Kruijssen J.~M.~D., Mieske S., 2009, A\&A, 500, 785

\bibitem[\protect\citeauthoryear{Lai et al.}{2008}] {lai2008}
Lai D. K., Bolte M., Johnson J. A., Lucatello S., Heger A., Woosley S.~E., 2008, ApJ, 681, 1524

\bibitem[\protect\citeauthoryear{Lambert \& Reddy}{2004}]{lambert2004}
Lambert D.~L., Reddy B.~E., 2004, MNRAS, 349, 757

\bibitem[\protect\citeauthoryear{Larsen al.}{2017}] {larsen2017}
Larsen S.~S., Brodie J.~P., Strader J., 2017, A\&A, 601, A96

\bibitem[\protect\citeauthoryear{Lawler et al}{2006}] {lawler2006}
Lawler J.~E., Den Hartog E.~A., Sneden C., Cowan J.~J., 2006, ApJS, 162, 227

\bibitem[\protect\citeauthoryear{Lawler et al}{2009}] {lawler2009}
Lawler J.~E., Sneden C., Cowan J.~J., Ivans I.~I., Den Hartog E.~A., 2009, ApJS, 182, 51

\bibitem[\protect\citeauthoryear{Lawler et al.}{2013}] {lawler2013}
Lawler J.~E., Guzman A., Wood M.~P., Sneden C., Cowan J.~J., 2013, ApJS, 205, 11

\bibitem[\protect\citeauthoryear{Lawler et al.}{2014}] {lawler2014}
Lawler J.~E., Wood M.~P., Den Hartog E.~A., Feigenson T., Sneden C., Cowan J.~J., 2014, ApJS, 215, 20

\bibitem[\protect\citeauthoryear{Lawler et al.}{2015}] {lawler2015}
Lawler J.~E., Sneden C., Cowan J.~J., 2015, ApJS, 220, 13

\bibitem[\protect\citeauthoryear{Leigh \& Sills}{2011}] {leigh2011}
Leigh N.~W.~C., Sills A., 2011, MNRAS, 410, 2370

\bibitem[\protect\citeauthoryear{Leigh et al.}{2016}] {leigh2016}
Leigh N.~W.~C., Stone N.~C., Geller A.~M., Shara M.~M., Muddu H., Solano--Oropeza D., Thomas Y., 2016, MNRAS, 463, 3311

\bibitem[\protect\citeauthoryear{Ljung et al.}{2006}] {ljung2006}
Ljung G., Nilsson H., Asplund M., Johansson S., 2006, A\&A, 456, 1181

\bibitem[\protect\citeauthoryear{Li et al.}{2019}] {li2019}
Li H., Du C., Liu S., Donlon T., Newberg H.~J., 2019, ApJ, 874, 74

\bibitem[\protect\citeauthoryear{Lind et al.}{2012}] {lind2012}
Lind K., Bergemann M., Asplund M., 2012, MNRAS, 427, 50

\bibitem[\protect\citeauthoryear{Luck}{2017}] {luck2017}
Luck R.~E., 2017, AJ, 153, 21

\bibitem[\protect\citeauthoryear{Majevski}{1993}] {majevski1993}
Majewski S.~R., 1993, ARA\&A, 31, 575

\bibitem[\protect\citeauthoryear{Mashonkina \& Gehren}{2000}] {mashonkina2000}
Mashonkina L., Gehren T., 2000, A\&A, 364, 249

\bibitem[\protect\citeauthoryear{Mashonkina et al.}{2008}] {mashonkina2008}
Mashonkina L., Zhao G., Gehren T. et al., 2008, A\&A, 478, 529

\bibitem[\protect\citeauthoryear{Massari, et al.}{2016}]{massari2016} 
Massari D., et al., 2016, A\&A, 586, A51

\bibitem[\protect\citeauthoryear{McLaughlin \& van der Marel}{2005}] {mclaughlin2005}
McLaughlin D.~E., van der Marel R. P., 2005, ApJS, 161, 304

\bibitem[\protect\citeauthoryear{Melendez et al.}{2010}] {melendez2010}
Melendez J., Casagrande L., Ramirez I., Asplund M., Schuster W.~J., 2010, A\&A, 515, L3

\bibitem[\protect\citeauthoryear{Mihalas \& Binney}{1981}] {mihalas1981} 
Mihalas D., Binney J. 1981, Galactic Astronomy: Structure and Kinematics (2nd ed.; San Francisco, CA: Freeman), 608 

\bibitem[\protect\citeauthoryear{Milone et al.}{2012}] {milone2012}
Milone A. P., Piotto G., Bedin L. R. et al., 2012, A\&A, 540, A16

\bibitem[\protect\citeauthoryear{Mints \& Hekker}{2017}] {mints2017}
Mints A., Hekker S., 2017, A\&A, 604, 108

\bibitem[\protect\citeauthoryear{Mishenina \& Kovtyukh}{2001}] {mishenina2001}
Mishenina T. V., Kovtyukh V.~V., 2001, A\&A, 370, 951

\bibitem[\protect\citeauthoryear{Mishenina et al.}{2003}] {mishenina2003}
Mishenina T.~V., Kovtyukh V.~V., Korotin S.~A., Soubiran C., 2003, Azh, 80, 5

\bibitem[\protect\citeauthoryear{Mishenina et al.}{2004}] {mishenina2004}
Mishenina T.~V., Soubiran C., Kovtyukh V.~V., Korotin S.~A., 2004, A\&A, 418, 551

\bibitem[\protect\citeauthoryear{Mishenina et al.}{2011}] {mishenina2011}
Mishenina T.~V., Gorbaneva, T.~I., Basak N.~Yu., Soubiran C., Kovtyukh V.~V., 2011, Azh, 88, 8

\bibitem[\protect\citeauthoryear{Miyamoto \& Nagai}{1975}] {miyamoto1975}
Miyamoto M., Nagai R., 1975, PASJ, 27, 533

\bibitem[\protect\citeauthoryear{Moultaka et al.}{2004}] {moultaka2004}
Moultaka J., Ilovaisky S.~A., Prugniel P., Soubiran C., 2004, PASP, 116, 693

\bibitem[\protect\citeauthoryear{Navarro et al.}{1996}] {navarro1996}
Navarro J.~F., Frenk C.~S., White S.~D.~M., 1996, ApJ, 462, 563

\bibitem[\protect\citeauthoryear{Nissen et al.}{1994}] {nissen1994}
Nissen P.~E., Gustafsson B., Edvardsson B., Gilmore G., 1994, A\&A, 285, 440

\bibitem[\protect\citeauthoryear{Nissen et al.}{2000}] {nissen2000}
Nissen P.~E., Chen Y.~Q., Schuster W.~J., Zhao G., 2000, A\&A, 353, 722

\bibitem[\protect\citeauthoryear{Nissen et al.}{2002}] {nissen2002}
Nissen P.~E., Primas F., Asplund M., Lambert D.~L., 2002, A\&A, 390, 235

\bibitem[\protect\citeauthoryear{Nissen}{2004}] {nissen2004}
Nissen P.~E., 2004, in Carnegie Observatories Astrophysics Ser. Vol. 4: Origin and Evolution of the Elements, ed. McWilliam A. \& Rauch M. (Cambridge:
Cambridge Univ. Press)

\bibitem[\protect\citeauthoryear{Nissen \& Schuster}{2010}] {nissen2010}
Nissen P.~E., Schuster W.~J., 2010, A\&A, 511, L10

\bibitem[\protect\citeauthoryear{Nissen \& Schuster}{2011}] {nissen2011}
Nissen P.~E., Schuster W.~J., 2011, A\&A, 530, A15

\bibitem[\protect\citeauthoryear{\"{O}nal Ta\c{s} et al.}{2018}] {onaltas2018}
\"{O}nal Ta\c{s} \"{O}, Bilir S., Plevne O., 2018, Ap\&SS, 363, 35

\bibitem[\protect\citeauthoryear{Pace}{2013}] {pace2013}
Pace G., 2013, A\&A, 551, 8

\bibitem[\protect\citeauthoryear{Pehlivan Rhodin et al.}{2017}] {pehlivan2017}
Pehlivan Rhodin, A., Hartman, H., Nilsson, H., Jonsson, P. 2017, A\&A, 598, A102

\bibitem[\protect\citeauthoryear{Pereira et al.}{2017}] {pereira2017} 
Pereira, C.~B., Smith, V.~V., Drake, N.~A., et al.\ 2017, \mnras, 469, 774

\bibitem[\protect\citeauthoryear{Pereira et al.}{2019}] {pereira2019}
Pereira C.~B., Drake N.~A., Roig F., 2019, MNRAS, 488, 482

\bibitem[\protect\citeauthoryear{Peterson}{2013}] {peterson2013}
Peterson R.~C., 2013, ApJ, 768, L13

\bibitem[\protect\citeauthoryear{Pilachowski et al.}{1996}] {pilachowski1996}
Plachowski C.~A., Sneden C., Kraft R.~P., 1996, AJ, 111, 1689

\bibitem[\protect\citeauthoryear{Plevne et al.}{2020}] {Plevne20}
Plevne O., \"{O}nal Ta\c{s} \"{O}, Bilir S., Seabroke G.M, 2020, ApJ, 893, 108, 21

\bibitem[\protect\citeauthoryear{Porachaska et al.}{2000}] {porachaska2000}
Prochaska J.~X., McWilliam A., 2000, ApJ, 537, L57

\bibitem[\protect\citeauthoryear{Prugniel \& Soubiran}{2001}] {prugniel2001}
Prugniel P., Soubiran C., 2001, A\&A, 369, 1048

\bibitem[\protect\citeauthoryear{Prugniel et al.}{2007}]{prugniel2007} 
Prugniel P., Soubiran C., Koleva M., Le Borgne D., 2007, yCat, III/251

\bibitem[\protect\citeauthoryear{Ramirez et al.}{2012}] {ramirez2012}
Ramirez I., Fish J. R., Lambert D. L., Allende Prieto C., 2012, ApJ, 756, 46 

\bibitem[\protect\citeauthoryear{Ramirez et al.}{2013}] {ramirez2013}
Ramirez I., Allende Prieto C., Lambert D.~L., 2013, ApJ, 763, 78

\bibitem[\protect\citeauthoryear{Reddy et al.}{2003}] {reddy2003}
Reddy B.~E., Tomkin J., Lambert D.~L., Allende Prieto C.~A., 2003, MNRAS, 340, 304

\bibitem[\protect\citeauthoryear{Reddy et al.}{2006}] {reddy2006}
Reddy B.~E., Lambert D.~L., Allende Prieto C., 2006, MNRAS, 367, 1329

\bibitem[\protect\citeauthoryear{Robin et al.}{1996}] {robin1996}
Robin A.~C., Haywood M., Creze M., Ojha D.~K., Bienayme O., 1996, A\&A, 305, 125

\bibitem[\protect\citeauthoryear{Roederer et al.}{2014}] {roederer2014}
Roederer I.~U., Preston G.~W., Thompson I.~B. et al., 2014, AJ, 147, 136

\bibitem[\protect\citeauthoryear{Roederer \& Thompson}{2015}] {roederer2015}
Roederer I.~U., Thompson I.~B., 2015, MNRAS, 449, 3889

\bibitem[\protect\citeauthoryear{Sahin \& Lambert}{2009}] {sahin2009}
\c{S}ahin T., Lambert D.~L., 2009, MNRAS, 398, 1730

\bibitem[\protect\citeauthoryear{Sahin et al.}{2011}] {sahin2011}
\c{S}ahin T., Lambert D.~L., Klochkova V.~G., Tavolganskaya N.~S., 2011, MNRAS, 410, 612

\bibitem[\protect\citeauthoryear{Sahin et al.}{2016}] {sahin2016}
\c{S}ahin T., Lambert D.~L., Klochkova V.~G., Panchuk V.~E., 2016, MNRAS, 461, 4071

\bibitem[\protect\citeauthoryear{Sahin}{2017}] {sahin2017}
\c{S}ahin T., 2017, Turkish Journal of Physics, 41, 367

\bibitem[\protect\citeauthoryear{Sahlholdt et al.}{2018}] {sahlholdt2018}
Sahlholdt C.~L., Feltzing S., Lindegre L., Church R.~P., 2018, MNRAS, 482, 895

\bibitem[\protect\citeauthoryear{Sakari et al.}{2018}] {sakari2018} 
Sakari, C.~M., Placco, V.~M., Farrell, E.~M., et al.\ 2018, \apj, 868, 110

\bibitem[\protect\citeauthoryear{Salaris \& Weiss}{1998}] {salaris1998}
Salaris M., Weiss A., 1998, A\&A, 335, 943 

\bibitem[\protect\citeauthoryear{Sesar et al.}{2007}] {sesar2007}
Sesar B. et al., 2007, AJ, 134, 2236

\bibitem[\protect\citeauthoryear{Schonrich \& Bergemann}{2014}] {schonrich2014}
Sch\"{o}nrich R., Bergemann M., 2014, MNRAS, 443, 698

\bibitem[\protect\citeauthoryear{Sch{\"o}nrich et al.}{2019}]{schonrich2019} Sch{\"o}nrich R., McMillan P., Eyer, L.\ 2019, \mnras, 487, 3568

\bibitem[\protect\citeauthoryear{Schuster \& Nissen}{1989}] {schuster1989}
Schuster W. J., Nissen P. E., 1989, A\&A, 222, 69

\bibitem[\protect\citeauthoryear{Shi et al.}{2009}] {shi2009}
Shi J.~R., Gehren T., Mashonkina L., Zhao G., 2009, A\&A, 503, 533

\bibitem[\protect\citeauthoryear{Sitnova et al.}{2015}] {sitnova2015}
Sitnova T., Zhao G., Mashonkina L. et al., 2015, ApJ, 808, 148

\bibitem[\protect\citeauthoryear{Sneden}{1973}] {sneden1973}
Sneden C., 1973, Ph.D. thesis, Univ. Texas, Austin

\bibitem[\protect\citeauthoryear{Sobeck et al.}{2007}] {sobeck2007}
Sobeck J.~S., Lawler J. E., Sneden C., 2007, ApJ, 667, 1242

\bibitem[\protect\citeauthoryear{Saubiran et al.}{2003}] {soubiran200}
Soubiran C., Bienayme O., Siebert A., 2003, A\&A, 398, 151

\bibitem[\protect\citeauthoryear{Spite et al.}{2017}] {spite2017}
Spite M., Peterson R. C., Gallagher A. J., Barbuy B., Spite F., 2017, A\&A, 600, A26

\bibitem[\protect\citeauthoryear{Steinmetz et al.}{2006}] {steinmetz2006}
Steinmetz M., Zwitter T., Siebert A. et al., 2006, AJ, 132, 1645

\bibitem[\protect\citeauthoryear{Valenti \& Fischer}{2005}] {valenti2005}
Valenti J.~A., Fischer D.~A., 2005, ApJS, 159, 141

\bibitem[\protect\citeauthoryear{VandenBerg et al.}{2013}] {vandenberg2013}
VandenBerg D.~A., Brogaard K., Leaman R., Casagrande L., 2013, ApJ, 775, 134 

\bibitem[\protect\citeauthoryear{VandenBerg et al.}{2014}] {vandenberg2014}
VandenBerg D.~A., Bond H.~E., Nelan E.~P., Nissen P.~E., Schaefer G.~H., Harmer D., 2014, ApJ, 792, 110

\bibitem[\protect\citeauthoryear{Vasiliev}{2019}] {vasiliev2019}
Vasiliev E., 2019, MNRAS, 484, 2832

\bibitem[\protect\citeauthoryear{van Leeuwen}{2007}] {leeuwen2007}
van Leeuwen F., 2007, A\&A, 474, 653

\bibitem[\protect\citeauthoryear{Ventura et al.}{2001}]{ventura2001} 
Ventura, P., D'Antona, F., Mazzitelli, I., et al.\ 2001, \apjl, 550, L65

\bibitem[\protect\citeauthoryear{Webb \& Leigh}{2015}] {webb2015}
Webb J.~J., Leigh N.~W.~C., 2015, MNRAS, 453, 3278

\bibitem[\protect\citeauthoryear{Wood et al.}{2013}] {wood2013}
Wood M.~P., Lawler J.~E., Sneden C., Cowan J.~J., 2013, ApJS, 208, 27

\bibitem[\protect\citeauthoryear{Wood et al.}{2014}] {wood2014}
Wood M.~P., Lawler J.~E., Sneden C., Cowan J.~J., 2014, ApJS, 211, 20

\bibitem[\protect\citeauthoryear{Worley \& Cottrel}{2010}] {worley2010}
Worley C.~C., Cottrell P.~L., 2010, MNRAS, 406, 2504

\bibitem[\protect\citeauthoryear{Yan et al.}{2016}] {yan2016}
Yan H.~L., Shi J.~R., Nissen P.~E., Zhao G., 2016, A\&A 585, A102

\bibitem[\protect\citeauthoryear{Yanni et al.}{2009}] {yanni2009}
Yanny B., Rockosi C., Newberg H.~J. et al., 2009, AJ, 137, 4377

\bibitem[\protect\citeauthoryear{Yaz G{\"o}k{\c{c}}e, et al.}{2017}] {Yaz17} 
Yaz G{\"o}k{\c{c}}e E., Bilir S., Karaali S., Plevne O., 2017, Ap\&SS, 362, 185

\bibitem[\protect\citeauthoryear{Zhao \& Gehren}{2000}]{zhao2000}
Zhao G., Gehren T., 2000, A\&A, 362, 1077

\bibitem[\protect\citeauthoryear{Zhang \& Zhao}{2005}] {zhang2005}
Zhang H.~W., Zhao G., 2005, MNRAS, 364, 712

\bibitem[\protect\citeauthoryear{Zhang et al.}{2008}] {zhang2008}
Zhang H.~W., Gehren T, Zhao G., 2008, preprint (arXiv:0802.2609)

\bibitem[\protect\citeauthoryear{Zhao et al.}{2012}] {zhao2012}
Zhao G., Zhao Y.~H., Chu Y.~Q., Jing Y.~P., Deng L.~C., 2012, Research in Astronomy and Astrophysics, 12, 723

\bibitem[\protect\citeauthoryear{Zhao et al.}{2016}] {zhao2016}
Zhao G., Mashonkina L., Yan H.~L. et al., 2016, ApJ, 833, 225 


\end{thebibliography}
\end{document}